\documentclass[aps,prx,reprint,preprintnumbers,superscriptaddress,nofootinbib,longbibliography,floatfix]{revtex4-2}
\pdfoutput=1
\usepackage{rotating}
\usepackage{array}
\usepackage{amsmath}
\usepackage[normalem]{ulem}
\usepackage{slashed}
\usepackage{booktabs}
\usepackage[pdftex,table]{xcolor}
\usepackage{units}
\usepackage{xfrac}
\usepackage{mathtools}
\usepackage{empheq}
\usepackage[]{units}
\usepackage{multirow}
\usepackage{amssymb}
\usepackage{url}
\usepackage{comment}
\usepackage{physics}
\usepackage{color,soul}
\usepackage{bbm}
\usepackage[caption=false]{subfig}
\usepackage{adjustbox}

\usepackage{hyperref}
\hypersetup{
  colorlinks=true,
  citecolor=blue,
  linkcolor=blue,
  urlcolor=blue
}

\newcommand{\xbf}{\textbf{x}}

\begin{document}

\title{Improving Generative Model-based Unfolding with Schr\"{o}dinger Bridges}

\author{Sascha Diefenbacher}
\email{sdiefenbacher@lbl.gov}
\affiliation{Physics Division, Lawrence Berkeley National Laboratory, Berkeley, CA 94720, USA}

\author{Guan-Horng Liu}
\email{ghliu@gatech.edu}
\affiliation{Autonomous Control and Decision Systems Laboratory, Georgia Institute of Technology, Atlanta, GA 30332, USA}

\author{Vinicius Mikuni}
\email{vmikuni@lbl.gov}
\affiliation{National Energy Research Scientific Computing Center, Berkeley Lab, Berkeley, CA 94720, USA}

\author{Benjamin Nachman}
\email{bpnachman@lbl.gov}
\affiliation{Physics Division, Lawrence Berkeley National Laboratory, Berkeley, CA 94720, USA}
\affiliation{Berkeley Institute for Data Science, University of California, Berkeley, CA 94720, USA}

\author{Weili Nie}
\email{wnie@nvidia.com}
\affiliation{Machine Learning Research Group, NVIDIA Research}

\begin{abstract}
    Machine learning-based unfolding has enabled unbinned and high-dimensional differential cross section measurements.  Two main approaches have emerged in this research area: one based on discriminative models and one based on generative models.  The main advantage of discriminative models is that they learn a small correction to a starting simulation while generative models scale better to regions of phase space with little data.  We propose to use Schr\"{o}dinger Bridges and diffusion models to create \textsc{SBUnfold}, an unfolding approach that combines the strengths of both discriminative and generative models.  The key feature of \textsc{SBUnfold} is that its generative model maps one set of events into another without having to go through a known probability density as is the case for normalizing flows and standard diffusion models. We show that \textsc{SBUnfold} achieves excellent performance compared to state of the art methods on a synthetic $Z$+jets dataset.
\end{abstract}

\maketitle


\section{Introduction}
\label{sec:intro}

Correcting detector effects -- called \textit{deconvolution} or \textit{unfolding} -- is the central statistical task in differential cross section measurements in particle, nuclear, and astrophysics.  Classical unfolding methods are based on histograms, which result in binned measurements in a small number of dimensions.  Machine learning has the potential to revolutionize differential cross section measurements by enabling unbinned and high-dimensional measurements.  A number of machine learning-based unfolding techniques have been proposed~\cite{Glazov:2017vni,Datta:2018mwd,bunse2018unification,Ruhe2019MiningFS,Andreassen:2019cjw,Bellagente:2019uyp,1800956,Vandegar:2020yvw,Andreassen:2021zzk,Howard:2021pos,Backes:2022vmn,Arratia:2022wny,Chan:2023tbf,Shmakov:2023kjj,Alghamdi:2023emm} (see also Ref.~\cite{Arratia:2021otl} for an overview) and the OmniFold method~\cite{Andreassen:2019cjw,Andreassen:2021zzk} has recently been applied to studies of hadronic final states with data from H1~\cite{H1:2021wkz,H1prelim-22-031,H1:2023fzk,H1prelim-21-031}, LHCb~\cite{LHCb:2022rky}, CMS~\cite{Komiske:2022vxg}, and STAR~\cite{Song:2023sxb}.

Let $X$ represent\footnote{Upper case letters denote random variables and lower case letters correspond to realizations of those random variables.} an event at detector-level and $Z$ represent the same event at particle-level.  The goal of unfolding is to infer the most likely density $p(z)$ using simulated pairs $(Z,X)$ and observations $x$ from data.  In classical approaches, $X$ and $Z$ are discretized and unfolding proceeds via regularized matrix inversion to approximate the maximum likelihood solution.  The likelihood is a product of Poisson probability mass functions, although most measurements do not directly maximize this likelihood.  For example, one of the most common classical unfolding algorithms is called Lucy-Richardson deconvolution~\cite{1974AJ.....79..745L,Richardson:72} (also known as Iterative Bayesian Unfolding~\cite{DAGOSTINI1995487}) which uses an Expectation-Maximization (EM) algorithm to converge to a maximum likelihood estimator.

In the unbinned case, the likelihood is not known.  One solution is to use the EM algorithm, which is at the core of two maximum likelihood machine learning approaches.  The first is OmniFold, which proceeds as follows\footnote{This integral format is a continuum limit representation.  In practice, the integrals are replaces with sums over examples.}:

\begin{description}
    \item[E step] $\omega_{i+1}(x)=p_\text{data}(x)/\tilde{p}_\text{sim.}(x)$
    \end{description}
    \hspace{2cm}{\color{gray}$\tilde{p}_\text{sim.}(x) \equiv \int \text{d}z \, p_\text{sim.}(x,z)\,\nu_i(z)$}
    \begin{description}
    \item[M step] $\nu_{i+1}(z)=\bar{p}_\text{sim.}(z)/p_\text{sim.}(z)$
\end{description}
\hspace{2cm}{\color{gray}$\bar{p}_\text{sim.}(z)\equiv\int \text{d}x\, p_\text{sim.}(x,z)\,\omega_{i+1}(x)$}\\

\noindent Both the Expectation (E) and Maximization (M) steps are achieved in practice by training classifiers and interpreting the resulting score as the target likelihood ratio, e.g. take samples from $p_\text{data}$ and from $\tilde{p}_\text{sim.}$ (samples from $p_\text{sim.}$ weighted by $\nu$) and train a classifier to distinguish them for the E step.  The final result is the set of simulated events weighted by $\nu$ with probability density $p_\text{sim.}(z)\,\nu(z)$ in the continuum limit.  An alternative approach called IcINN~\cite{Backes:2022vmn} is instead based on generative models:
\begin{description}
    \item[E step] $p_i(z|x)\propto p_\text{sim.}(x|z)\,\nu_i(z)$
    \item[M step] $\nu_{i+1}(z)=\check{p}_\text{sim.}(z)/p_\text{sim.}(z)$
\end{description}
\hspace{2cm}{\color{gray}$\check{p}_\text{sim.}(z)\equiv\int \text{d}x\, p_i(z|x)\,p_\text{data}(x)$}\\

\noindent The E step of the IcINN is achieved by training a generative model (a normalizing flow~\cite{rezende2016variational}) to emulate $p_\text{sim.}(x|z)\,\nu_i(z)$ while the M step uses a classifier as in OmniFold.  The groundwork for the IcINN was laid in Ref.~\cite{1800956}, which focused on the E step, as have other papers using normalizing flows, diffusion models, and generative adversarial networks~\cite{Datta:2018mwd,Bellagente:2019uyp,Shmakov:2023kjj}.  Comparing both the E and M steps of OmniFold and the IcINN, one can see that they are formally the same and thus the EM proof in Ref.~\cite{Andreassen:2019cjw} applies to both approaches.  

While formally both algorithms achieve the same EM algorithm step by step, they have complementary strengths in practice.  One of OmniFold's strengths is that it starts from an existing simulation and if that simulation is close to nature, then the neural networks only have to learn a small correction.  In contrast\footnote{This is true for normalizing flows and traditional diffusion models.  Generative Adversarial Networks (GANs)~\cite{Goodfellow:2014upx} do not require a tractable noise distribution, but are less stable.  Reference~\cite{Howard:2021pos} used a physical latent space for a Varitional Autoencoder (VAE), but VAEs are not state of the art in generative modeling.}, the normalizing flows in IcINN have to map a known probability density (e.g. a multidimensional Gaussian) to the data and these can be quite different.  However, the E step of OmniFold trains directly on data and thus its performance degrades when there are not many events.  In contrast, the E step of the IcINN is trained using simulation and so can better handle the case of fewer events.

We propose an approach called \textsc{SBUnfold} that incorporates strengths of both OmniFold and the IcINN.  We employ a technique called a Schr\"{o}dinger Bridge (SB) using a diffusion model to learn the generative model in the IcINN workflow.  In contrast to a normalizing flow or standard diffusion model, a Schr\"{o}dinger Bridge learns to map one dataset into another without needing to know the probability density of one of the datasets.  Thus, the Schr\"{o}dinger Bridge should ideally learn a small correction while also preserving the E step learning with simulation and not data.  The latter property means that \textsc{SBUnfold} should outperform OmniFold when there are few events in data.  Since OmniFold and the IcINN differ only in the E step, we focus exclusively on the E step of the first iteration for all methods and will refer to IcINN as cINN for disambiguation.

This paper is organized as follows.  Section~\ref{sec:method} introduces Schr\"{o}dinger Bridges and Sec.~\ref{sec:dataset} describes how this can be used for unfolding.  Numerical results from the OmniFold public $Z$+jets dataset are presented in Sec.~\ref{sec:results}.  The paper ends with conclusions and outlook in Sec.~\ref{sec:conclusions}.

\section{Schrodinger Bridge and the connection with Diffusion Models}
\label{sec:method}
Diffusion models have become popular choices for generative modeling due to their large degree of flexibility, stable training, and often competitive results compared to other approaches. The core idea is to design a time-dependent perturbation process that slowly perturbs data towards a tractable noise distribution. The goal of the generative model is to then reverse this process, starting from a noise sample and denoising towards new data observations. From a fixed choice of stochastic differential equation (SDE)
\begin{equation}
    \mathrm{d}\xbf = f(\xbf,t)\mathrm{d}t + g(t)\mathrm{dw},
\end{equation}
described by parameters $f(\xbf,t) \in \mathbb{R}^d$ and $g(t) \in \mathbb{R}$, the evolution over time of the data observation $\xbf \in \mathbb{R}^{d}$ is determined. The same initial data point can undergo different paths due to the additional stochastic term identified by the Wiener process, or Brownian motion,  w$(t)\in\mathbb{R}^d$, often sampled from a normal distribution with the same dimension as the data. The reverse process follows the reverse SDE equation that reads:
\begin{equation}
    \mathrm{d}\xbf = [f(\xbf,t)-g(t)^2\nabla\log p(\xbf,t)]\mathrm{d}t + g(t)\mathrm{d\bar{w}}.
    \label{eq:rsde}
\end{equation}
The only unknown term in the Eq.~\ref{eq:rsde} is the score function, $\nabla\log p(\xbf,t)$, which can be approximated  using denoising score matching~\cite{Song2021ScoreBasedGM} and minimizing the loss function
\begin{equation}
    \mathcal{L}_{SGM} = \frac{1}{2}\mathbb{E}_{\xbf_t,t}\left  \| \epsilon_\theta(\xbf_t,t) -  \sigma_t\nabla_{\xbf_t} \log q(\xbf_t|\xbf)\right\| ^2_2,
    \label{eq:loss_sgm}
\end{equation}
where a neural network with trainable parameters $\theta$ is optimized to approximate the score function of data that have been perturbed by a Gaussian distribution with time-dependent parameters $\alpha_t$ and $\sigma_t$. For this Gaussian perturbation, $q(\xbf_t|\xbf) = \mathcal{N}(\xbf_t,\alpha_t\xbf,\sigma_t^2\textbf{I})$, requiring $f(\xbf,t)$  to be affine with respect $\xbf$ and resulting in a Gaussian noise at the end of the diffusion process. For an arbitrary noise distribution, the corresponding choice of $f(\xbf,t)$ cannot be easily identified, often restricting the family of distributions one is allowed to choose for the diffusion process.  A more general framework that allows the mapping between general distributions corresponds to the solution of the Schr\"{o}dinger Bridge (SB) problem. Initially proposed by Erwin Schr\"{o}dinger~\cite{schrodinger1931umkehrung}, the problem concerned the inference of the trajectories from particles undergoing a diffusion process  where experimental observations of the particle's trajectory are available only at specific time values, fixing the boundary conditions of the problem. The connection with diffusion generative models becomes more clear by considering the following forward and backward SDEs:
\begin{subequations}
    \begin{align}
        \mathrm{d}\xbf &= [f(\xbf,t) + g(t)^2\nabla\log {\Psi}(\xbf, t) ] \mathrm{d}t  + g(t)\mathrm{dw}, 
        \label{eq:fsb}
        \\
        \mathrm{d}\xbf &= [f(\xbf,t) - g(t)^2\nabla\log \hat{\Psi}(\xbf, t) ] \mathrm{d}t  + g(t)\mathrm{d\bar{w}}.
        \label{eq:rsb}
    \end{align} \label{eq:sb-sde}%
\end{subequations}
The wave-functions $\Psi$ and $\hat{\Psi}$ satisfy 
\begin{align*}
    \frac{\partial{\Psi(\xbf,t)}}{\partial t}    &= - \nabla \Psi(\xbf,t)^\intercal f(\xbf,t) - \frac{1}{2} g(t)^2 \Delta \Psi(\xbf,t) \\
    \frac{\partial{\hat{\Psi}(\xbf,t)}}{\partial t}    &= - \nabla \cdot (\hat{\Psi}(\xbf,t) f(\xbf,t)) + \frac{1}{2} g(t)^2 \Delta \hat{\Psi}(\xbf,t)  
    \label{eq:sb-pde}
\end{align*}
with $\Psi(\xbf,0) \hat{\Psi}(\xbf,0) =p_A(\xbf)$ and $\Psi(\xbf,1) \hat{\Psi}(\xbf,1) =p_B(\xbf)$ for  densities $p_A(\xbf)$ and $p_B(\xbf)$. Compared to Eq.~\ref{eq:rsde}, the presence of an additional non-linear term to the drift function, $g(t)^2\nabla\log \hat{\Psi}(\xbf, t)$, enables the diffusion between densities that are not necessarily represented by a standard normal distributions at the end of the diffusion process. Additionally, $\nabla\log \hat{\Psi}(\xbf, t)$ no longer represents the score function of the perturbed data, but is related to it since
\begin{equation}
    \Psi(\xbf,t) \hat{\Psi}(\xbf,t) =q(\xbf,t),
\end{equation}
hence
\begin{equation}
    \nabla\log\Psi(\xbf,t) +  \nabla\log\hat{\Psi}(\xbf,t) = \nabla\log q(\xbf,t).
\end{equation}

While the equations describing the general SB problem have similarities with the standard framework for diffusion generative models, a general strategy to solve the problem is not immediately obvious. The authors of Ref.~\cite{SB_nvidia} propose a tractable solution named \textsc{I$^2$SB}, by considering the presence of pairs of observations in the dataset such that $p(\xbf_a,\xbf_b) = p_A(\xbf_a)p_B(\xbf_b|\xbf_a)$. This assumption also holds for the unfolding methodology where pairs of particle collisions before and after detector interactions are always available in the simulation. With this approximation, the authors of Ref.~\cite{SB_nvidia} have shown that by setting the linear drift term $f(\xbf,t){:=} 0$, the posterior of Eq.~\ref{eq:sb-sde}, $q(\xbf|\xbf_a, \xbf_b)$, has the analytic form:
\begin{align}
    &q(\xbf|\xbf_a, \xbf_b) = \mathcal{N}(\xbf_t;\mu_t(\xbf_a,\xbf_b),\Sigma_t),\\
    \mu_t &= \frac{\bar{\sigma}_t^2}{\bar{\sigma}_t^2 + \sigma^2_t} \xbf_a +
              \frac{\sigma^2_t    }{\bar{\sigma}_t^2 + \sigma^2_t} \xbf_b,\quad
    \Sigma_t = \frac{\sigma_t^2 \bar{\sigma}_t^2}{\bar{\sigma}_t^2 + \sigma^2_t} \cdot I,
    \label{eq:perturb}
\end{align}
with $\sigma^2_t {:=} \int_0^t g^2(\tau) \mathrm{d}\tau$ and $\bar{\sigma}^2_t {:=} \int_t^1 g^2(\tau) \mathrm{d}\tau$. From this expression, given pairs $(\xbf_a,\xbf_b)$, we can directly determine $\xbf_t = \mu_t + \Sigma_t\epsilon$, $\epsilon\sim\mathcal{N}(0,1)^d$ for any time step $t$. The loss function is then identified similar to Eq.~\ref{eq:loss_sgm} as:
\begin{equation}
    \mathcal{L}_\mathrm{I^2SB} = \frac{1}{2}\mathbb{E}_{\xbf_t,t}\left \| \epsilon_\theta(\xbf_t,t) -  \frac{\xbf_t - \xbf_0}{\sigma_t}\right\| ^2_2,
    \label{eq:loss_sb}
\end{equation}
where the right term of the loss function approximates the score function of the backward drift $\nabla\log \hat{\Psi}(\xbf, t)$, which can then be used during sampling to transport samples from $p_B$ to $p_A$.

During sampling, standard recursive samplers like DDPM~\cite{ddpm} can be used with the prediction of the denoised data $\xbf_0^\theta = \xbf_{n+1} - \sigma_{n+1}\epsilon_\theta(\xbf_{n+1},n+1)$ at time step $n < N$ defined as:
\begin{equation}
    \xbf_n \sim q(\xbf_n| \xbf_0^\theta, \xbf_{n+1}), \xbf_N\sim p_B.
\end{equation}

The stochastic solver is able to produce different observations given the same exact inputs, which can then be used for different coverage tests regarding the validity of the generated outputs. A second option is to consider the deterministic case where the posterior distributions are effectively replaced by their means. This particular case is described by the solution of the following ODE:
\begin{equation}
    \mathrm{d}\xbf_t = \mathbf{v}_t(\xbf_t|\xbf_0)\mathrm{d}t = \frac{\beta_t}{\sigma_t^2}(\xbf_t - \xbf_0)\mathrm{d}t \,,
\end{equation}
which describes an optimal transport plan~\cite{peyre2019computational}. In our studies, we observe similar sample quality between the stochastic and deterministic settings and report the results based on the deterministic case for simplicity.

We set $f(\xbf,t){:=}0$ and $g(t)=\sqrt{\beta(t)}$, with $\beta(t)$ the triangular function:
\begin{equation}
    \beta(t) = \begin{cases}
    \beta_{0} + 2 (\beta_{1} - \beta_{0}) t, & 0 \leq t < \frac{1}{2}, \\
    \beta_{1} - 2 (\beta_{1} - \beta_{0}) (t - \frac{1}{2}), & \frac{1}{2} \leq t \leq 1.
\end{cases}
\end{equation}
with $\beta_0 = 10^{-5}$ and $\beta_1 = 10^{-4}$.

\section{Unfolding Methodology}
\label{sec:dataset}

We test \textsc{SBUnfold} using the public dataset from Ref.~\cite{Andreassen:2019cjw}, available on Zenodo~\cite{andreassen_anders_2019_3548091} and briefly summarized in the following.  Proton-proton collisions producing a $Z$ boson are generated at a center-of-mass energy of $\sqrt{s}=14$ TeV.  A non-trivial test of unfolding requires at least two datasets, one that acts as the `data' and one that is the `simulation'.  For the `data', collisions are simulated with the default tune of Herwig 7.1.5~\cite{Bahr:2008pv,Bellm:2015jjp,Bellm:2017bvx}.  We only make use of the reconstructed events from Herwig as would be the case with real data.  The Herwig particle-level events are only used when evaluating the performance of different methods.  For the `simulation', events are simulated with Tune 26~\cite{ATL-PHYS-PUB-2014-021} of Pythia~8.243~\cite{Sjostrand:2007gs,Sjostrand:2006za,Sjostrand:2014zea}.  Detector distortions are simulated with Delphes~3.4.2~\cite{deFavereau:2013fsa} and the CMS tune that uses a particle flow reconstruction.  In future work, we will investigate the performance of \textsc{SBUnfold} on the full phase space of particles (particle-level) and particle flow objects (detector-level).  For this study, we focus on a fixed set of six observables that serve as the benchmark for fixed-dimension unfolding. These observables are computed from the substructure of the leading jet.  The jets are clustered using all particle flow objects at detector level and all stable non-neutrino truth particles at particle level.  They are defined by the anti-$k_T$ algorithm~\cite{Cacciari:2008gp} with radius parameter $R=0.4$ as implemented in FastJet~3.3.2~\cite{Cacciari:2011ma,Cacciari:2005hq}. The first four observables are the jet mass $m$, constituent multiplicity $M$, the $N$-subjettiness ratio $\tau_{21}=\tau_2^{(\beta=1)}/\tau_1^{(\beta=1)}$~\cite{Thaler:2010tr,Thaler:2011gf}, and the jet width $w$ (implemented as $\tau_1^{(\beta=1)}$).  The remaining two observables are the jet mass $\ln\rho = \ln m_\text{SD}^2/p_T^2$ and momentum fraction $z_g$ after Soft Drop grooming~\cite{Larkoski:2014wba,Dasgupta:2013ihk} with $z_\text{cut}=0.1$ and $\beta=0$.  Many of the observables are computed with \textsc{FastJet Contrib 1.042}~\cite{fjcontrib}.

 The $Z$ bosons are required to have $p_T>200$ GeV in order to render acceptance effects negligible\footnote{See Ref.~\cite{Andreassen:2021zzk} for how to include acceptance effects.}.  Each dataset consists of about 1.6 millon events.

The noise prediction model of \textsc{SBUnfold} is implemented using a fully-connected architecture incorporating multiple skip connections. Specifically, the model employs four \textsc{ResNet}~\cite{he2016deep} blocks, where each residual layer is connected to the output of a single-layer network through a skip connection. The activation function used is \textsc{LeakyRelu}~\cite{DBLP:journals/corr/XuWCL15} with a slope of $\alpha=0.01$, and all layer sizes are set to 32. The implementation of the model is carried out using \textsc{Pytorch}~\cite{pytorch}. The generation is carried out by starding from data observations at reconstruction level and using the SB to move towards the generator level events using 1000 time steps to improve accuracy.
The \textsc{cINN} implementation uses the normalizing flow model presented in Ref.~\cite{1800956}  while the first step of the first iteration of \textsc{OmniFold} uses a fully-connected model with three hidden layers, \textsc{ReLu} activation function, and node sizes set to 50, 150, 50, before the output layer with a single node and \textsc{sigmoid} activation function.  Both \textsc{cINN} and \textsc{OmniFold} are implemented in \textsc{TensorFlow}~\cite{tensorflow}. For all models, the number of trainable parameters are kept similar at around 13k. The initial learning rate is set to $10^{-3}$ and all models are trained for 30 epochs with batch size of 128. A total of 1M training events are used to train all models taken from the \textsc{Pythia} simulations. The pseudo-data used to evaluate the performance of the unfolding is taken from \textsc{Herwig}. Both \textsc{SBUnfold} and \textsc{cINN} do not have access to the data during training while \textsc{OmniFold} uses a total of 2M training events (1M from \textsc{Pythia} and 1M from \textsc{Herwig}). We investigate the impact of the pseudo-data training size in Sec.~\ref{sec:results} when determining the unfolded results.  All trainings are carried out using the Perlmutter supercomputer with a single NVIDIA A100 GPU. 

\begin{figure*}[ht]
\centering
    \includegraphics[width=0.3\textwidth]{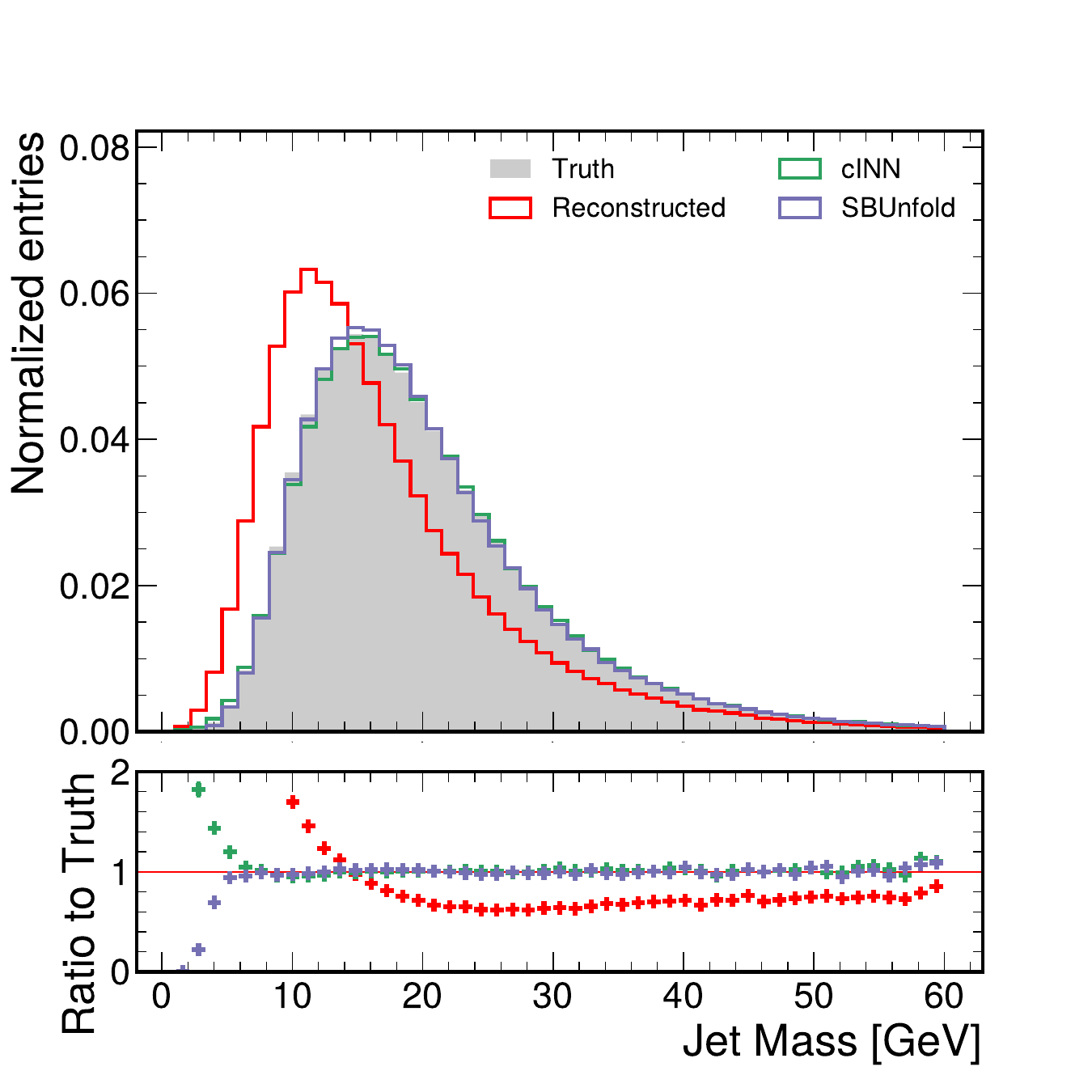}
    \includegraphics[width=0.3\textwidth]{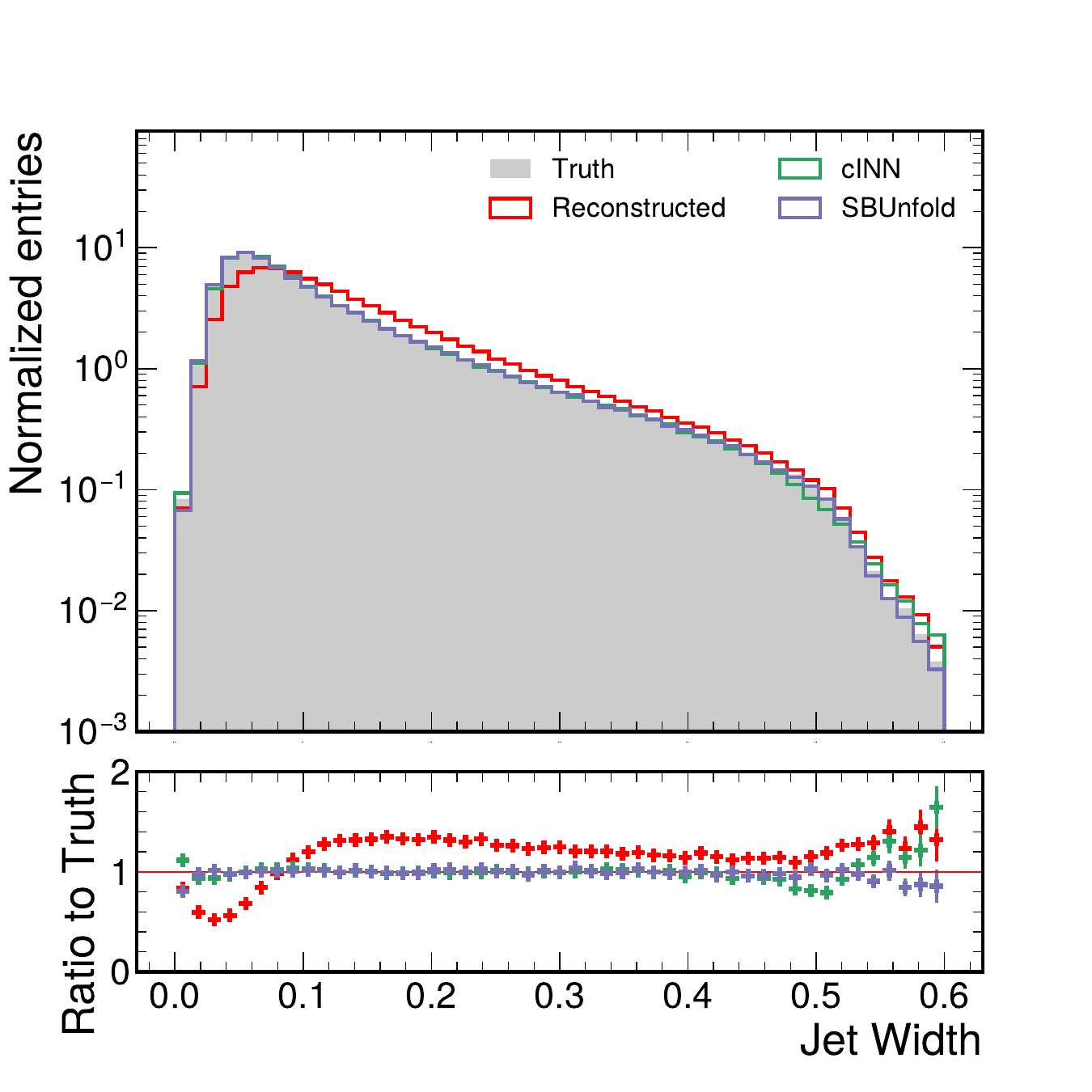}
    \includegraphics[width=0.3\textwidth]{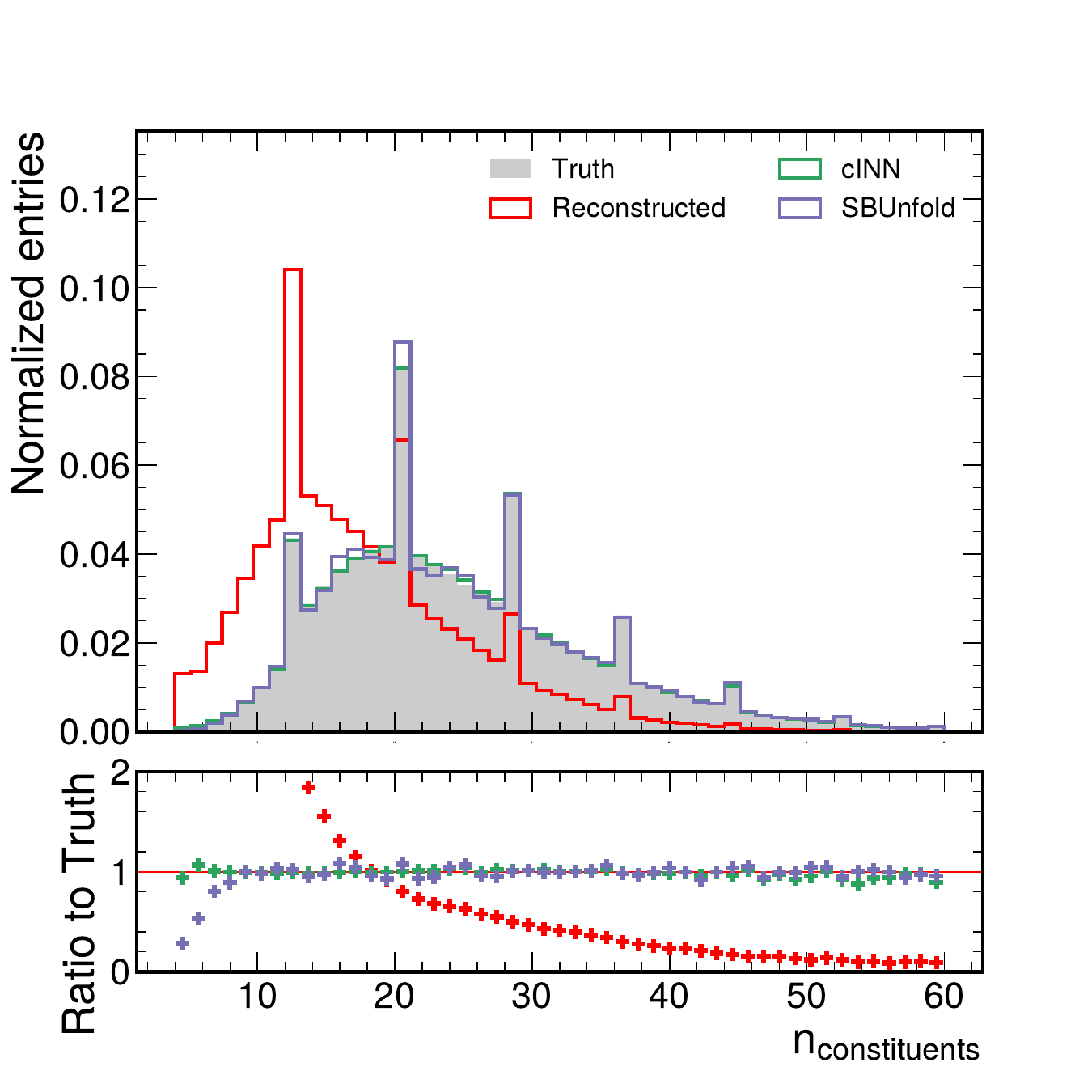}
    \includegraphics[width=0.3\textwidth]{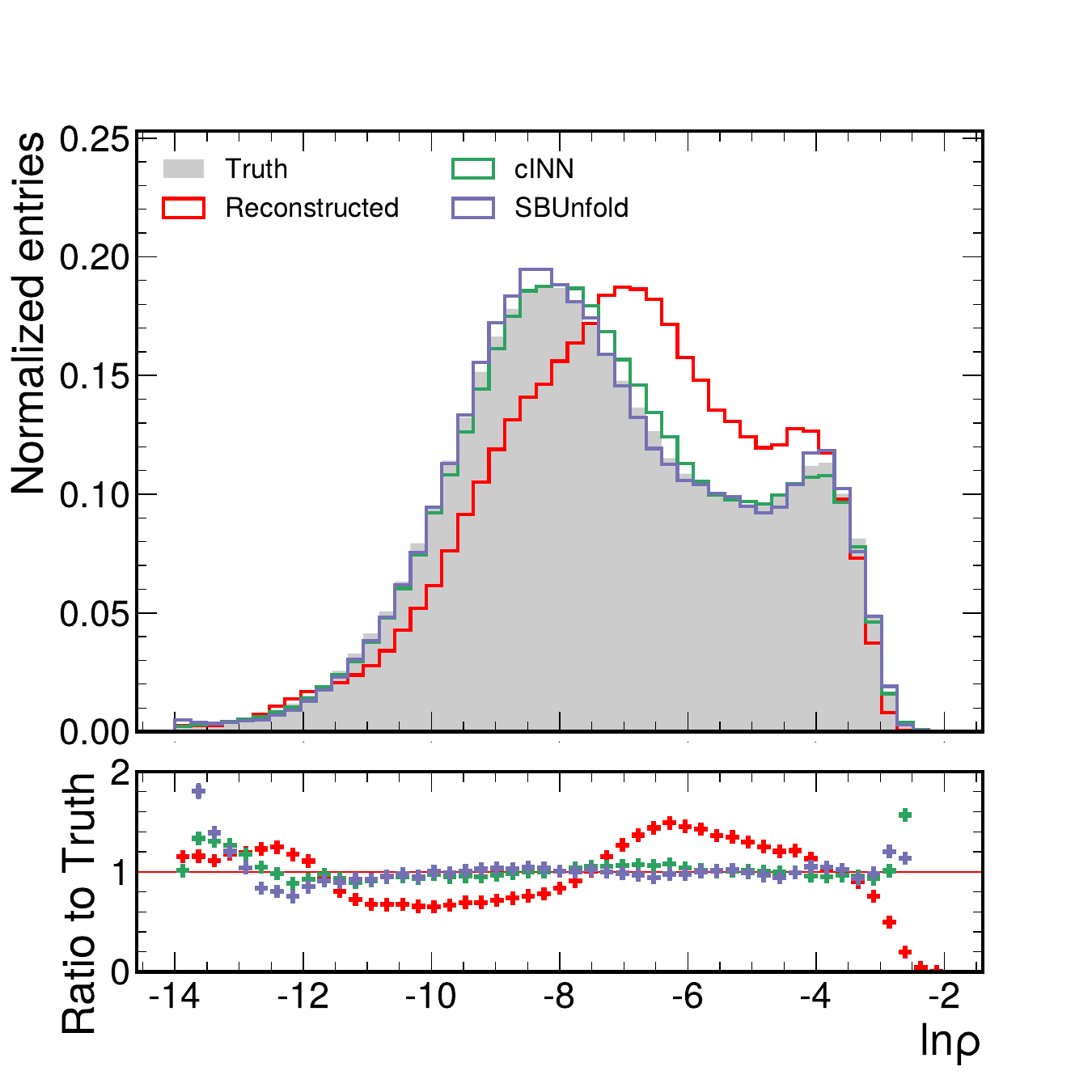}
    \includegraphics[width=0.3\textwidth]{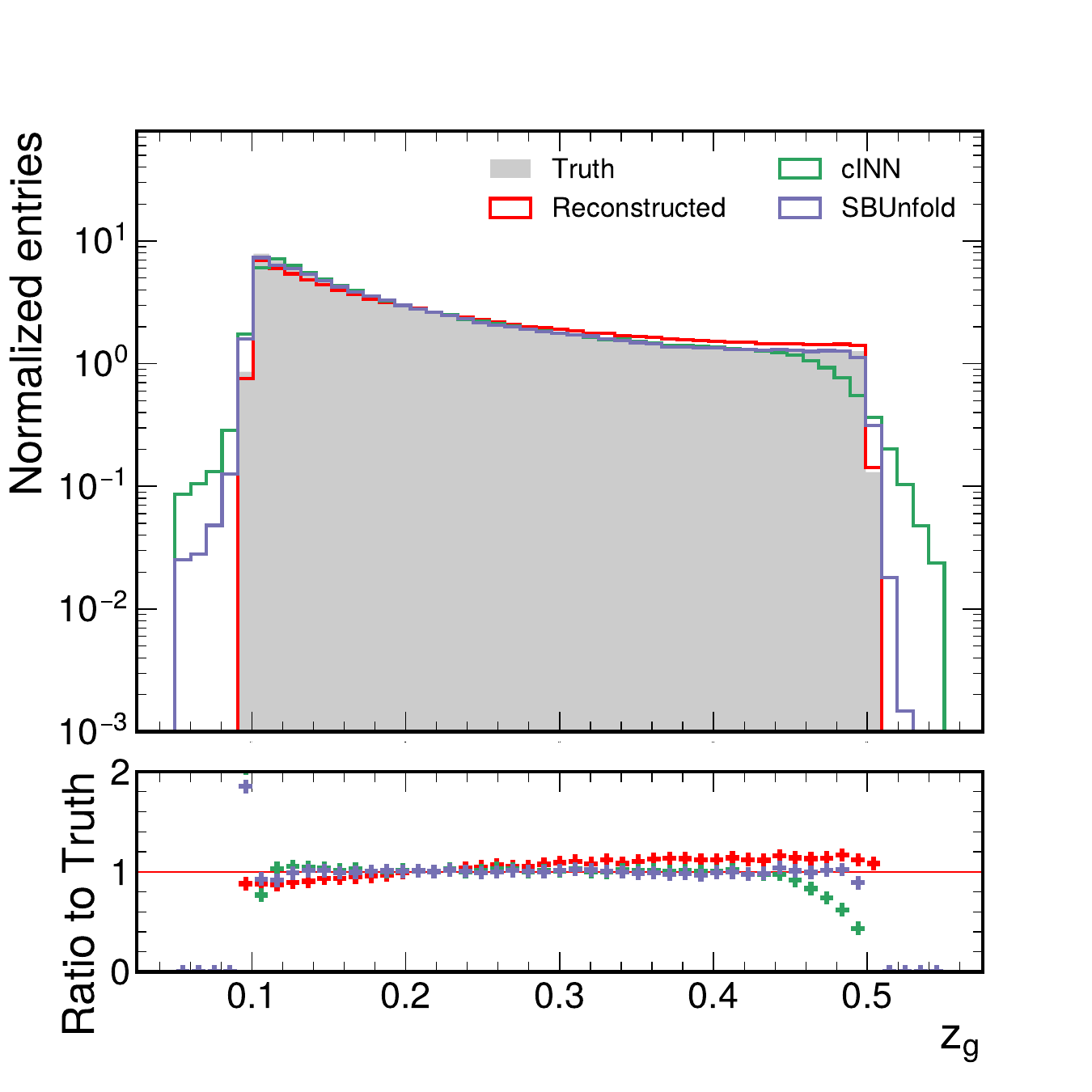}
    \includegraphics[width=0.3\textwidth]{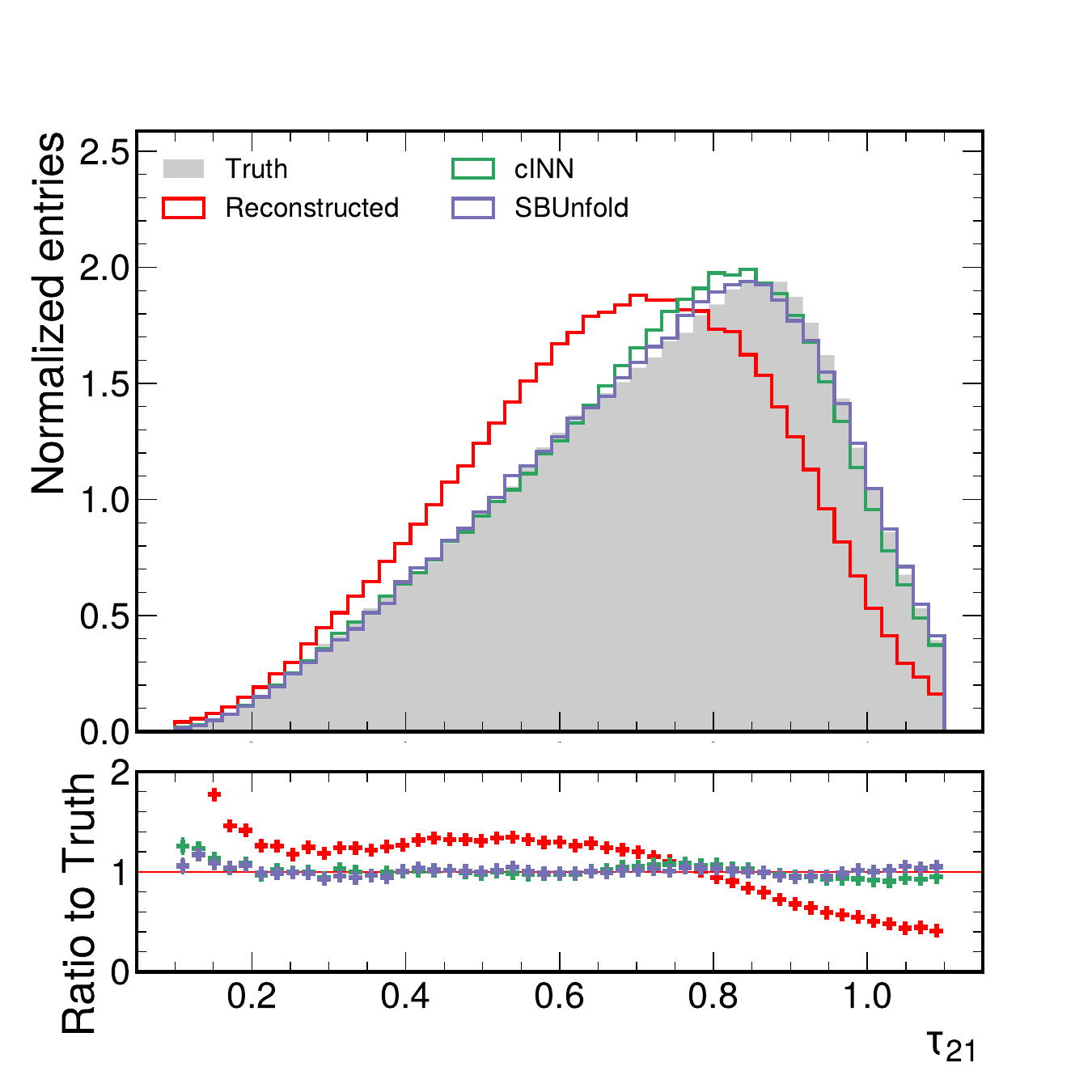}
\caption{Comparison between different unfolding algorithms for six different physics observables unfolded. All observables are unfolded simultaneously without binning, with histograms shown only for evaluation. Results are evaluated over 600'000 pseudo-data points. Statistical uncertainties are shown only in the ratio panel. Pseudo-data and simulation are described by Pythia.}
\label{fig:unfold_600k_pythia}
\end{figure*}

\section{Results}
\label{sec:results}
We evaluate the performance of \textsc{SBUnfold} compared to other unfolding methodologies by first looking at unfolded distributions where both pseudo-data and samples used to train the different unfolding methods originate from the \textsc{Pythia} simulation. The results showing the unfolded distributions are shown in Fig.~\ref{fig:unfold_600k_pythia}. In App.~\ref{app:diffusion}, we also provide the comparison of the results obtained by a standard diffusion model.

\begin{figure*}[ht]
\centering
    \includegraphics[width=0.45\textwidth]{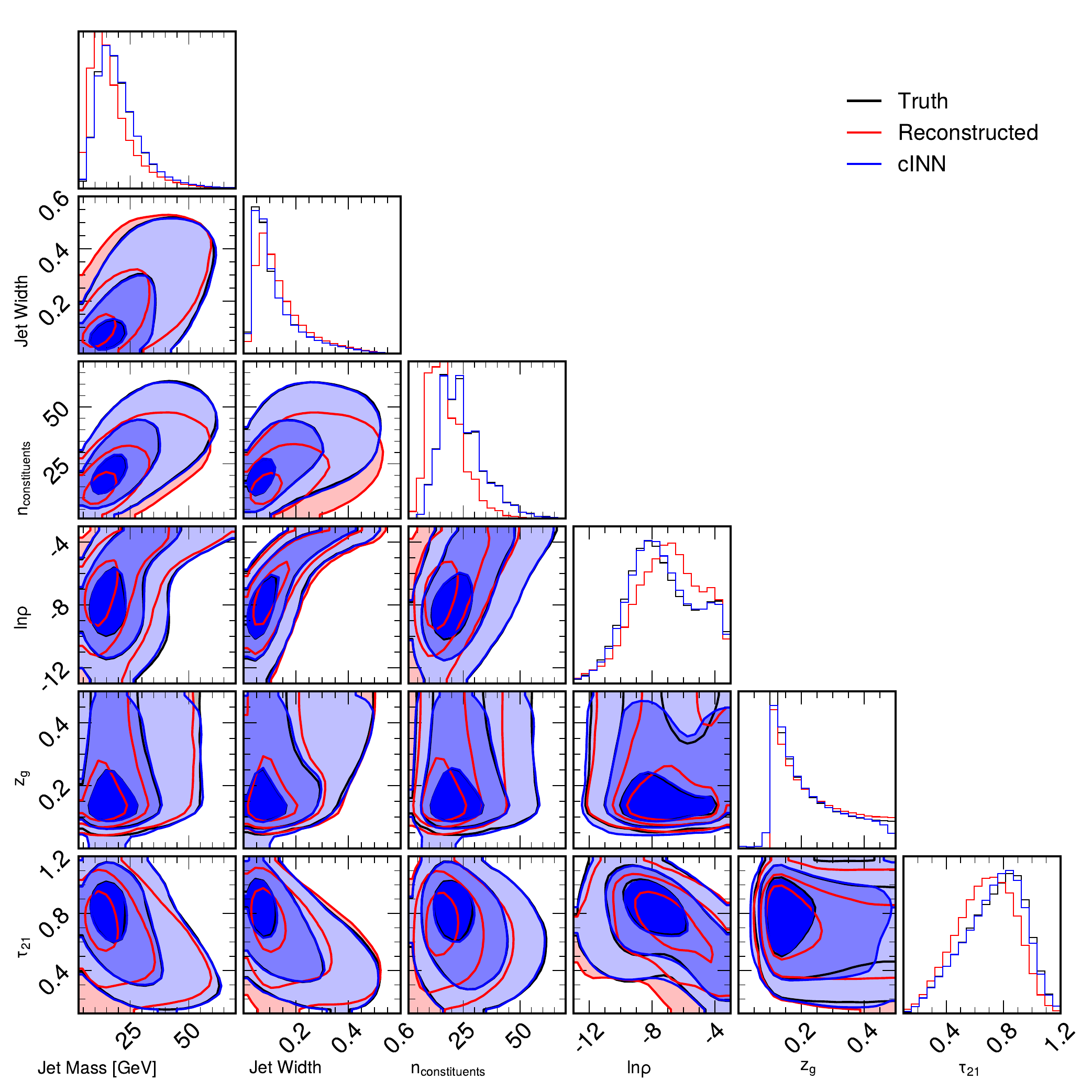}
    \includegraphics[width=0.45\textwidth]{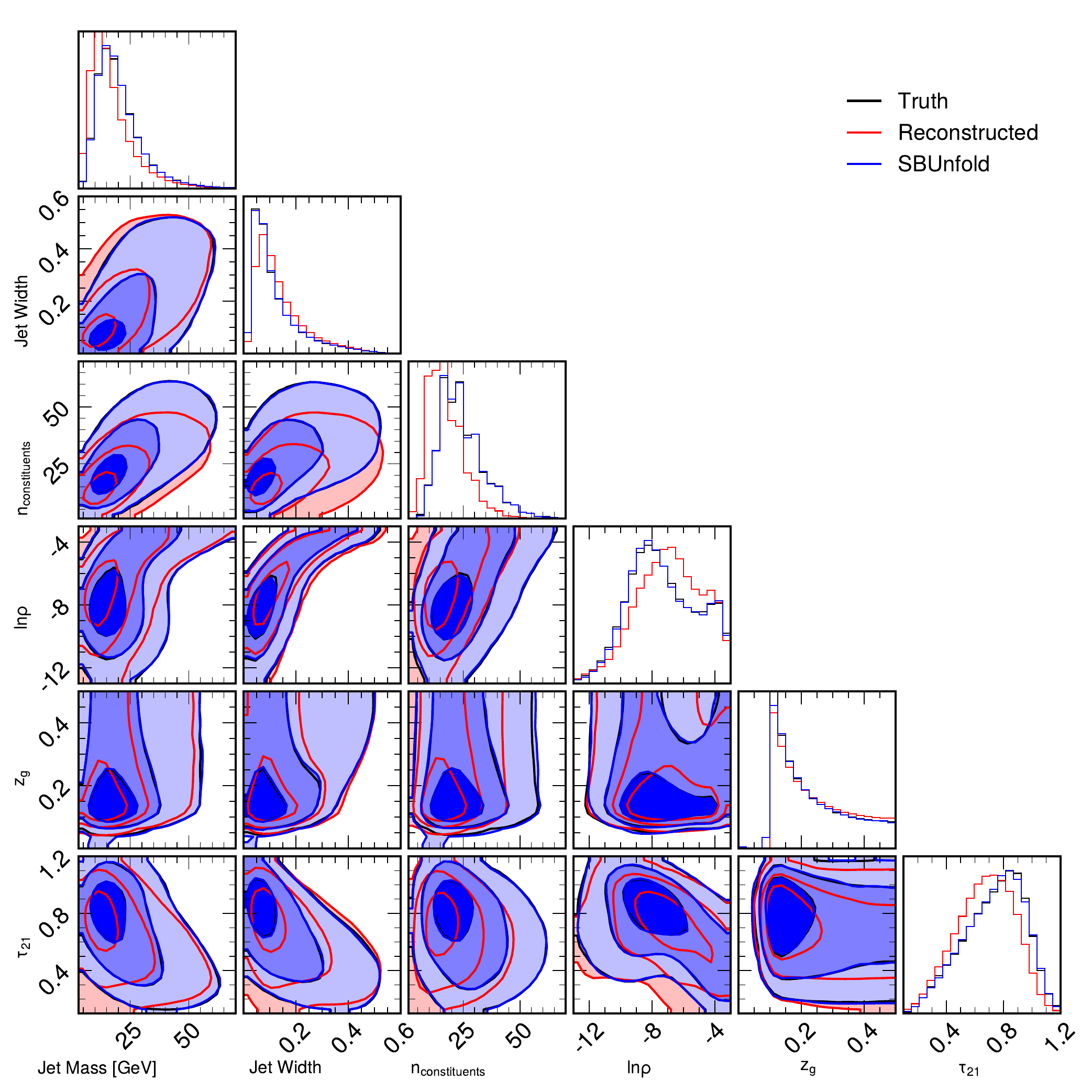}
\caption{Pairwise correlation plot between all unfolded variables using the \textsc{cINN} (left) or \textbf{SBUnfold} (right) algorithms.}
\label{fig:corner}
\end{figure*}

We observe a good agreement with the expected distributions from \textsc{Pythia} at generator level and the responses of the generative models for unfolding. We omit the \textsc{OmniFold} results since the weighting function becomes trivial when both data and simulation are statistically identical. We also investigate the distribution of the pairwise correlation between each pair of unfolded features to verify that \textsc{SBUnfold} is also capable of learning the correct correlations between distributions. The results are shown in Fig.~\ref{fig:corner}.

\begin{table}[ht]
    \centering
	\small
    \caption{Comparison of the earth mover's distance (EMD) and triangular discriminator between different algorithms. EMD is calculated over unbinned distributions while triangular discriminator uses histograms as inputs. Uncertainties from EMD are derived using 100 bootstraps with replacement taken from the unfolded data. Results are evaluated using 600'000 pseudo-data points sampled from Pythia. Quantities in bold represent the method with best performance.}
    \label{tab:EMD_600k_pythia}
	\begin{tabular}{l|c|c|c|c|c|cc}
        Model &   \multicolumn{2}{c}{EMD($\times 10$)/Triangular Discriminator($\times10^3$)}\\
        &   {\scriptsize \textsc{cINN}} & {\scriptsize \textsc{SBUnfold}} \\
        \hline     
        Jet mass &  1.4$\pm$0.2/\textbf{0.29} & \textbf{0.70$\pm$0.06}/0.30 \\
        Jet Width &    0.013$\pm$0.002/0.25 & \textbf{0.0029$\pm$0.0005}/\textbf{0.06} \\
        N & 2.3$\pm$0.8/\textbf{0.09} & \textbf{0.57$\pm$0.04}/0.9 \\ 
        $\log\rho$ &  1.1$\pm$0.3/\textbf{0.64} & \textbf{0.27$\pm$0.01}/0.68 \\ 
        $z_g$ &  0.095$\pm$0.003/10.9 & \textbf{0.009$\pm$0.001}/\textbf{3.1} \\
        $\tau_{21}$  & 0.2$\pm$0.1/0.6 & \textbf{0.016$\pm$0.001}/\textbf{0.2} \\
	\end{tabular}
\end{table}

We again observe a good agreement between the generated distributions and expected distributions from the simulation for both generative models, with \textsc{SBUnfold} showing improved description compared to the \textsc{cINN} for a few distributions such as $z_g$ vs. $\log\rho$ and $z_g$ vs $\tau_{21}$. The distribution of $z_g$ shows a sharp cutoff at $z_g=0$ which is hard to reproduce with generative models. Since the same sharp distribution is observed at reconstructed level events, \textsc{SBUnfold} can take advantage of the reconstructed prior when starting the diffusion process, contrary to the \textsc{cINN} that always requires samples from the standard normal distribution for the prior transformation. We also quantify the agreement between unfolded distributions with expected values derived from \textsc{Pythia} by calculating both the triangular discriminator metric~\cite{850703,Gras:2017jty,Bright-Thonney:2018mxq} over histograms as presented in Fig.~\ref{fig:unfold_600k_pythia}, as well as using the Earth mover's distance (EMD) directly on the marginalized one-dimensional features generated by the unfolding methods. The results are listed in Table~\ref{tab:EMD_600k_pythia}.

\begin{figure*}[ht]
\centering
    \includegraphics[width=0.3\textwidth]{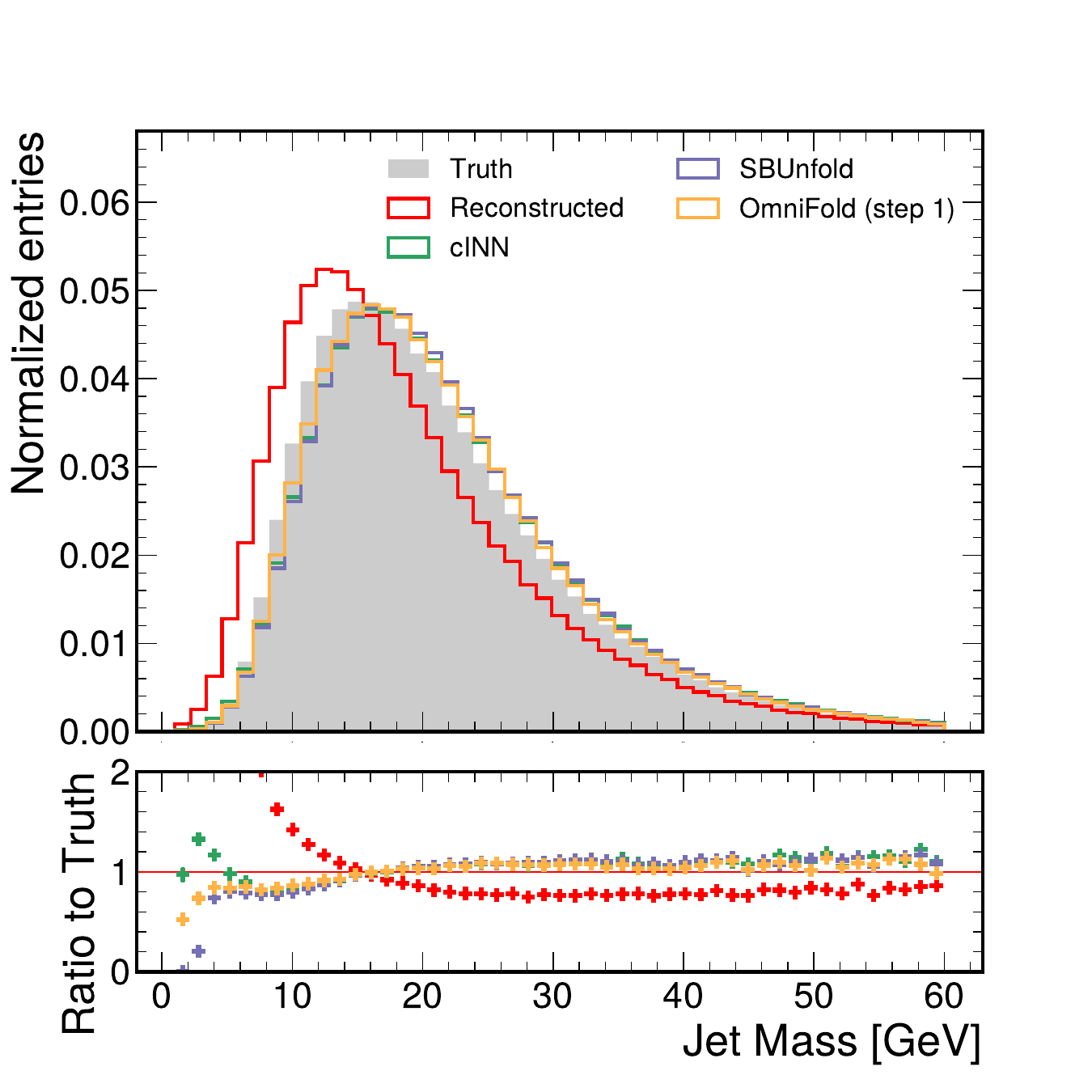}
    \includegraphics[width=0.3\textwidth]{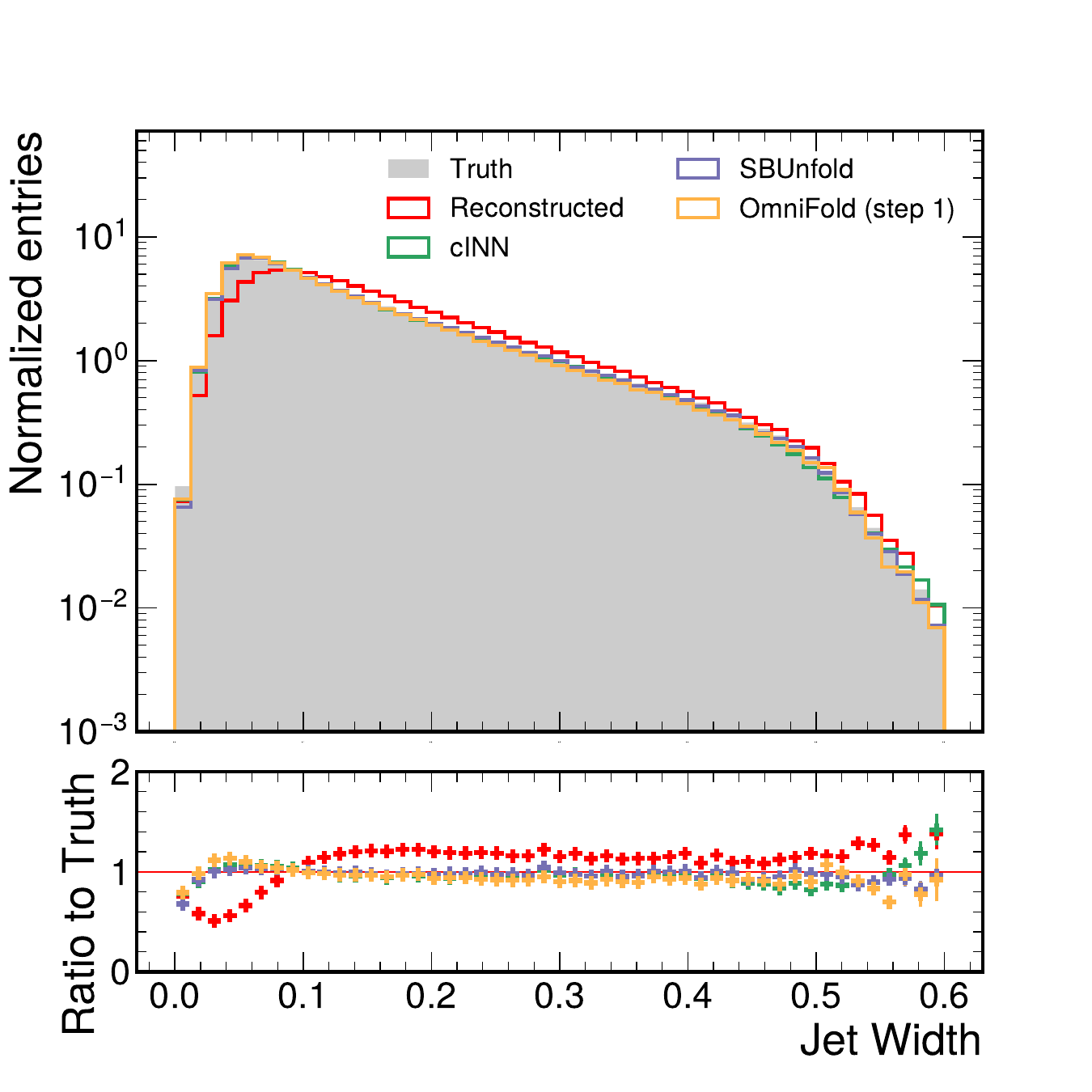}
    \includegraphics[width=0.3\textwidth]{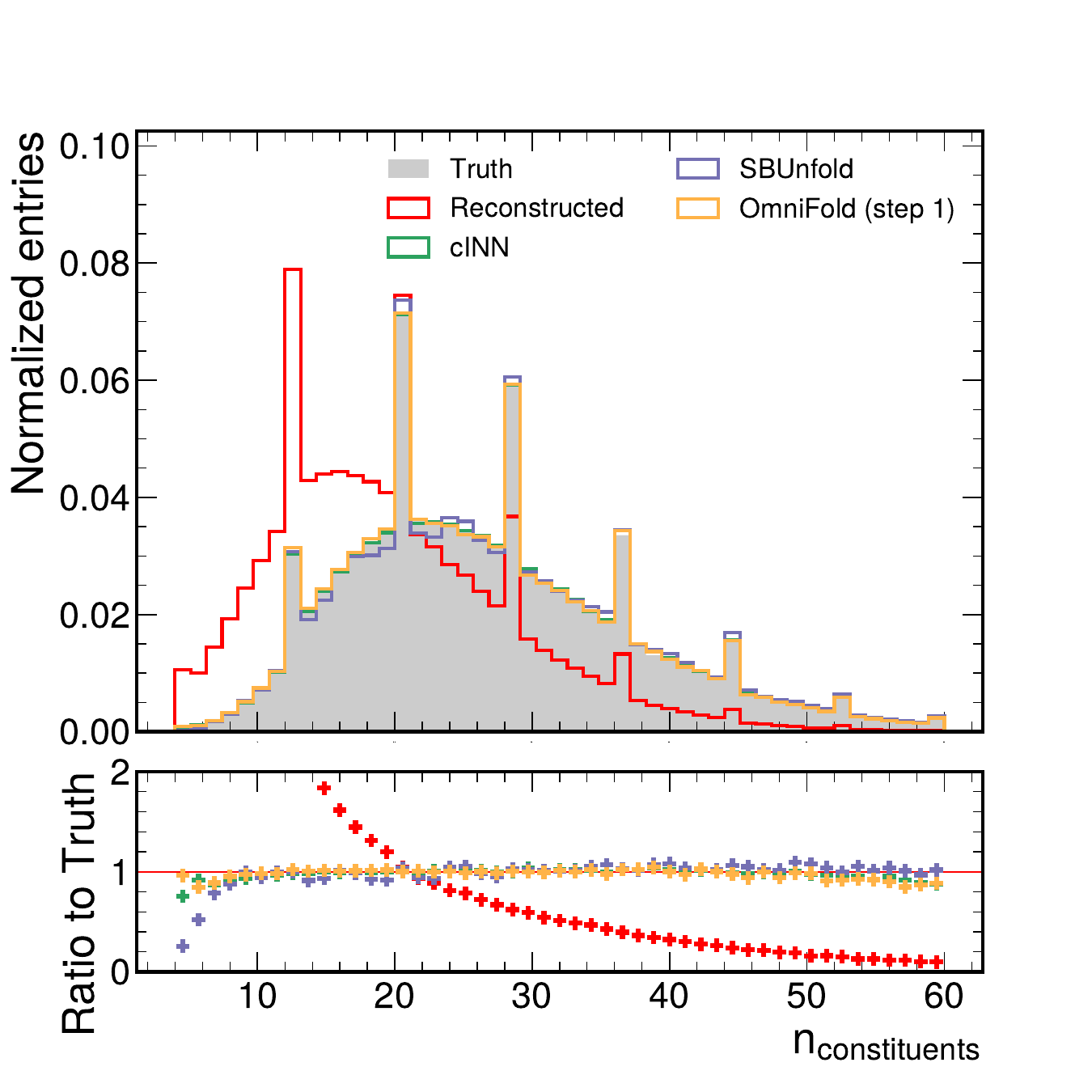}
    \includegraphics[width=0.3\textwidth]{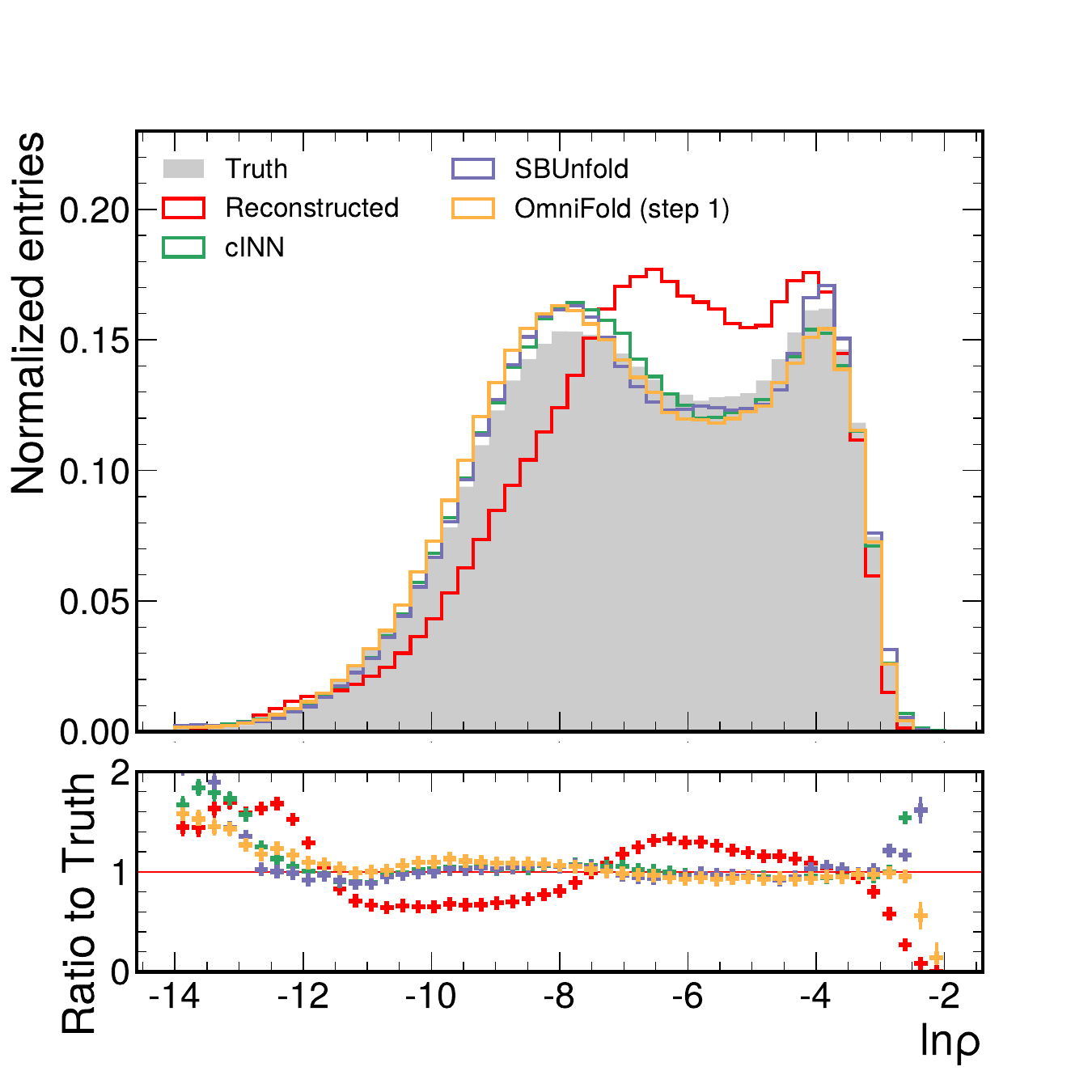}
    \includegraphics[width=0.3\textwidth]{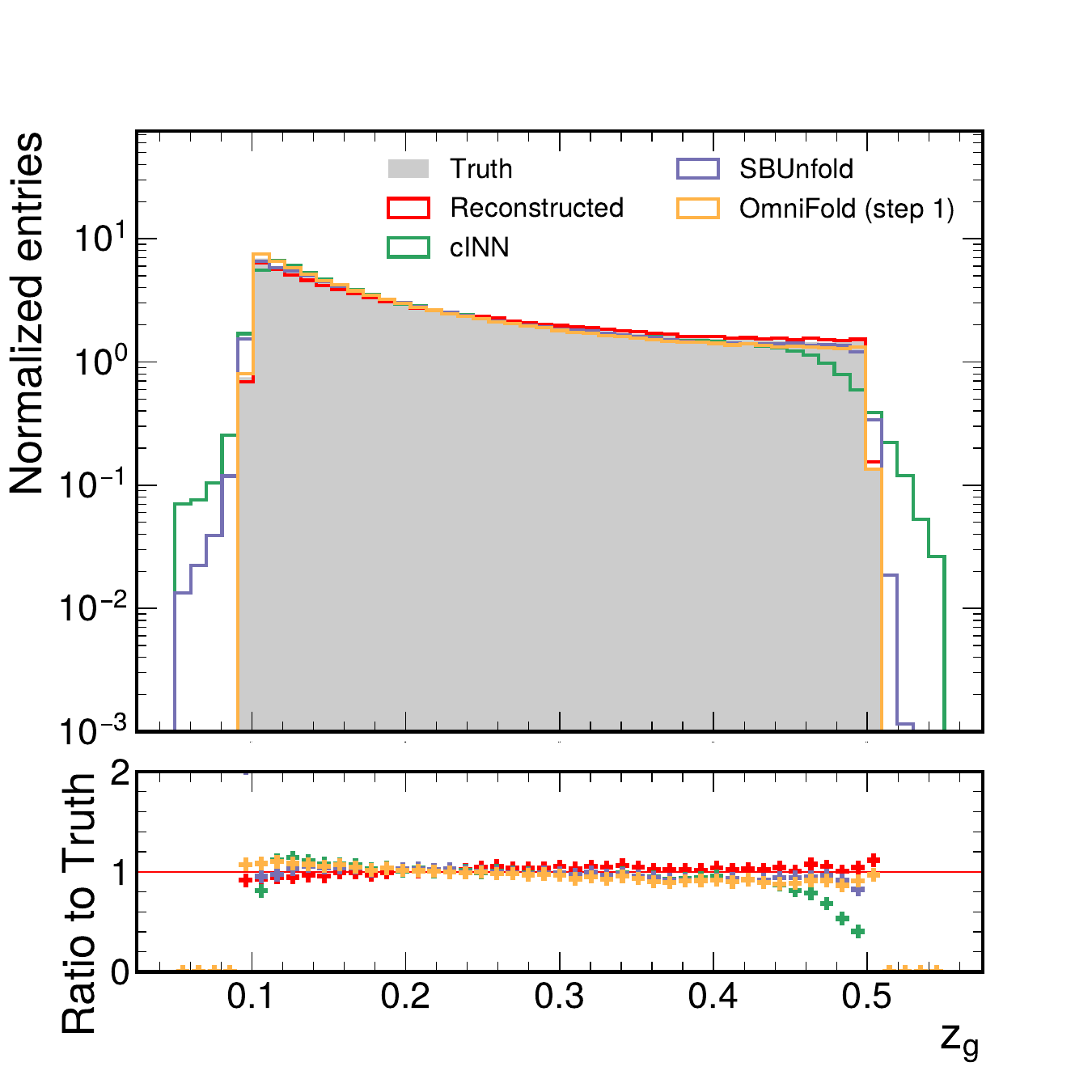}
    \includegraphics[width=0.3\textwidth]{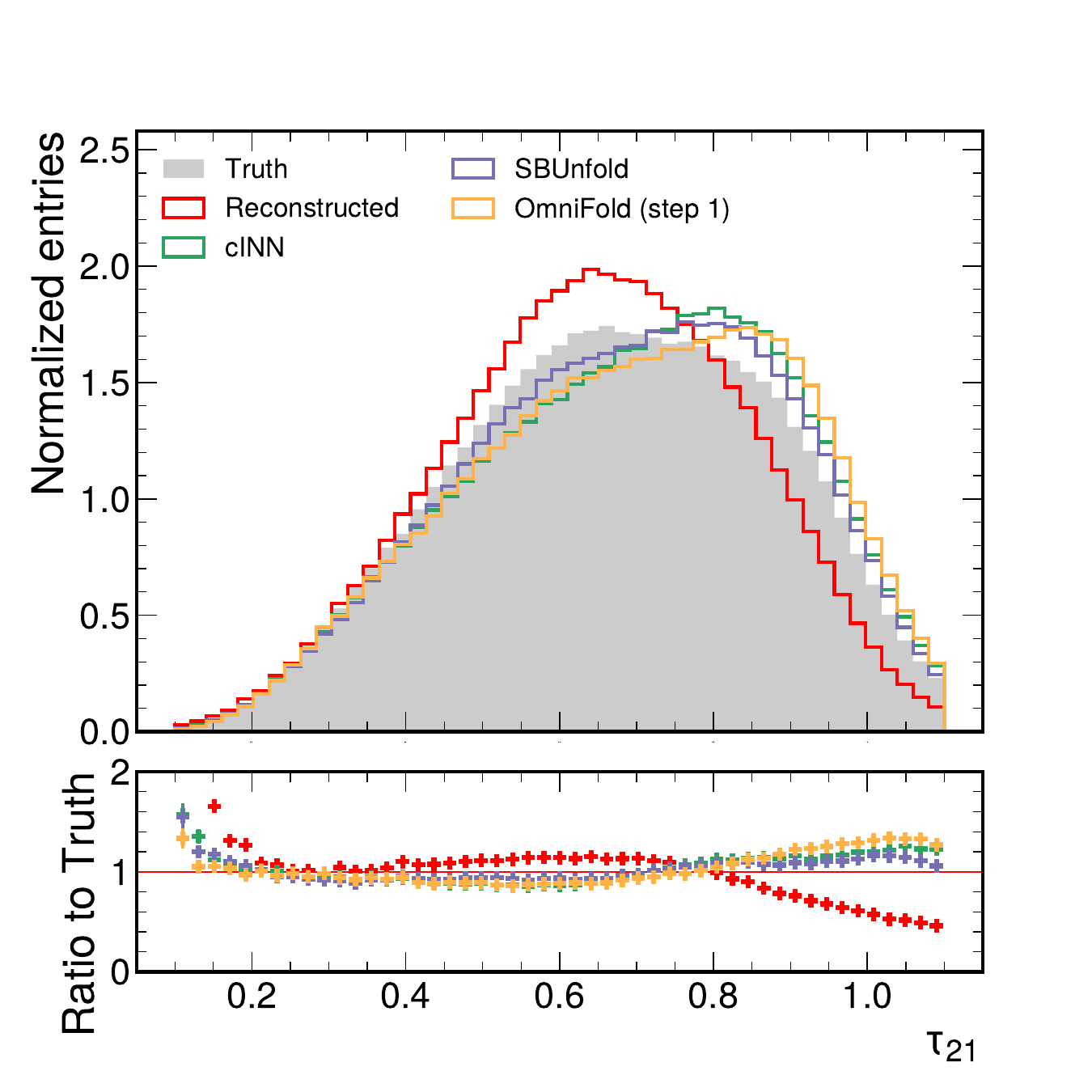}        
\caption{Comparison between different unfolding algorithms for six different physics observables unfolded. All observables are unfolded simultaneously without binning, with histograms shown only for evaluation. Results are evaluated over 600'000 pseudo-data points. Statistical uncertainties are shown only in the ratio panel. Pseudo-data is taken from Herwig while Pythia is taken as the main simulator.}
\label{fig:unfold_600k}
\end{figure*}

For all distributions we observe significantly lower EMD values for \textsc{SBUnfold} compared to \textsc{cINN} with four out of six also showing lower values of the triangular discriminator. Uncertainties from the EMD calculation are derived from the standard deviation of 100 bootstraps with replacement taken from the unfolded data. We observe the EMD uncertainties from \textsc{cINN} to be often higher than \textsc{SBUnfold}, which points to improved stability from \textsc{SBUnfold} due the more informative prior. 

Next, we keep the same unfolding methodology from \textsc{SBUnfold} and \textsc{cINN} trained over \textsc{Pythia} samples but instead use reconstructed events from \textsc{Herwig} as the data representative. We compare the unfolded results between the different distributions in Fig.~\ref{fig:unfold_600k}.

We observe improved agreement in all unfolded distributions compared to the distributions of reconstructed events, while systematic shifts are observed in all unfolding methodologies for the jet mass and $\tau_{21}$ distributions. These features highlight the prior dependence of all models which cannot be excluded without the presence of the M step and additional iterations. In Table~\ref{tab:EMD_600k} we calculate the EMD and triangular discriminator using the unfolded distributions with \textsc{Herwig} as pseudo-data.

\begin{table}[ht]
    \centering
	\small
    \caption{Comparison of the earth mover's distance (EMD) and triangular discriminator between different algorithms. EMD is calculated over unbinned distributions while triangular discriminator uses histograms as inputs. Uncertainties from EMD are derived using 100 bootstraps with replacement taken from the unfolded data. Results are evaluated using 600,000 pseudo-data points sampled from Herwig. Quantities in bold represent the method with best performance.}
    \label{tab:EMD_600k}
	\begin{tabular}{l|c|c|c|c|c|cc}
        Model &   \multicolumn{3}{c}{EMD($\times 10$)/Triangular Discriminator($\times10^3$)}\\
        &  {\scriptsize \textsc{OmniFold} Step 1} & {\scriptsize \textsc{cINN}} & {\scriptsize \textsc{SBUnfold}} \\
        \hline     
        Jet mass &   \textbf{6.1$\pm$0.1}/\textbf{1.5} & 10.1$\pm$1.2/2.4 & 9.0$\pm$0.1/3.1 \\
        Jet Width &   0.06$\pm$0.001/1.1 & 0.05$\pm$0.003/0.7 &\textbf{ 0.02$\pm$0.001}/\textbf{0.2} \\
        N & \textbf{1.7$\pm$0.1}/\textbf{0.2}  & 6.1$\pm$4.0/0.2 & 3.0$\pm$0.1/0.6 \\ 
        $\log\rho$ & 1.35$\pm$0.03/1.1  & 3.1$\pm$2.1/0.8 & \textbf{0.4$\pm$0.1}/\textbf{0.7} \\ 
        $z_g$ & 0.086$\pm$0.001/\textbf{1.2 } & 0.3$\pm$0.1/12.7 & \textbf{0.049$\pm$0.001}/3.5 \\
        $\tau_{21}$ & 0.23$\pm$0.02/4.6  & 0.7$\pm$0.4/3.5 & \textbf{0.12$\pm$0.02}/\textbf{1.4} \\
	\end{tabular}
\end{table}

Once again we observe a good performance of \textsc{SBUnfold}, achieving the lowest EMD values for four out of six distributions and lowest triangular discriminator values for three of the six. We also observe \textsc{OmniFold} achieving similar performance for all observables, in particular \textsc{OmniFold} shows improved performance for the particle multiplicity, which is not a continuous distribution and hence harder for generative models to determine precisely. We also study how \textsc{SBUnfold} corrects physics observables back to generator level events by calculating the migration matrix between reconstructed and unfolded observables. Results showing that \textsc{SBUnfold} learns to apply small corrections to reconstructed events are shown in App.~\ref{app:migration}.  

\begin{figure*}[ht]
\centering
    \includegraphics[width=0.3\textwidth]{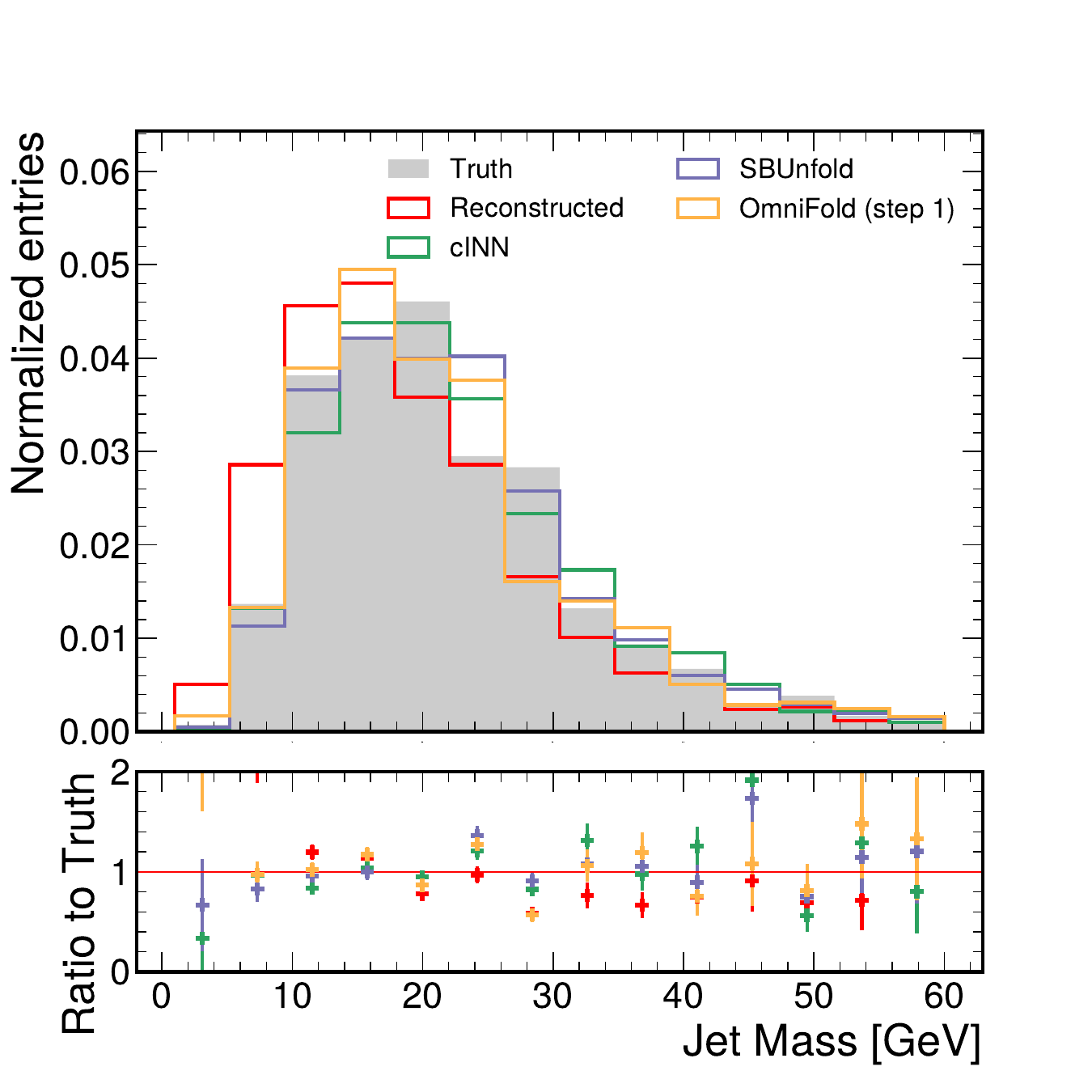}
    \includegraphics[width=0.3\textwidth]{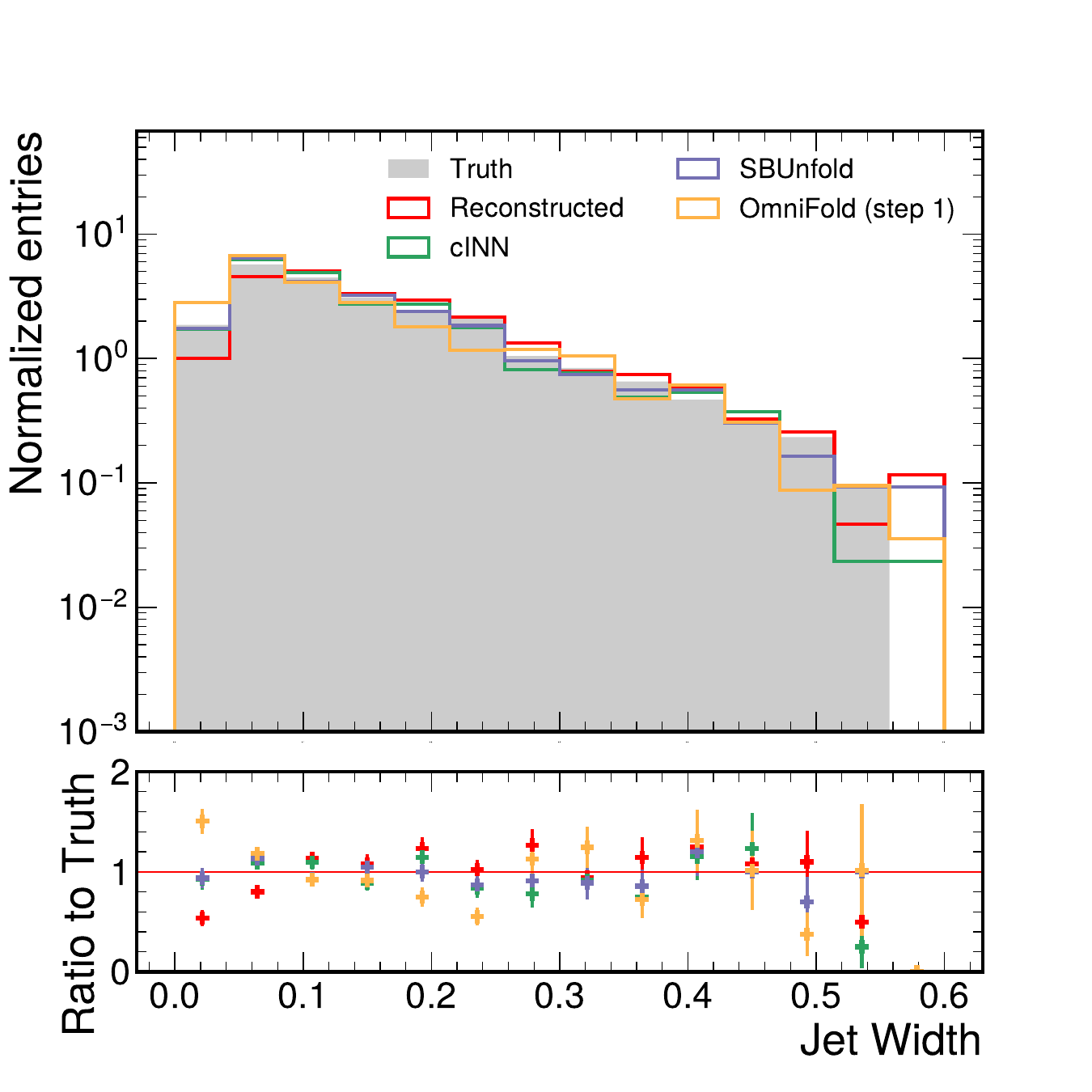}
    \includegraphics[width=0.3\textwidth]{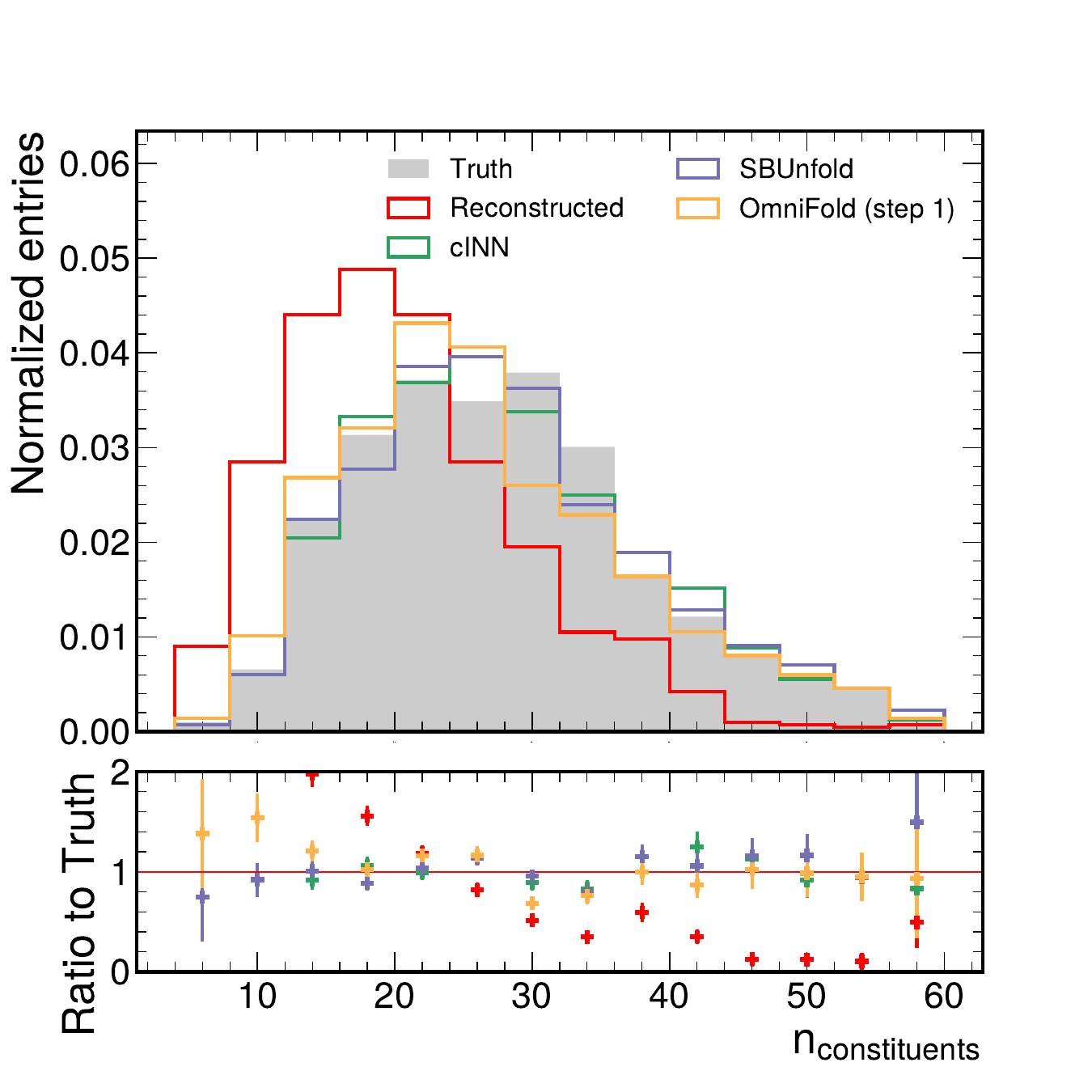}
    \includegraphics[width=0.3\textwidth]{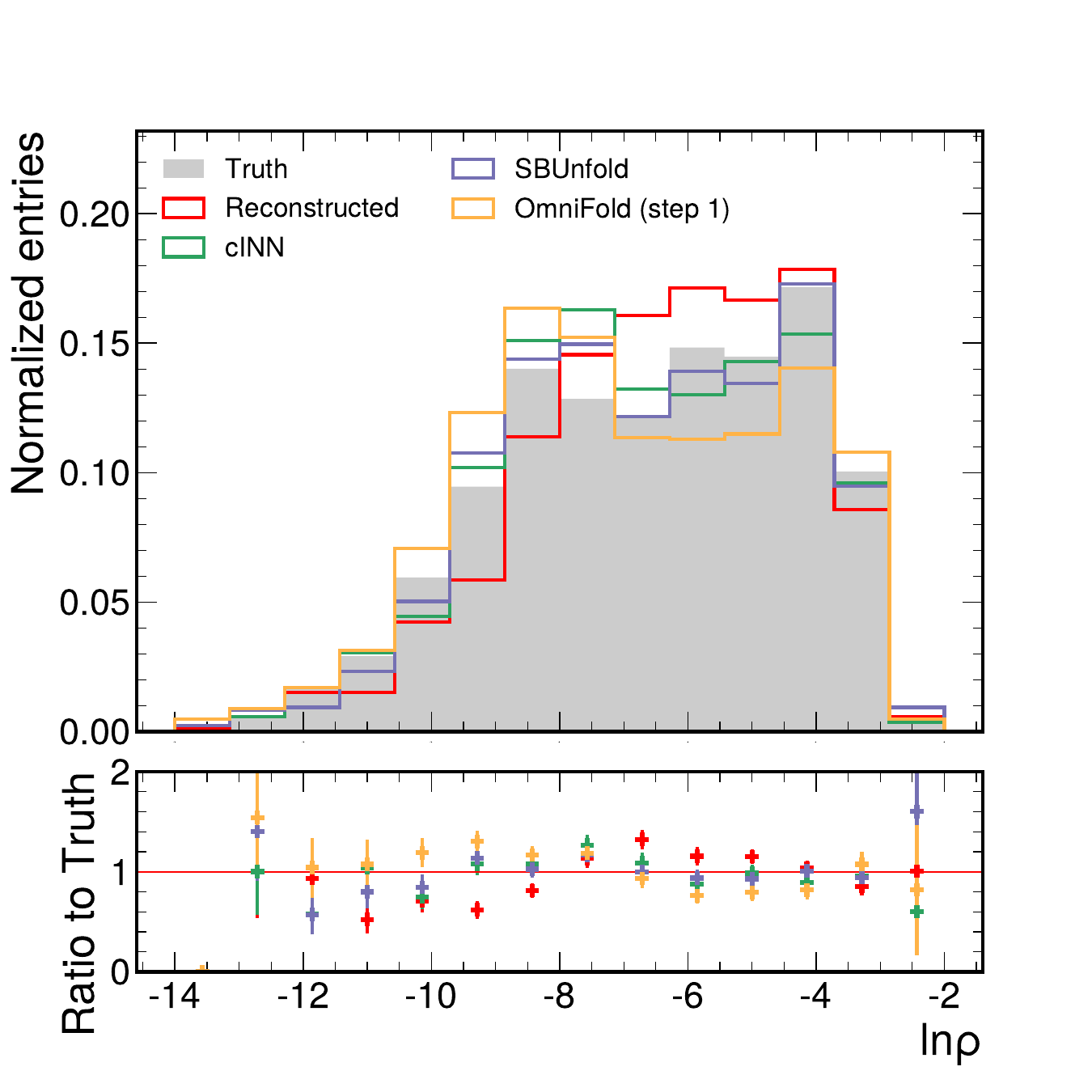}
    \includegraphics[width=0.3\textwidth]{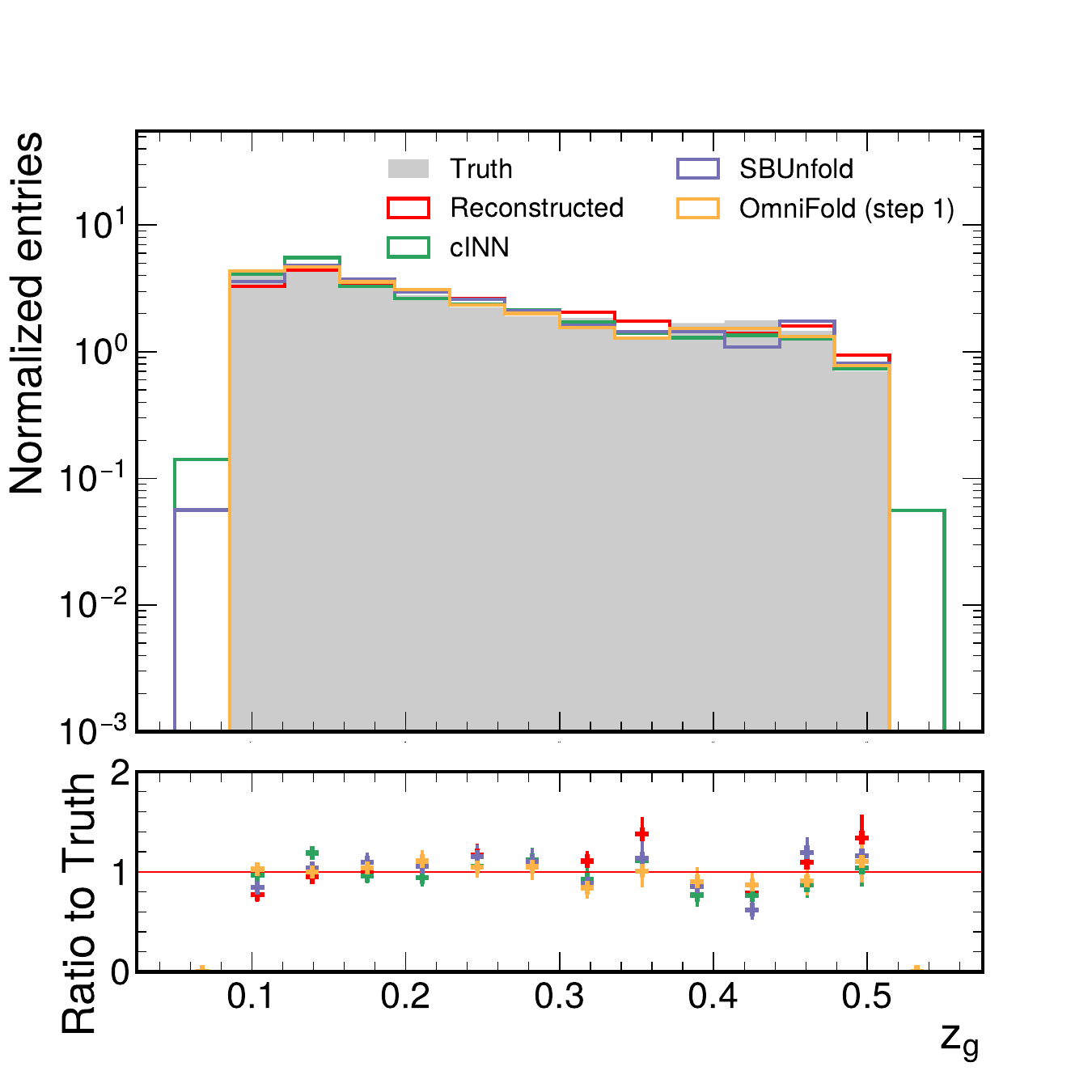}
    \includegraphics[width=0.3\textwidth]{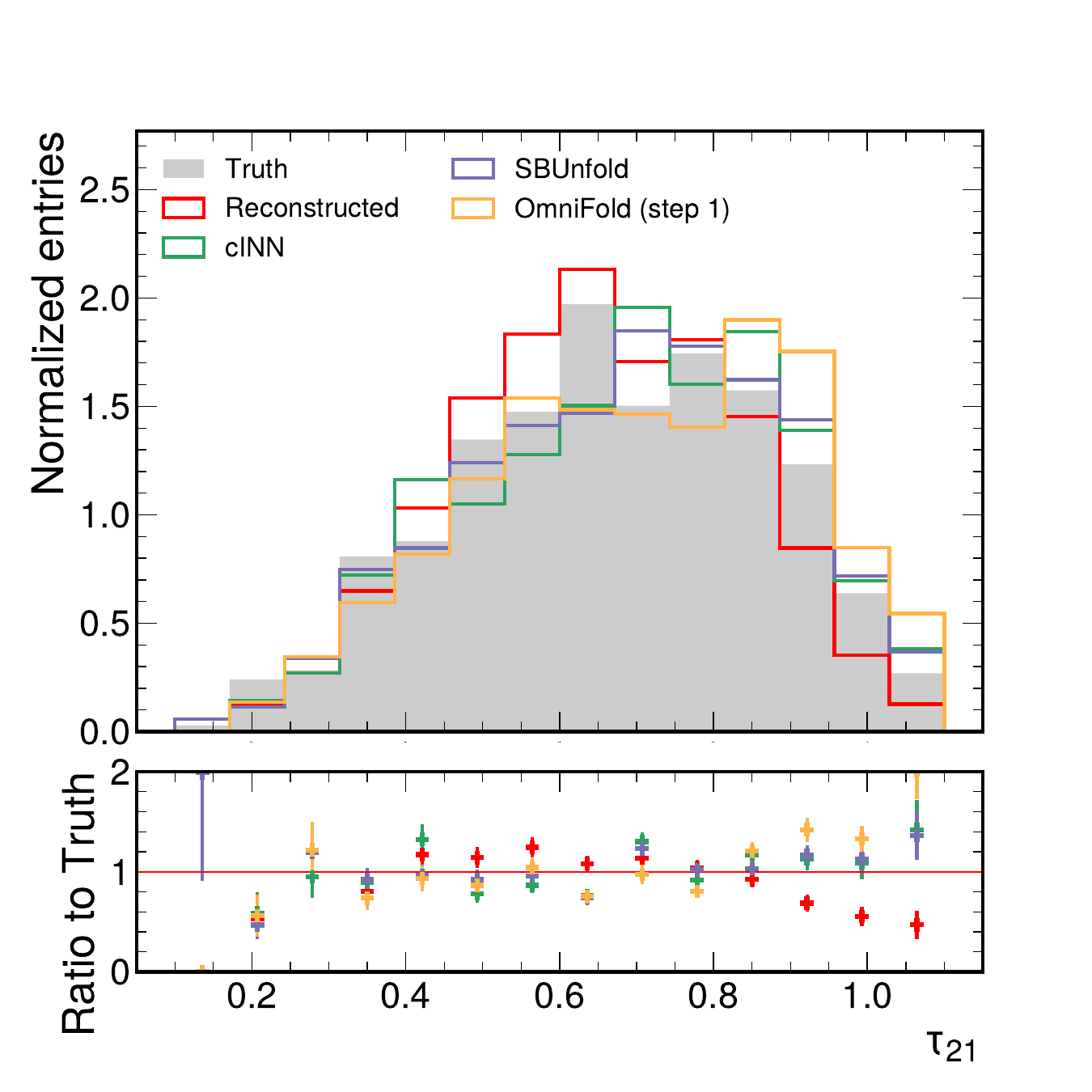}        
\caption{Comparison between different unfolding algorithms for six different physics observables unfolded. All observables are unfolded simultaneously without binning, with histograms shown only for evaluation.  Results are evaluated over 1'000 pseudo-data points, the same number was used to train \textsc{OmniFold} while other algorithms only rely on simulation. Statistical uncertainties are shown only in the ratio panel. Pseudo-data is taken from Herwig while Pythia is taken as the main simulator.}
\label{fig:unfold_1k}
\end{figure*}

\textsc{OmniFold} is also the only algorithm that uses the statistical power of the data to determine the unfolded distributions. We investigate how the unfolding results change when the available number of data entries is reduced to only 1,000 instead of the 600,000 thousands we used previously. For both \textsc{SBUnfold} and \textsc{cINN} this change only modifies the number of generated samples from the trained model while \textsc{OmniFold} only has access to the 1,000 data examples during training while the number of \textsc{Pythia} samples is not changed and kept at 1,000,000. The unfolded results are shown in Fig.~\ref{fig:unfold_1k}.

Due to the lower number of data entries we observe larger statistical fluctuations from the unfolded results which still shows a good agreement compared to the expected quantities. We evaluate further the unfolded results by calculating the EMD and triangular discriminator, listed in Table~\ref{tab:EMD_1k}.

\begin{table}[ht]
    \centering
	\small
    \caption{Comparison of the earth mover's distance (EMD) and triangular discriminator between different algorithms. EMD is calculated over unbinned distributions while triangular discriminator uses histograms as inputs. Uncertainties from EMD are derived using 100 bootstraps with replacement taken from the unfolded data. Results are evaluated using 1'000 pseudo-data points sampled from Herwig. Quantities in bold represent the method with best performance.}
    \label{tab:EMD_1k}
	\begin{tabular}{l|c|c|c|c|c|cc}
        Model &   \multicolumn{3}{c}{EMD($\times10$)/Triangular Discriminator($\times10^3$)}\\
        &  {\scriptsize \textsc{OmniFold} Step 1} & {\scriptsize \textsc{cINN}} & {\scriptsize \textsc{SBUnfold}} \\
        \hline     
        Jet mass &  \textbf{ 8.7$\pm$1.8}/13.6 & 9.2$\pm$3.0/8.4 &\textbf{ 7.7$\pm$2.5}/\textbf{6.9} \\
        Jet Width &   0.14$\pm$0.02/18 & 0.07$\pm$0.02/5.7 &\textbf{ 0.05$\pm$0.02}/\textbf{4.6}\\
        N & 12$\pm$3/10.9  & \textbf{5.4$\pm$1.3}/\textbf{3.8} & \textbf{5.8 $\pm$1.6}/3.7 \\ 
        $\log\rho$ & 4.0$\pm$0.8/11  & 1.6$\pm$0.5/6.2 & \textbf{1.2$\pm$0.3}/\textbf{4.4} \\ 
        $z_g$ & 0.08$\pm$0.02/\textbf{1.5 } & 0.08$\pm$0.03/7.2 & \textbf{0.06$\pm$0.01}/7.1 \\
        $\tau_{21}$ & 0.4$\pm$0.07/16  & 0.2$\pm$0.05/12 &\textbf{ 0.1$\pm$0.04}/\textbf{8} \\
	\end{tabular}
\end{table}

The reported results for \textsc{SBUnfold} and \textsc{cINN} are less affected by the reduced data sizes, even though the uncertainties from EMD have greatly increased due to limited pseudo-data examples. On the other hand, \textsc{OmniFold} shows a worse degradation compared to previous results, with EMD values often disagreeing by more than two standard deviations.

\section{Conclusion and Outlook}
\label{sec:conclusions}
In this paper we presented \textsc{SBUnfold}, an unfolding algorithm that uses a Schr\"{o}dinger bridge (SB) to design a stochastic mapping between distributions. Since the SB allows the transport between arbitrary distributions\footnote{Our Schr\"{o}dinger bridge is not a universal function approximator; it will map the distributions to each other, but may not preserve the conditional probability density~\cite{SB_nvidia}.  Our numerical results indicate that this is not an issue and we expect that monotonic detector distortions should be in the class of functions that SBs are able to accommodate, but we leave further investigations to future work.  We thank Jesse Thaler for useful discussions on this point.}, \textsc{SBUnfold} does not require a tractable prior similar to other generative models, but instead uses a diffusion process to directly denoise physics observables from detector effects. We have demonstrated the performance of \textsc{SBUnfold} using synthetic $Z$+jets samples and compared with both state-of-the-art methods using generative models (\textsc{cINN}) and reweighting (\textsc{OmniFold}). Using different metrics, we observe an excellent performance of \textsc{SBUnfold}. Compared to \textsc{cINN}, \textsc{SBUnfold} shows improved fidelity for distributions with sharp features and smaller uncertainties overall likely due to more informative priors from reconstructed level events. Compared to \textsc{OmniFold}, \textsc{SBUnfold} shows a more robust performance against variations of the number of available data observations, showing promising results for cases where the amount of available data is reduced.

We have only investigated the expectation step of different unfolding algorithms, leaving the maximization step for future study. We notice that the maximization step based on classifiers used by \textsc{IcINN} is also applicable for \textsc{SBUnfold}, and would be interesting to compare how the unfolding results improve with subsequent iterations between both algorithms.

Finally, similar to \textsc{OmniFold}, \textsc{SBUnfold} can be readily adapted to different data structures such as image-like datasets. On the other hand, the diffusion process currently relies on the specific ordering of the features used during unfolding (see Eq.~\ref{eq:perturb}), making the application of \textsc{SBUnfold} with low-level features, such as particles, not trivial. Following studies will investigate how to accommodate permutation invariance within \textsc{SBUnfold}.

\section*{Code Availability}

The code for this paper can be found at \url{https://github.com/ViniciusMikuni/SBUnfold}.

\section*{Acknowledgments}
We thank Daniel Whiteson, Tilman Plehn, and Jesse Thaler for thoughtful discussions and feedback on the manuscript. VM, SD, and BN are supported by the U.S. Department of Energy (DOE), Office of Science under contract DE-AC02-05CH11231. GHL would like to thank Evangelos Theodorou for helpful discussions. This research used resources of the National Energy Research Scientific Computing Center, a DOE Office of Science User Facility supported by the Office of Science of the U.S. Department of Energy under Contract No. DE-AC02-05CH11231 using NERSC award HEP-ERCAP0021099.

\appendix

\section{Comparison with standard diffusion models}
\label{app:diffusion}
\begin{figure*}[ht]
\centering
    \includegraphics[width=0.3\textwidth]{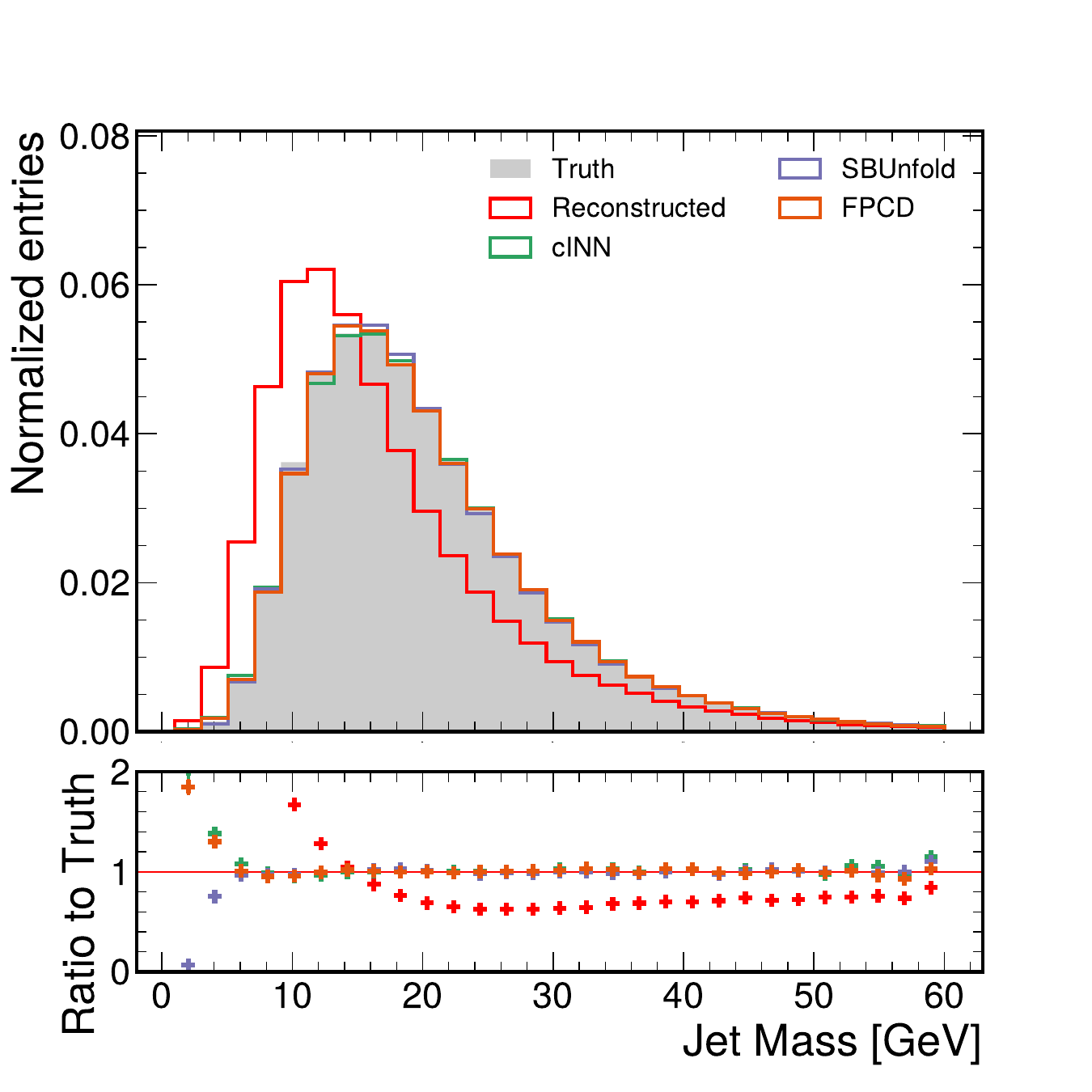}
    \includegraphics[width=0.3\textwidth]{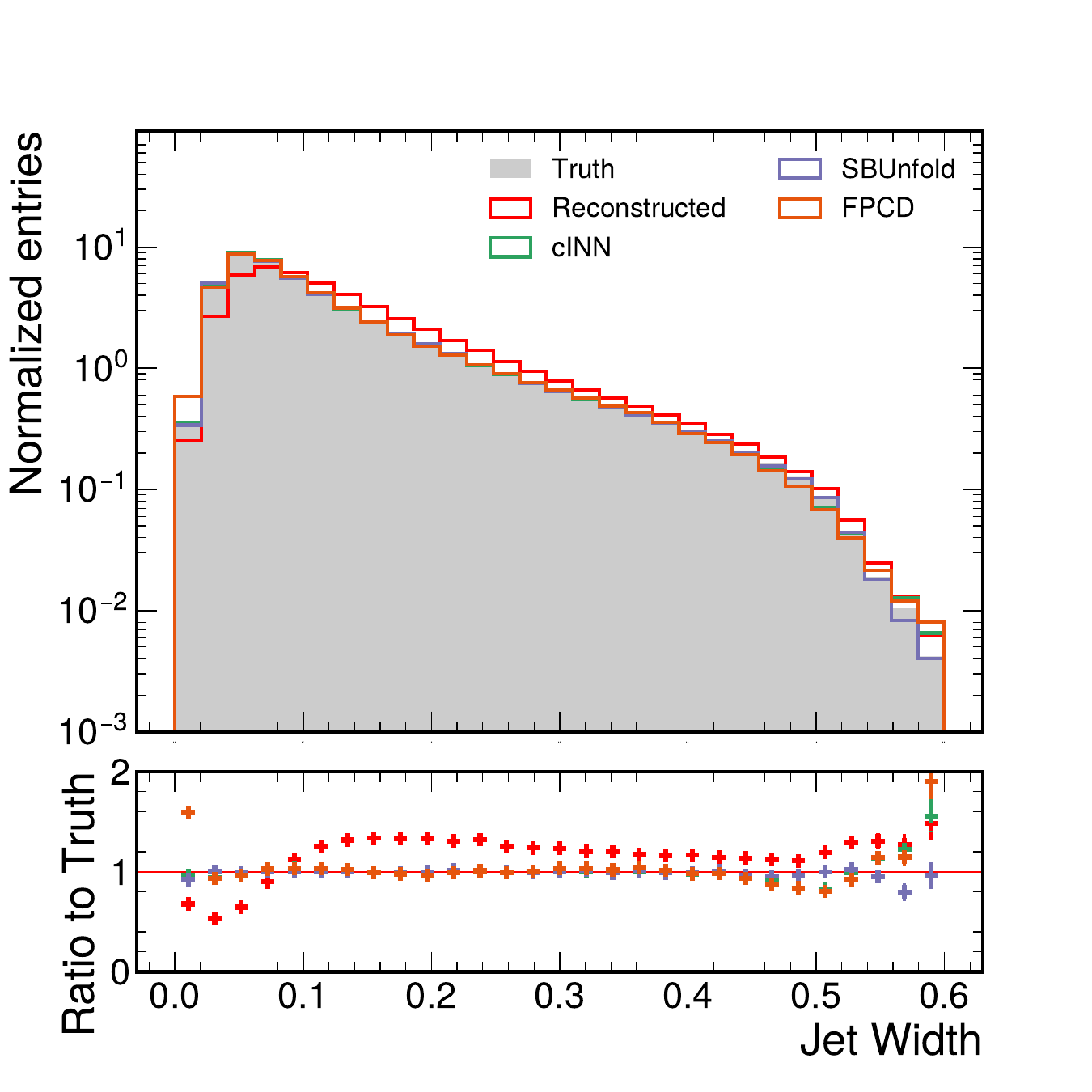}
    \includegraphics[width=0.3\textwidth]{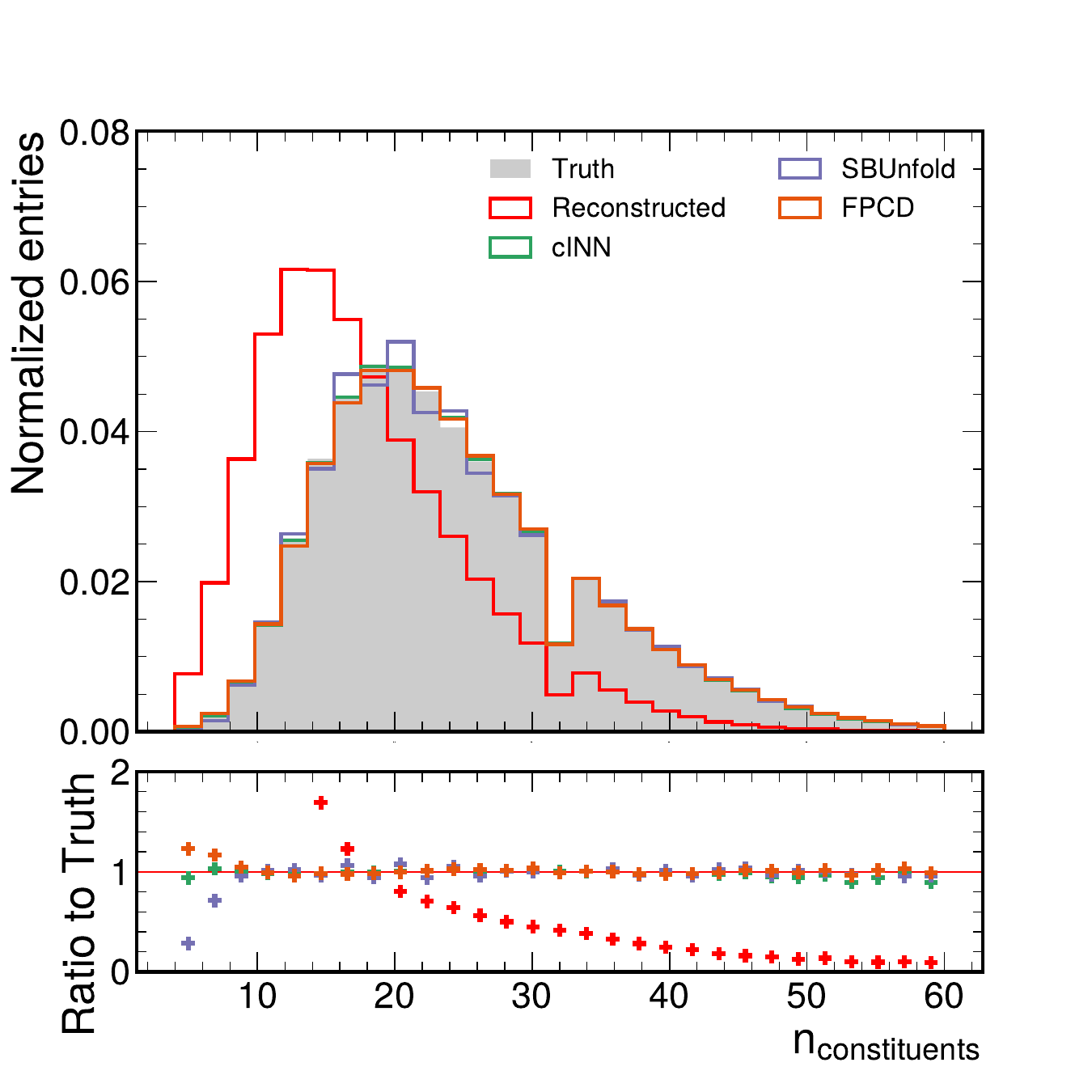}
    \includegraphics[width=0.3\textwidth]{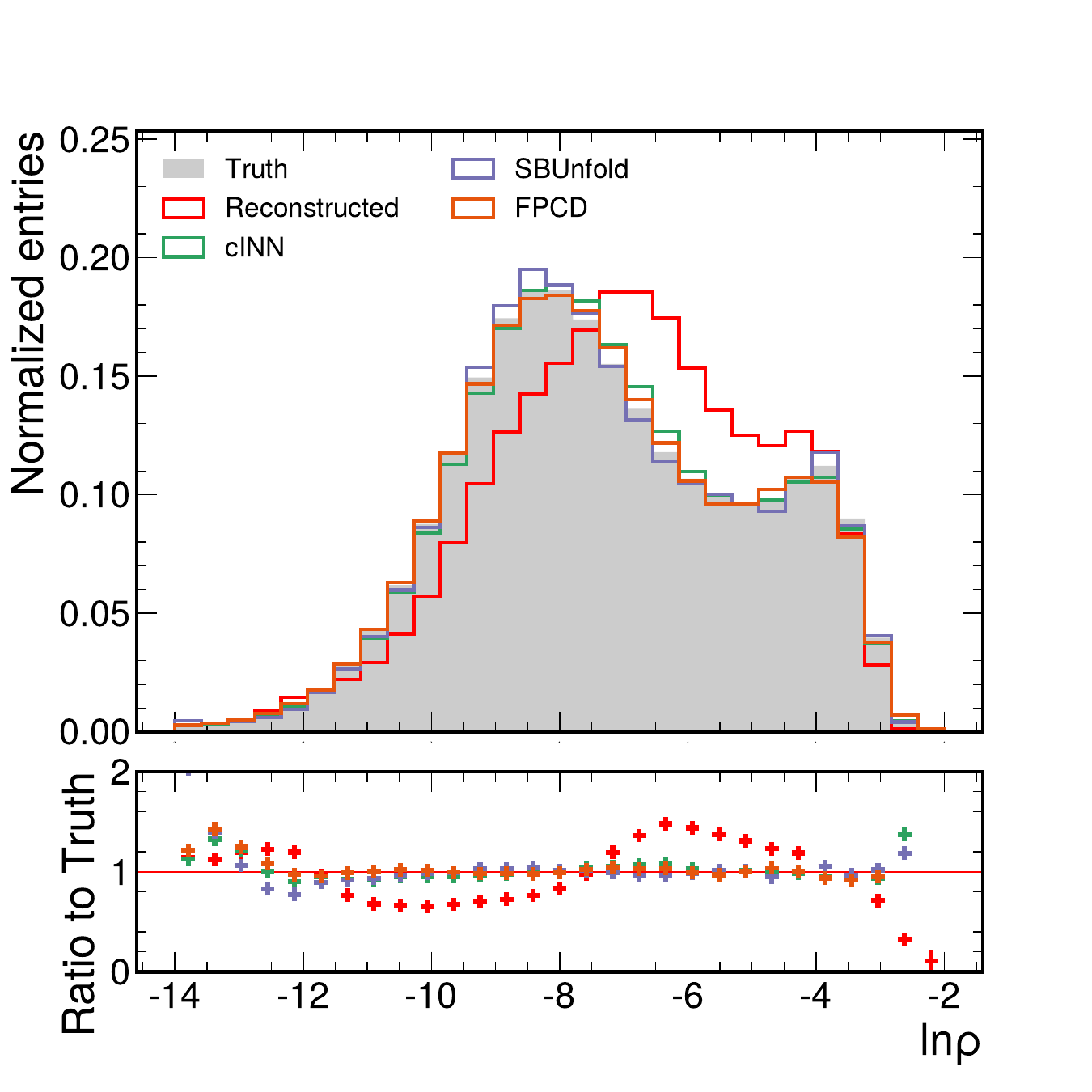}
    \includegraphics[width=0.3\textwidth]{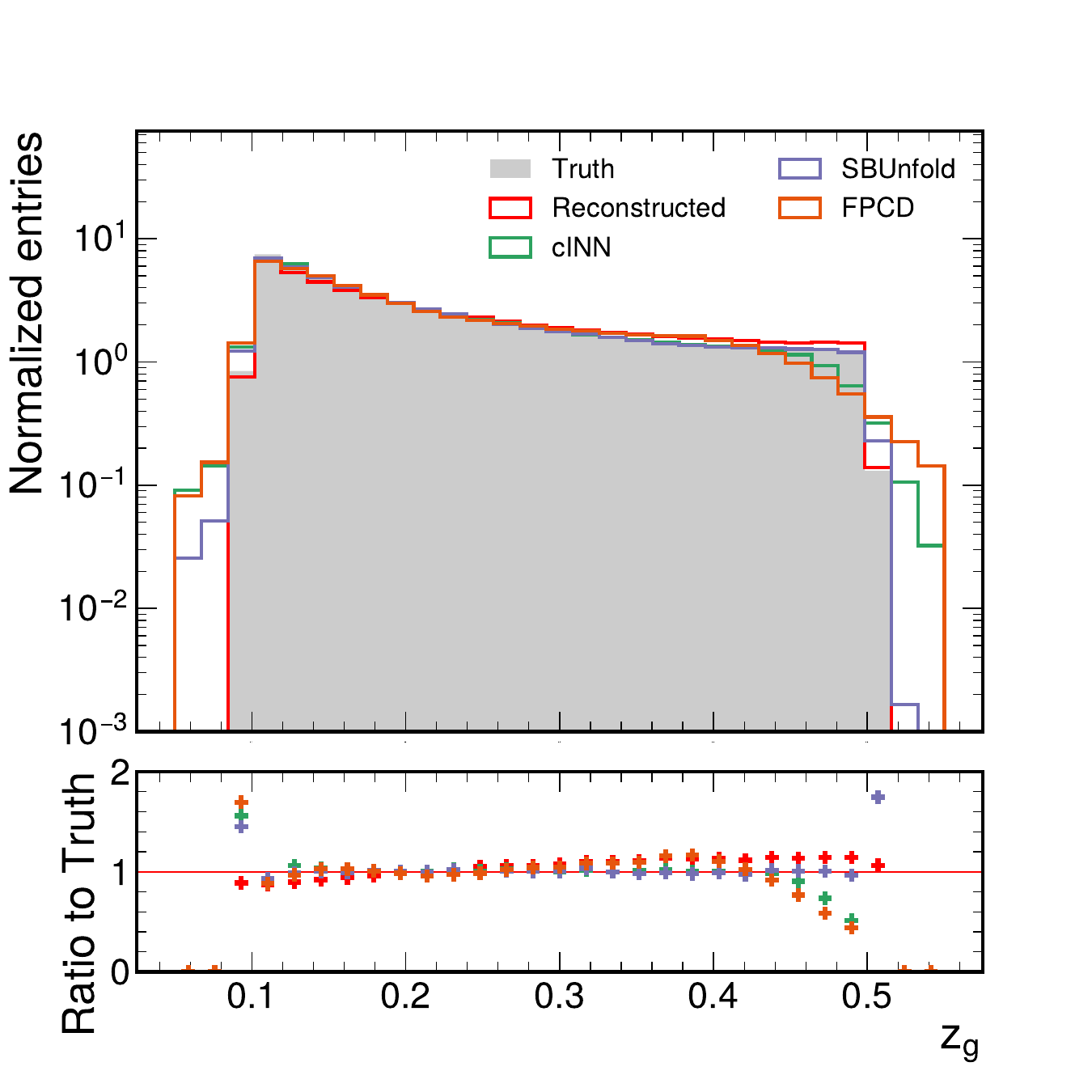}
    \includegraphics[width=0.3\textwidth]{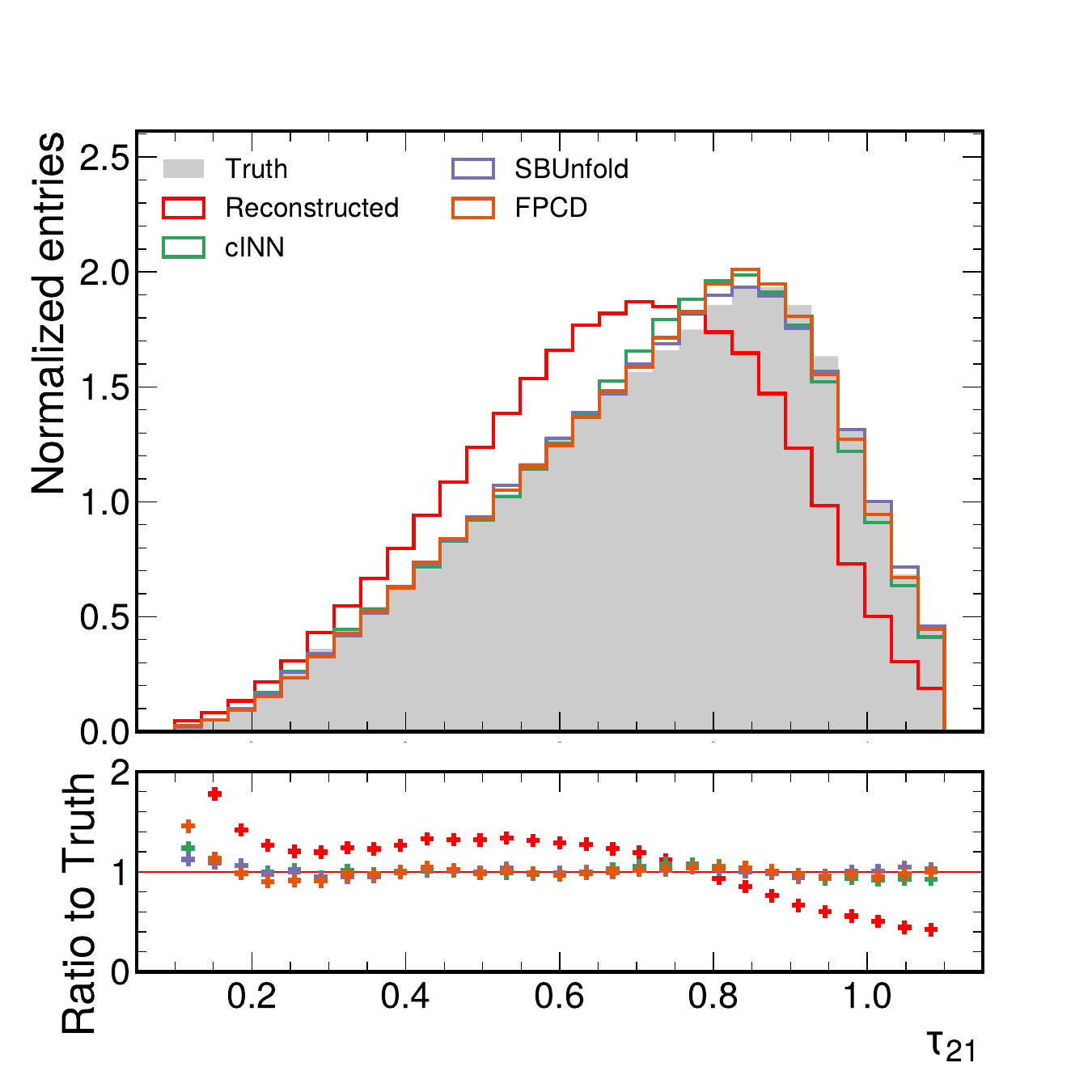}
\caption{Comparison between different unfolding algorithms for six different physics observables unfolded.  Results are evaluated over 600'000 pseudo-data points. Statistical uncertainties are shown only in the ratio panel. Pseudo-data and simulation are described by Pythia.}
\label{fig:unfold_600k_pythia_diffusion}
\end{figure*}

We investigate the benefits of using \textsc{SBUnfold} as opposed to a standard diffusion model. The diffusion model implementation follows the \textsc{FPCD} model proposed in~\cite{PhysRevD.108.036025} for jet kinematic generation. The backbone neural network architecture used is the same as the one used in \textsc{SBUnfold}. Similarly, the learning rate, number of training epochs, and training samples are kept the same between the two models. The main conceptual difference between the implementations is that \textsc{SBUnfold} starts from the reconstruction level events instead of a Gaussian prior used in FPCD. Similarly to the \textsc{cINN} implementation, the FPCD model is only conditioned on reconstruction level events to determine the unfolded response. Results of the unfolded distributions using \textsc{Pythia} both as pseudo-data and simulation are shown in Fig.~\ref{fig:unfold_600k_pythia_diffusion}.

For all distributions we observe a good agreement between the unfolded results and the pseudo-data. In Table~\ref{tab:EMD_600k_pythia_diffusion} we calculate the EMD and triangular discriminator using the unfolded distributions obtained by the different methodologies. 

\begin{table*}[ht]
    \centering
	\small
    \caption{Comparison of the earth mover's distance (EMD) and triangular discriminator between different unfolding methodologies. EMD is calculated over unbinned distributions while triangular discriminator uses histograms as inputs. Uncertainties from EMD are derived using 100 bootstraps with replacement taken from the unfolded data. Results are evaluated using 600'000 pseudo-data points sampled from Pythia. Quantities in bold represent the method with best performance.}
    \label{tab:EMD_600k_pythia_diffusion}
	\begin{tabular}{l|c|c|c|c|c|cc}
        Model &   \multicolumn{3}{c}{EMD($\times 10$)/Triangular Discriminator($\times10^3$)}\\
        & {\scriptsize \textsc{FPCD}}  & {\scriptsize \textsc{cINN}} & {\scriptsize \textsc{SBUnfold}} \\
        \hline     
        Jet mass &  0.74$\pm$0.08/\textbf{0.19} & 1.4$\pm$0.2/0.29 & \textbf{0.70$\pm$0.06}/0.30 \\
        Jet Width &  0.0087$\pm$0.0006/0.9 &  0.013$\pm$0.002/0.25 & \textbf{0.0029$\pm$0.0005}/\textbf{0.04} \\
        N & 0.81$\pm$0.06/0.1 & 2.3$\pm$0.8/\textbf{0.09} & \textbf{0.57$\pm$0.04}/0.9 \\ 
        $\log\rho$ &  0.34$\pm$0.01/0.77 &1.1$\pm$0.3/\textbf{0.64} & \textbf{0.27$\pm$0.01}/0.68 \\ 
        $z_g$ & 0.035$\pm$0.007/12.4 &  0.095$\pm$0.003/10.9 & \textbf{0.009$\pm$0.001}/\textbf{3.1} \\
        $\tau_{21}$  &  0.024$\pm$0.002/0.3 & 0.2$\pm$0.1/0.6 & \textbf{0.016$\pm$0.001}/\textbf{0.2} \\
	\end{tabular}
\end{table*}

From the results we observe that \textsc{FPCD} often outperforms the \textsc{cINN} in the EMD metric while still not achieving the same level of performance as \textsc{SBUnfold}. These results indicate that standard diffusion models may be more expressive compared to the normalizing flow model implemented in the original \textsc{cINN}, but still not as precise as \textsc{SBUnfold} which leverages the more informative prior distribution from reconstruction level objects rather than the Gaussian prior required by both \textsc{cINN} and \textsc{FPCD}.

\section{Migration matrix of unfolded events}
\label{app:migration}
\begin{figure*}[ht]
\centering
    \includegraphics[width=0.3\textwidth]{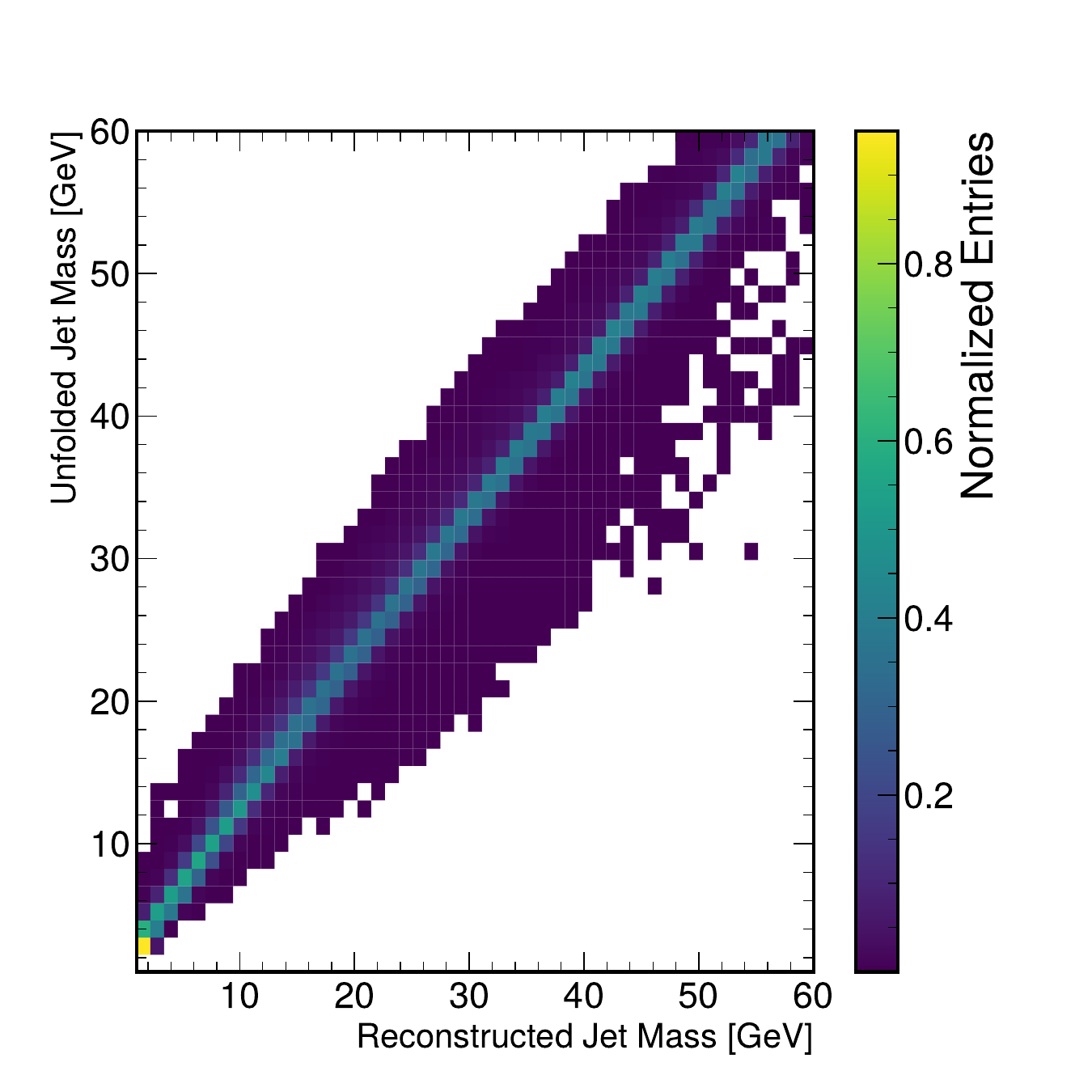}
    \includegraphics[width=0.3\textwidth]{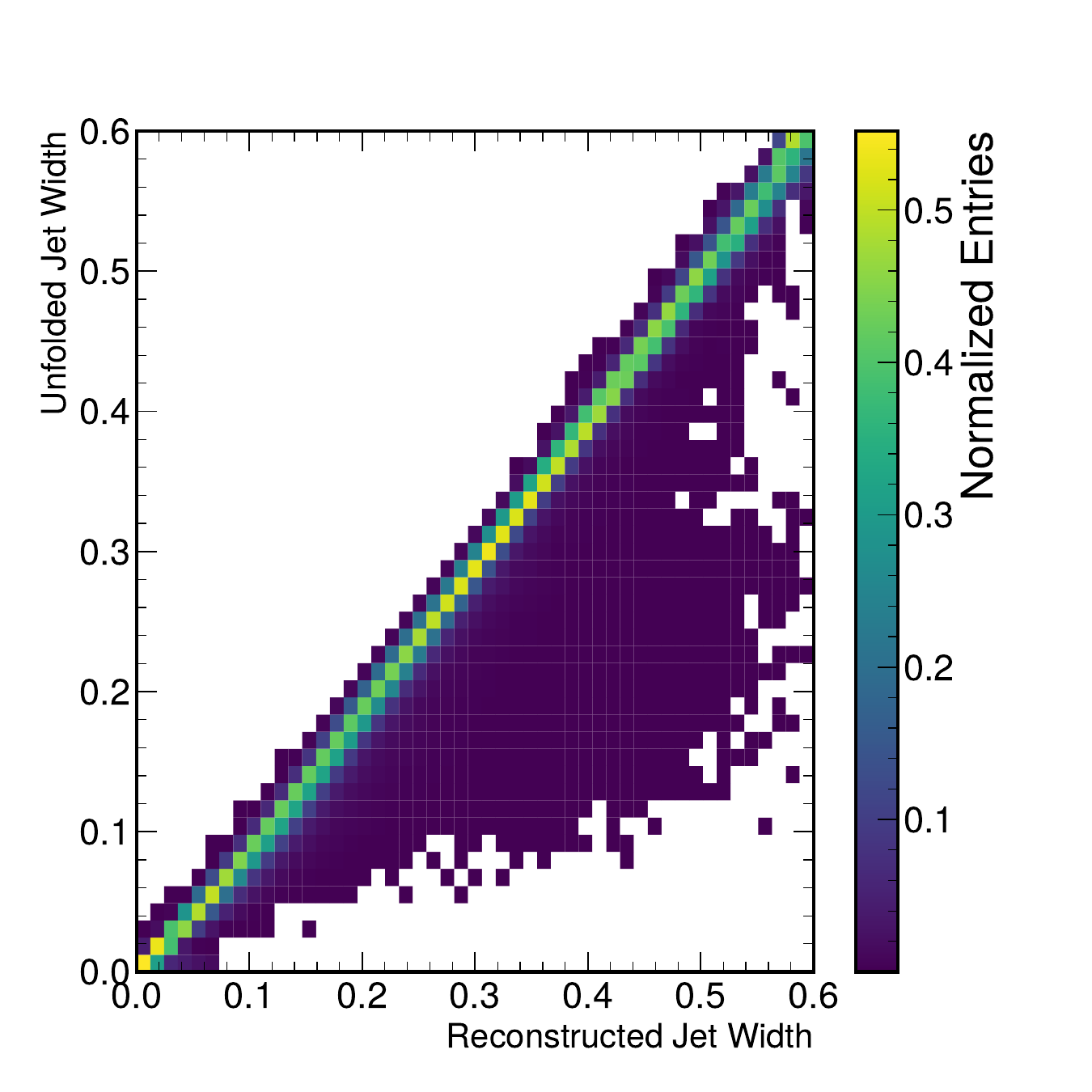}
    \includegraphics[width=0.3\textwidth]{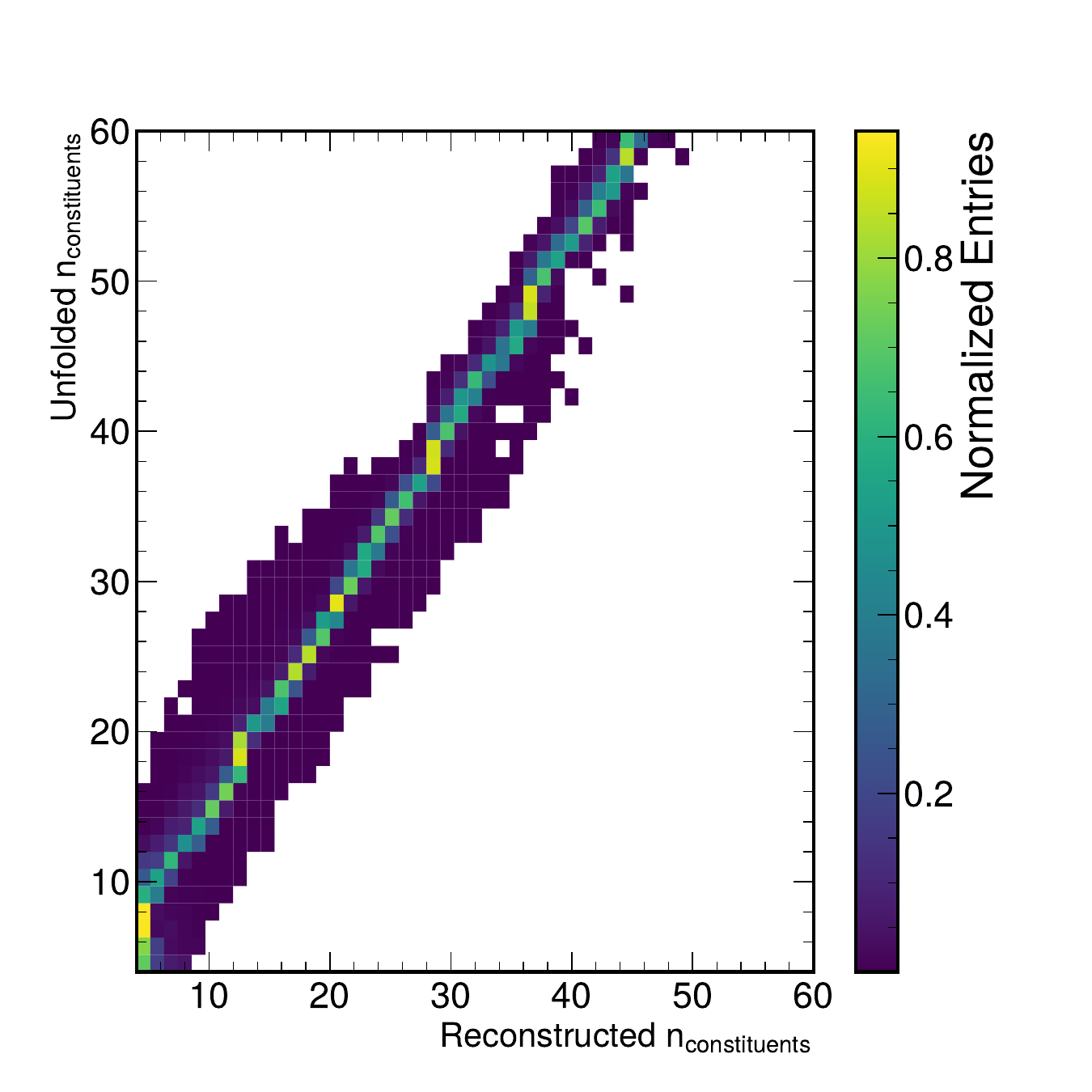}
    \includegraphics[width=0.3\textwidth]{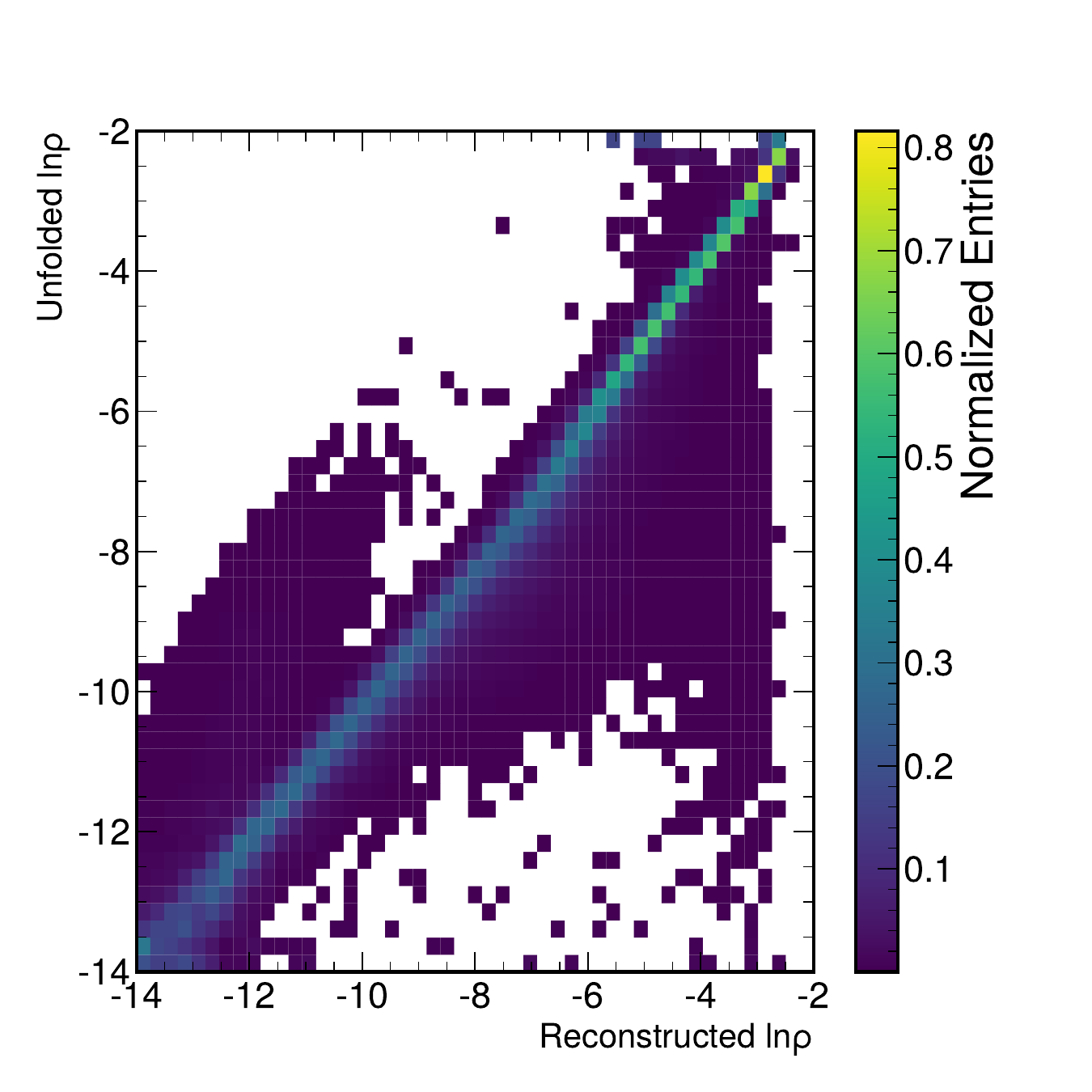}
    \includegraphics[width=0.3\textwidth]{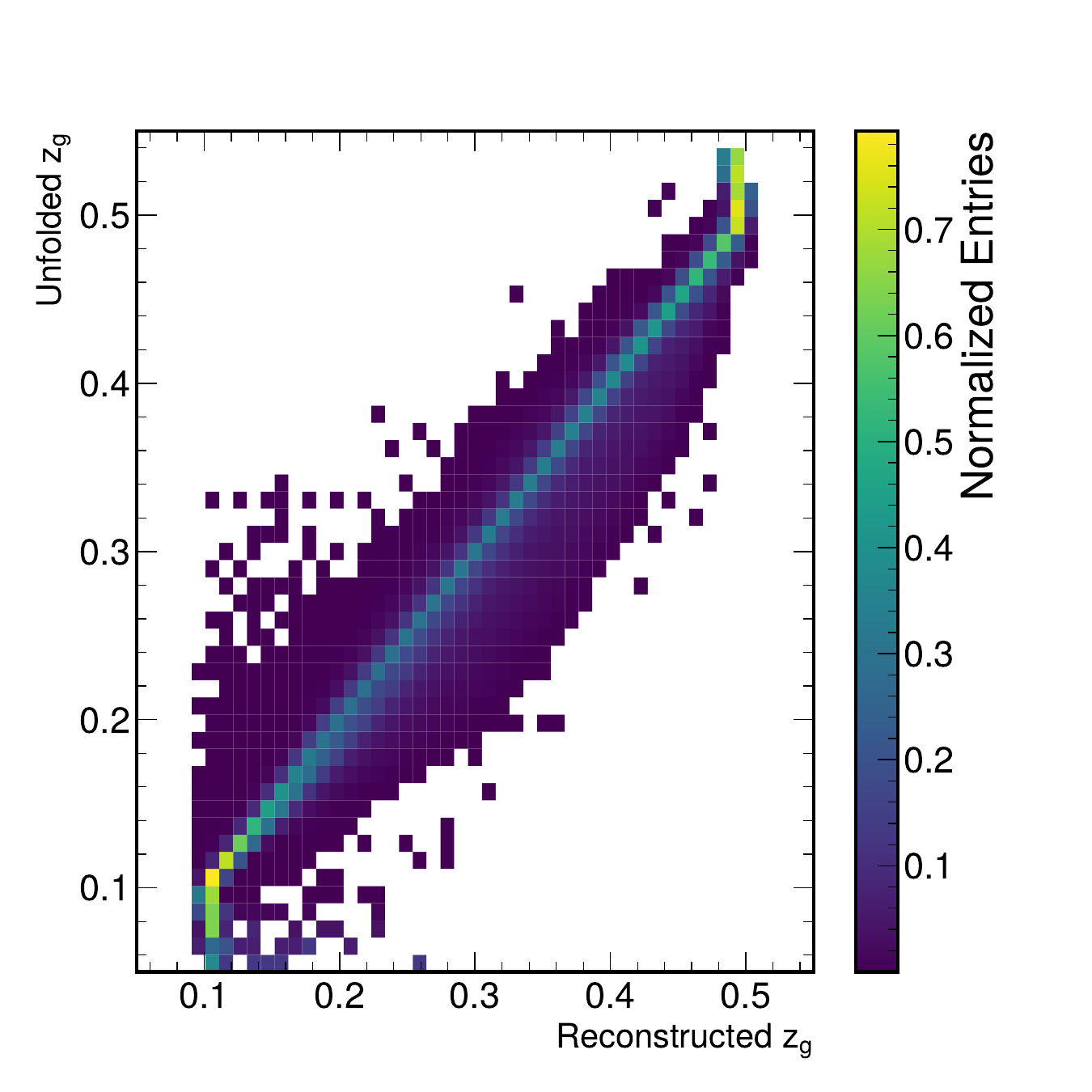}
    \includegraphics[width=0.3\textwidth]{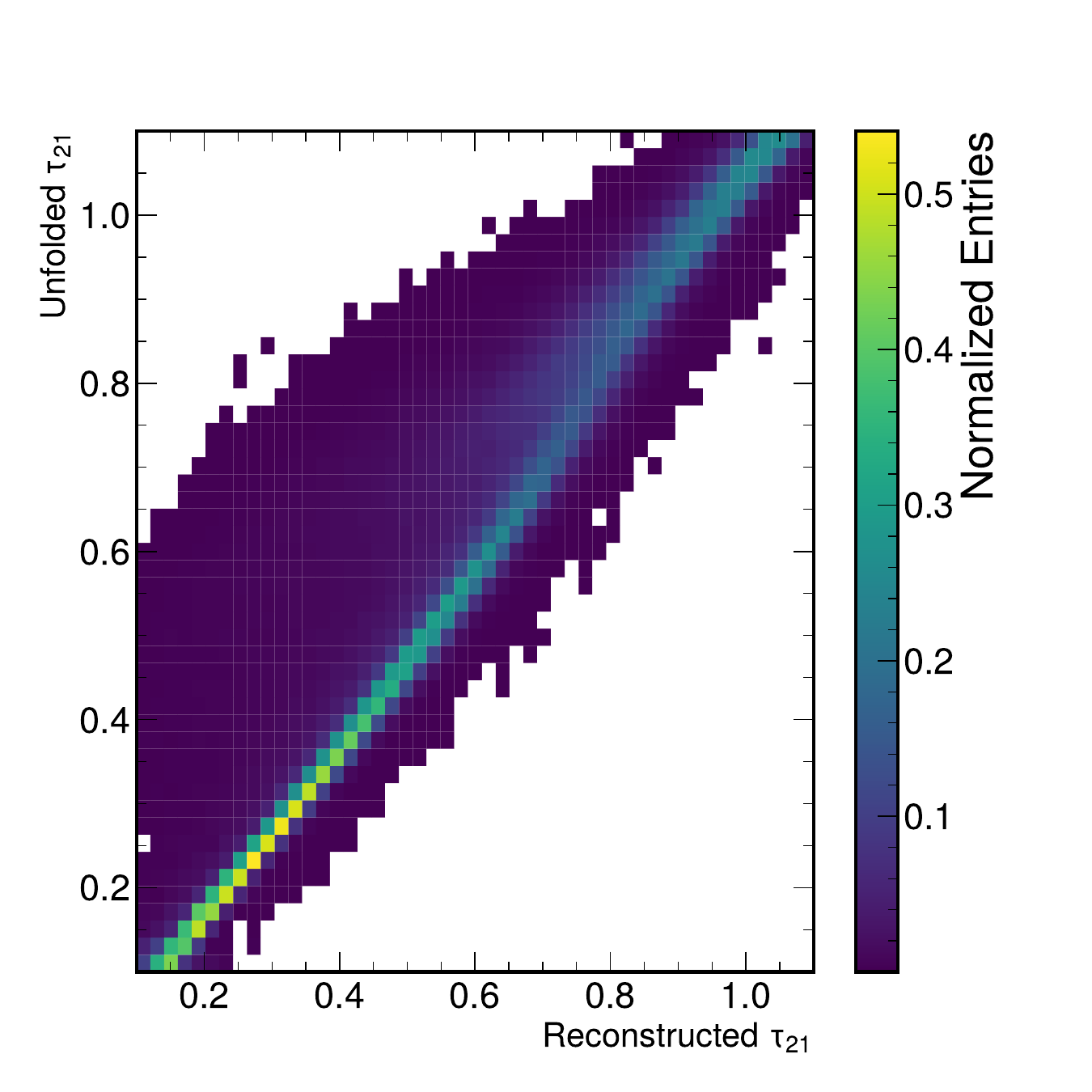}        
\caption{Migration distribution between reconstructed level events and generator level distribution after unfolding with \textsc{SBUnfold}. Results are evaluated over 600'000 pseudo-data points. Pseudo-data is taken from Herwig while Pythia is taken as the main simulator.}
\label{fig:unfold_migration}
\end{figure*}

In this appendix we investigate how \textsc{SBUnfold} maps reconstructed level distributions back to generator level samples by calculating the corresponding migration matrix using Herwig as the pseudo-data while the detector response is derived from the Pythia simulation. The distributions are shown in Fig.~\ref{fig:unfold_migration}.

The migration matrix for all observables has a high fraction of events present close to the diagonal, consistent with \textsc{SBunfold} learning to apply only a small correction to the inputs.

\bibliography{HEPML,other}

\begin{thebibliography}{58}%
\makeatletter
\providecommand \@ifxundefined [1]{%
 \@ifx{#1\undefined}
}%
\providecommand \@ifnum [1]{%
 \ifnum #1\expandafter \@firstoftwo
 \else \expandafter \@secondoftwo
 \fi
}%
\providecommand \@ifx [1]{%
 \ifx #1\expandafter \@firstoftwo
 \else \expandafter \@secondoftwo
 \fi
}%
\providecommand \natexlab [1]{#1}%
\providecommand \enquote  [1]{``#1''}%
\providecommand \bibnamefont  [1]{#1}%
\providecommand \bibfnamefont [1]{#1}%
\providecommand \citenamefont [1]{#1}%
\providecommand \href@noop [0]{\@secondoftwo}%
\providecommand \href [0]{\begingroup \@sanitize@url \@href}%
\providecommand \@href[1]{\@@startlink{#1}\@@href}%
\providecommand \@@href[1]{\endgroup#1\@@endlink}%
\providecommand \@sanitize@url [0]{\catcode `\\12\catcode `\$12\catcode
  `\&12\catcode `\#12\catcode `\^12\catcode `\_12\catcode `\%12\relax}%
\providecommand \@@startlink[1]{}%
\providecommand \@@endlink[0]{}%
\providecommand \url  [0]{\begingroup\@sanitize@url \@url }%
\providecommand \@url [1]{\endgroup\@href {#1}{\urlprefix }}%
\providecommand \urlprefix  [0]{URL }%
\providecommand \Eprint [0]{\href }%
\providecommand \doibase [0]{http://dx.doi.org/}%
\providecommand \selectlanguage [0]{\@gobble}%
\providecommand \bibinfo  [0]{\@secondoftwo}%
\providecommand \bibfield  [0]{\@secondoftwo}%
\providecommand \translation [1]{[#1]}%
\providecommand \BibitemOpen [0]{}%
\providecommand \bibitemStop [0]{}%
\providecommand \bibitemNoStop [0]{.\EOS\space}%
\providecommand \EOS [0]{\spacefactor3000\relax}%
\providecommand \BibitemShut  [1]{\csname bibitem#1\endcsname}%
\let\auto@bib@innerbib\@empty
\bibitem [{\citenamefont {Glazov}(2017)}]{Glazov:2017vni}%
  \BibitemOpen
  \bibfield  {author} {\bibinfo {author} {\bibfnamefont {A.}~\bibnamefont
  {Glazov}},\ }\href@noop {} {\  (\bibinfo {year} {2017})},\ \Eprint
  {http://arxiv.org/abs/1712.01814} {arXiv:1712.01814 [physics.data-an]}
  \BibitemShut {NoStop}%
\bibitem [{\citenamefont {Datta}\ \emph {et~al.}(2018)\citenamefont {Datta},
  \citenamefont {Kar},\ and\ \citenamefont {Roy}}]{Datta:2018mwd}%
  \BibitemOpen
  \bibfield  {author} {\bibinfo {author} {\bibfnamefont {K.}~\bibnamefont
  {Datta}}, \bibinfo {author} {\bibfnamefont {D.}~\bibnamefont {Kar}}, \ and\
  \bibinfo {author} {\bibfnamefont {D.}~\bibnamefont {Roy}},\ }\href@noop {} {\
   (\bibinfo {year} {2018})},\ \Eprint {http://arxiv.org/abs/1806.00433}
  {arXiv:1806.00433 [physics.data-an]} \BibitemShut {NoStop}%
\bibitem [{\citenamefont {Bunse}\ \emph {et~al.}(2018)\citenamefont {Bunse},
  \citenamefont {Piatkowski}, \citenamefont {Ruhe}, \citenamefont {Rhode},\
  and\ \citenamefont {Morik}}]{bunse2018unification}%
  \BibitemOpen
  \bibfield  {author} {\bibinfo {author} {\bibfnamefont {M.}~\bibnamefont
  {Bunse}}, \bibinfo {author} {\bibfnamefont {N.}~\bibnamefont {Piatkowski}},
  \bibinfo {author} {\bibfnamefont {T.}~\bibnamefont {Ruhe}}, \bibinfo {author}
  {\bibfnamefont {W.}~\bibnamefont {Rhode}}, \ and\ \bibinfo {author}
  {\bibfnamefont {K.}~\bibnamefont {Morik}},\ }in\ \href@noop {} {\emph
  {\bibinfo {booktitle} {5th International Conference on Data Science and
  Advanced Analytics (DSAA)}}}\ (\bibinfo {organization} {IEEE},\ \bibinfo
  {year} {2018})\ pp.\ \bibinfo {pages} {21--30}\BibitemShut {NoStop}%
\bibitem [{\citenamefont {Ruhe}\ \emph {et~al.}(2019)\citenamefont {Ruhe},
  \citenamefont {Voigt}, \citenamefont {Wornowizki}, \citenamefont
  {B{\"o}rner}, \citenamefont {Rhode},\ and\ \citenamefont
  {Morik}}]{Ruhe2019MiningFS}%
  \BibitemOpen
  \bibfield  {author} {\bibinfo {author} {\bibfnamefont {T.}~\bibnamefont
  {Ruhe}}, \bibinfo {author} {\bibfnamefont {T.}~\bibnamefont {Voigt}},
  \bibinfo {author} {\bibfnamefont {M.}~\bibnamefont {Wornowizki}}, \bibinfo
  {author} {\bibfnamefont {M.}~\bibnamefont {B{\"o}rner}}, \bibinfo {author}
  {\bibfnamefont {W.}~\bibnamefont {Rhode}}, \ and\ \bibinfo {author}
  {\bibfnamefont {K.}~\bibnamefont {Morik}}\ }(\bibinfo {year}
  {2019})\BibitemShut {NoStop}%
\bibitem [{\citenamefont {Andreassen}\ \emph {et~al.}(2020)\citenamefont
  {Andreassen}, \citenamefont {Komiske}, \citenamefont {Metodiev},
  \citenamefont {Nachman},\ and\ \citenamefont {Thaler}}]{Andreassen:2019cjw}%
  \BibitemOpen
  \bibfield  {author} {\bibinfo {author} {\bibfnamefont {A.}~\bibnamefont
  {Andreassen}}, \bibinfo {author} {\bibfnamefont {P.~T.}\ \bibnamefont
  {Komiske}}, \bibinfo {author} {\bibfnamefont {E.~M.}\ \bibnamefont
  {Metodiev}}, \bibinfo {author} {\bibfnamefont {B.}~\bibnamefont {Nachman}}, \
  and\ \bibinfo {author} {\bibfnamefont {J.}~\bibnamefont {Thaler}},\ }\href
  {\doibase 10.1103/PhysRevLett.124.182001} {\bibfield  {journal} {\bibinfo
  {journal} {Phys. Rev. Lett.}\ }\textbf {\bibinfo {volume} {124}},\ \bibinfo
  {pages} {182001} (\bibinfo {year} {2020})},\ \Eprint
  {http://arxiv.org/abs/1911.09107} {arXiv:1911.09107 [hep-ph]} \BibitemShut
  {NoStop}%
\bibitem [{\citenamefont {Bellagente}\ \emph {et~al.}(2019)\citenamefont
  {Bellagente}, \citenamefont {Butter}, \citenamefont {Kasieczka},
  \citenamefont {Plehn},\ and\ \citenamefont
  {Winterhalder}}]{Bellagente:2019uyp}%
  \BibitemOpen
  \bibfield  {author} {\bibinfo {author} {\bibfnamefont {M.}~\bibnamefont
  {Bellagente}}, \bibinfo {author} {\bibfnamefont {A.}~\bibnamefont {Butter}},
  \bibinfo {author} {\bibfnamefont {G.}~\bibnamefont {Kasieczka}}, \bibinfo
  {author} {\bibfnamefont {T.}~\bibnamefont {Plehn}}, \ and\ \bibinfo {author}
  {\bibfnamefont {R.}~\bibnamefont {Winterhalder}},\ }\href {\doibase
  10.21468/SciPostPhys.8.4.070} {\  (\bibinfo {year} {2019}),\
  10.21468/SciPostPhys.8.4.070},\ \Eprint {http://arxiv.org/abs/1912.00477}
  {arXiv:1912.00477 [hep-ph]} \BibitemShut {NoStop}%
\bibitem [{\citenamefont {Bellagente}\ \emph {et~al.}(2020)\citenamefont
  {Bellagente}, \citenamefont {Butter}, \citenamefont {Kasieczka},
  \citenamefont {Plehn}, \citenamefont {Rousselot},\ and\ \citenamefont
  {Winterhalder}}]{1800956}%
  \BibitemOpen
  \bibfield  {author} {\bibinfo {author} {\bibfnamefont {M.}~\bibnamefont
  {Bellagente}}, \bibinfo {author} {\bibfnamefont {A.}~\bibnamefont {Butter}},
  \bibinfo {author} {\bibfnamefont {G.}~\bibnamefont {Kasieczka}}, \bibinfo
  {author} {\bibfnamefont {T.}~\bibnamefont {Plehn}}, \bibinfo {author}
  {\bibfnamefont {A.}~\bibnamefont {Rousselot}}, \ and\ \bibinfo {author}
  {\bibfnamefont {R.}~\bibnamefont {Winterhalder}},\ }\href {\doibase
  10.21468/SciPostPhys.9.5.074} {\  (\bibinfo {year} {2020}),\
  10.21468/SciPostPhys.9.5.074},\ \Eprint {http://arxiv.org/abs/2006.06685}
  {arXiv:2006.06685 [hep-ph]} \BibitemShut {NoStop}%
\bibitem [{\citenamefont {Vandegar}\ \emph {et~al.}(2020)\citenamefont
  {Vandegar}, \citenamefont {Kagan}, \citenamefont {Wehenkel},\ and\
  \citenamefont {Louppe}}]{Vandegar:2020yvw}%
  \BibitemOpen
  \bibfield  {author} {\bibinfo {author} {\bibfnamefont {M.}~\bibnamefont
  {Vandegar}}, \bibinfo {author} {\bibfnamefont {M.}~\bibnamefont {Kagan}},
  \bibinfo {author} {\bibfnamefont {A.}~\bibnamefont {Wehenkel}}, \ and\
  \bibinfo {author} {\bibfnamefont {G.}~\bibnamefont {Louppe}},\ }\href@noop {}
  {\  (\bibinfo {year} {2020})},\ \Eprint {http://arxiv.org/abs/2011.05836}
  {arXiv:2011.05836 [stat.ML]} \BibitemShut {NoStop}%
\bibitem [{\citenamefont {Andreassen}\ \emph {et~al.}(2021)\citenamefont
  {Andreassen}, \citenamefont {Komiske}, \citenamefont {Metodiev},
  \citenamefont {Nachman}, \citenamefont {Suresh},\ and\ \citenamefont
  {Thaler}}]{Andreassen:2021zzk}%
  \BibitemOpen
  \bibfield  {author} {\bibinfo {author} {\bibfnamefont {A.}~\bibnamefont
  {Andreassen}}, \bibinfo {author} {\bibfnamefont {P.~T.}\ \bibnamefont
  {Komiske}}, \bibinfo {author} {\bibfnamefont {E.~M.}\ \bibnamefont
  {Metodiev}}, \bibinfo {author} {\bibfnamefont {B.}~\bibnamefont {Nachman}},
  \bibinfo {author} {\bibfnamefont {A.}~\bibnamefont {Suresh}}, \ and\ \bibinfo
  {author} {\bibfnamefont {J.}~\bibnamefont {Thaler}},\ }\href@noop {} {\
  (\bibinfo {year} {2021})},\ \Eprint {http://arxiv.org/abs/2105.04448}
  {arXiv:2105.04448 [stat.ML]} \BibitemShut {NoStop}%
\bibitem [{\citenamefont {Howard}\ \emph {et~al.}(2022)\citenamefont {Howard},
  \citenamefont {Mandt}, \citenamefont {Whiteson},\ and\ \citenamefont
  {Yang}}]{Howard:2021pos}%
  \BibitemOpen
  \bibfield  {author} {\bibinfo {author} {\bibfnamefont {J.~N.}\ \bibnamefont
  {Howard}}, \bibinfo {author} {\bibfnamefont {S.}~\bibnamefont {Mandt}},
  \bibinfo {author} {\bibfnamefont {D.}~\bibnamefont {Whiteson}}, \ and\
  \bibinfo {author} {\bibfnamefont {Y.}~\bibnamefont {Yang}},\ }\href {\doibase
  10.1038/s41598-022-10966-7} {\bibfield  {journal} {\bibinfo  {journal} {Sci.
  Rep.}\ }\textbf {\bibinfo {volume} {12}},\ \bibinfo {pages} {7567} (\bibinfo
  {year} {2022})},\ \Eprint {http://arxiv.org/abs/2101.08944} {arXiv:2101.08944
  [hep-ph]} \BibitemShut {NoStop}%
\bibitem [{\citenamefont {Backes}\ \emph {et~al.}(2022)\citenamefont {Backes},
  \citenamefont {Butter}, \citenamefont {Dunford},\ and\ \citenamefont
  {Malaescu}}]{Backes:2022vmn}%
  \BibitemOpen
  \bibfield  {author} {\bibinfo {author} {\bibfnamefont {M.}~\bibnamefont
  {Backes}}, \bibinfo {author} {\bibfnamefont {A.}~\bibnamefont {Butter}},
  \bibinfo {author} {\bibfnamefont {M.}~\bibnamefont {Dunford}}, \ and\
  \bibinfo {author} {\bibfnamefont {B.}~\bibnamefont {Malaescu}},\ }\href@noop
  {} {\  (\bibinfo {year} {2022})},\ \Eprint {http://arxiv.org/abs/2212.08674}
  {arXiv:2212.08674 [hep-ph]} \BibitemShut {NoStop}%
\bibitem [{\citenamefont {Arratia}\ \emph {et~al.}(2022)\citenamefont
  {Arratia}, \citenamefont {Britzger}, \citenamefont {Long},\ and\
  \citenamefont {Nachman}}]{Arratia:2022wny}%
  \BibitemOpen
  \bibfield  {author} {\bibinfo {author} {\bibfnamefont {M.}~\bibnamefont
  {Arratia}}, \bibinfo {author} {\bibfnamefont {D.}~\bibnamefont {Britzger}},
  \bibinfo {author} {\bibfnamefont {O.}~\bibnamefont {Long}}, \ and\ \bibinfo
  {author} {\bibfnamefont {B.}~\bibnamefont {Nachman}},\ }\href {\doibase
  10.1088/1748-0221/17/07/P07009} {\bibfield  {journal} {\bibinfo  {journal}
  {JINST}\ }\textbf {\bibinfo {volume} {17}},\ \bibinfo {pages} {P07009}
  (\bibinfo {year} {2022})},\ \Eprint {http://arxiv.org/abs/2203.16722}
  {arXiv:2203.16722 [hep-ex]} \BibitemShut {NoStop}%
\bibitem [{\citenamefont {Chan}\ and\ \citenamefont
  {Nachman}(2023)}]{Chan:2023tbf}%
  \BibitemOpen
  \bibfield  {author} {\bibinfo {author} {\bibfnamefont {J.}~\bibnamefont
  {Chan}}\ and\ \bibinfo {author} {\bibfnamefont {B.}~\bibnamefont {Nachman}},\
  }\href {\doibase 10.1103/PhysRevD.108.016002} {\bibfield  {journal} {\bibinfo
   {journal} {Phys. Rev. D}\ }\textbf {\bibinfo {volume} {108}},\ \bibinfo
  {pages} {016002} (\bibinfo {year} {2023})},\ \Eprint
  {http://arxiv.org/abs/2302.05390} {arXiv:2302.05390 [hep-ph]} \BibitemShut
  {NoStop}%
\bibitem [{\citenamefont {Shmakov}\ \emph {et~al.}(2023)\citenamefont
  {Shmakov}, \citenamefont {Greif}, \citenamefont {Fenton}, \citenamefont
  {Ghosh}, \citenamefont {Baldi},\ and\ \citenamefont
  {Whiteson}}]{Shmakov:2023kjj}%
  \BibitemOpen
  \bibfield  {author} {\bibinfo {author} {\bibfnamefont {A.}~\bibnamefont
  {Shmakov}}, \bibinfo {author} {\bibfnamefont {K.}~\bibnamefont {Greif}},
  \bibinfo {author} {\bibfnamefont {M.}~\bibnamefont {Fenton}}, \bibinfo
  {author} {\bibfnamefont {A.}~\bibnamefont {Ghosh}}, \bibinfo {author}
  {\bibfnamefont {P.}~\bibnamefont {Baldi}}, \ and\ \bibinfo {author}
  {\bibfnamefont {D.}~\bibnamefont {Whiteson}},\ }\href@noop {} {\  (\bibinfo
  {year} {2023})},\ \Eprint {http://arxiv.org/abs/2305.10399} {arXiv:2305.10399
  [hep-ex]} \BibitemShut {NoStop}%
\bibitem [{\citenamefont {Alghamdi}\ \emph {et~al.}(2023)\citenamefont
  {Alghamdi} \emph {et~al.}}]{Alghamdi:2023emm}%
  \BibitemOpen
  \bibfield  {author} {\bibinfo {author} {\bibfnamefont {T.}~\bibnamefont
  {Alghamdi}} \emph {et~al.},\ }\href@noop {} {\  (\bibinfo {year} {2023})},\
  \Eprint {http://arxiv.org/abs/2307.04450} {arXiv:2307.04450 [hep-ph]}
  \BibitemShut {NoStop}%
\bibitem [{\citenamefont {Arratia}\ \emph {et~al.}(2021)\citenamefont {Arratia}
  \emph {et~al.}}]{Arratia:2021otl}%
  \BibitemOpen
  \bibfield  {author} {\bibinfo {author} {\bibfnamefont {M.}~\bibnamefont
  {Arratia}} \emph {et~al.},\ }\href@noop {} {\  (\bibinfo {year} {2021})},\
  \Eprint {http://arxiv.org/abs/2109.13243} {arXiv:2109.13243 [hep-ph]}
  \BibitemShut {NoStop}%
\bibitem [{\citenamefont {Andreev}\ \emph {et~al.}(2022)\citenamefont {Andreev}
  \emph {et~al.}}]{H1:2021wkz}%
  \BibitemOpen
  \bibfield  {author} {\bibinfo {author} {\bibfnamefont {V.}~\bibnamefont
  {Andreev}} \emph {et~al.} (\bibinfo {collaboration} {H1}),\ }\href {\doibase
  10.1103/PhysRevLett.128.132002} {\bibfield  {journal} {\bibinfo  {journal}
  {Phys. Rev. Lett.}\ }\textbf {\bibinfo {volume} {128}},\ \bibinfo {pages}
  {132002} (\bibinfo {year} {2022})},\ \Eprint
  {http://arxiv.org/abs/2108.12376} {arXiv:2108.12376 [hep-ex]} \BibitemShut
  {NoStop}%
\bibitem [{\citenamefont {{H1 Collaboration}}(2022)}]{H1prelim-22-031}%
  \BibitemOpen
  \bibfield  {author} {\bibinfo {author} {\bibnamefont {{H1 Collaboration}}},\
  }\href
  {https://www-h1.desy.de/h1/www/publications/htmlsplit/H1prelim-22-031.long.html}
  {\bibfield  {journal} {\bibinfo  {journal} {H1prelim-22-031}\ } (\bibinfo
  {year} {2022})}\BibitemShut {NoStop}%
\bibitem [{\citenamefont {Andreev}\ \emph {et~al.}(2023)\citenamefont {Andreev}
  \emph {et~al.}}]{H1:2023fzk}%
  \BibitemOpen
  \bibfield  {author} {\bibinfo {author} {\bibfnamefont {V.}~\bibnamefont
  {Andreev}} \emph {et~al.} (\bibinfo {collaboration} {H1}),\ }\href@noop {} {\
   (\bibinfo {year} {2023})},\ \Eprint {http://arxiv.org/abs/2303.13620}
  {arXiv:2303.13620 [hep-ex]} \BibitemShut {NoStop}%
\bibitem [{\citenamefont {{H1 Collaboration}}(2023)}]{H1prelim-21-031}%
  \BibitemOpen
  \bibfield  {author} {\bibinfo {author} {\bibnamefont {{H1 Collaboration}}},\
  }\href
  {https://www-h1.desy.de/h1/www/publications/htmlsplit/H1prelim-23-031.long.html}
  {\bibfield  {journal} {\bibinfo  {journal} {H1prelim-23-031}\ } (\bibinfo
  {year} {2023})}\BibitemShut {NoStop}%
\bibitem [{LHC(2022)}]{LHCb:2022rky}%
  \BibitemOpen
  \href@noop {} {\  (\bibinfo {year} {2022})},\ \Eprint
  {http://arxiv.org/abs/2208.11691} {arXiv:2208.11691 [hep-ex]} \BibitemShut
  {NoStop}%
\bibitem [{\citenamefont {Komiske}\ \emph {et~al.}(2022)\citenamefont
  {Komiske}, \citenamefont {Kryhin},\ and\ \citenamefont
  {Thaler}}]{Komiske:2022vxg}%
  \BibitemOpen
  \bibfield  {author} {\bibinfo {author} {\bibfnamefont {P.~T.}\ \bibnamefont
  {Komiske}}, \bibinfo {author} {\bibfnamefont {S.}~\bibnamefont {Kryhin}}, \
  and\ \bibinfo {author} {\bibfnamefont {J.}~\bibnamefont {Thaler}},\ }\href
  {\doibase 10.1103/PhysRevD.106.094021} {\bibfield  {journal} {\bibinfo
  {journal} {Phys. Rev. D}\ }\textbf {\bibinfo {volume} {106}},\ \bibinfo
  {pages} {094021} (\bibinfo {year} {2022})},\ \Eprint
  {http://arxiv.org/abs/2205.04459} {arXiv:2205.04459 [hep-ph]} \BibitemShut
  {NoStop}%
\bibitem [{\citenamefont {Song}(2023)}]{Song:2023sxb}%
  \BibitemOpen
  \bibfield  {author} {\bibinfo {author} {\bibfnamefont {Y.}~\bibnamefont
  {Song}} (\bibinfo {collaboration} {STAR}),\ }\href@noop {} {\  (\bibinfo
  {year} {2023})},\ \Eprint {http://arxiv.org/abs/2307.07718} {arXiv:2307.07718
  [nucl-ex]} \BibitemShut {NoStop}%
\bibitem [{\citenamefont {{Lucy}}(1974)}]{1974AJ.....79..745L}%
  \BibitemOpen
  \bibfield  {author} {\bibinfo {author} {\bibfnamefont {L.~B.}\ \bibnamefont
  {{Lucy}}},\ }\href {\doibase 10.1086/111605} {\bibfield  {journal} {\bibinfo
  {journal} {Astronomical Journal}\ }\textbf {\bibinfo {volume} {79}},\
  \bibinfo {pages} {745} (\bibinfo {year} {1974})}\BibitemShut {NoStop}%
\bibitem [{\citenamefont {Richardson}(1972)}]{Richardson:72}%
  \BibitemOpen
  \bibfield  {author} {\bibinfo {author} {\bibfnamefont {W.~H.}\ \bibnamefont
  {Richardson}},\ }\href {\doibase 10.1364/JOSA.62.000055} {\bibfield
  {journal} {\bibinfo  {journal} {J. Opt. Soc. Am.}\ }\textbf {\bibinfo
  {volume} {62}},\ \bibinfo {pages} {55} (\bibinfo {year} {1972})}\BibitemShut
  {NoStop}%
\bibitem [{\citenamefont {D'Agostini}(1995)}]{DAGOSTINI1995487}%
  \BibitemOpen
  \bibfield  {author} {\bibinfo {author} {\bibfnamefont {G.}~\bibnamefont
  {D'Agostini}},\ }\href {\doibase
  https://doi.org/10.1016/0168-9002(95)00274-X} {\bibfield  {journal} {\bibinfo
   {journal} {Nucl. Instrum. Meth.}\ }\textbf {\bibinfo {volume} {A362}},\
  \bibinfo {pages} {487} (\bibinfo {year} {1995})}\BibitemShut {NoStop}%
\bibitem [{\citenamefont {Rezende}\ and\ \citenamefont
  {Mohamed}(2016)}]{rezende2016variational}%
  \BibitemOpen
  \bibfield  {author} {\bibinfo {author} {\bibfnamefont {D.~J.}\ \bibnamefont
  {Rezende}}\ and\ \bibinfo {author} {\bibfnamefont {S.}~\bibnamefont
  {Mohamed}},\ }\href@noop {} {\ \bibinfo {series} {Proceedings of Machine
  Learning Research},\ \textbf {\bibinfo {volume} {37}},\ \bibinfo {pages}
  {1530} (\bibinfo {year} {2016})},\ \Eprint {http://arxiv.org/abs/1505.05770}
  {arXiv:1505.05770 [stat.ML]} \BibitemShut {NoStop}%
\bibitem [{\citenamefont {Goodfellow}\ \emph {et~al.}(2014)\citenamefont
  {Goodfellow}, \citenamefont {Pouget-Abadie}, \citenamefont {Mirza},
  \citenamefont {Xu}, \citenamefont {Warde-Farley}, \citenamefont {Ozair},
  \citenamefont {Courville},\ and\ \citenamefont
  {Bengio}}]{Goodfellow:2014upx}%
  \BibitemOpen
  \bibfield  {author} {\bibinfo {author} {\bibfnamefont {I.~J.}\ \bibnamefont
  {Goodfellow}}, \bibinfo {author} {\bibfnamefont {J.}~\bibnamefont
  {Pouget-Abadie}}, \bibinfo {author} {\bibfnamefont {M.}~\bibnamefont
  {Mirza}}, \bibinfo {author} {\bibfnamefont {B.}~\bibnamefont {Xu}}, \bibinfo
  {author} {\bibfnamefont {D.}~\bibnamefont {Warde-Farley}}, \bibinfo {author}
  {\bibfnamefont {S.}~\bibnamefont {Ozair}}, \bibinfo {author} {\bibfnamefont
  {A.}~\bibnamefont {Courville}}, \ and\ \bibinfo {author} {\bibfnamefont
  {Y.}~\bibnamefont {Bengio}},\ }\href@noop {} {\  (\bibinfo {year} {2014})},\
  \Eprint {http://arxiv.org/abs/1406.2661} {arXiv:1406.2661 [stat.ML]}
  \BibitemShut {NoStop}%
\bibitem [{\citenamefont {Song}\ \emph {et~al.}(2021)\citenamefont {Song},
  \citenamefont {Sohl-Dickstein}, \citenamefont {Kingma}, \citenamefont
  {Kumar}, \citenamefont {Ermon},\ and\ \citenamefont
  {Poole}}]{Song2021ScoreBasedGM}%
  \BibitemOpen
  \bibfield  {author} {\bibinfo {author} {\bibfnamefont {Y.}~\bibnamefont
  {Song}}, \bibinfo {author} {\bibfnamefont {J.}~\bibnamefont
  {Sohl-Dickstein}}, \bibinfo {author} {\bibfnamefont {D.~P.}\ \bibnamefont
  {Kingma}}, \bibinfo {author} {\bibfnamefont {A.}~\bibnamefont {Kumar}},
  \bibinfo {author} {\bibfnamefont {S.}~\bibnamefont {Ermon}}, \ and\ \bibinfo
  {author} {\bibfnamefont {B.}~\bibnamefont {Poole}},\ }\href@noop {}
  {\bibfield  {journal} {\bibinfo  {journal} {ArXiv}\ }\textbf {\bibinfo
  {volume} {abs/2011.13456}} (\bibinfo {year} {2021})}\BibitemShut {NoStop}%
\bibitem [{\citenamefont {Schr{\"o}dinger}(1931)}]{schrodinger1931umkehrung}%
  \BibitemOpen
  \bibfield  {author} {\bibinfo {author} {\bibfnamefont {E.}~\bibnamefont
  {Schr{\"o}dinger}},\ }\href {\doibase
  https://doi.org/10.1002/ange.19310443014} {\bibfield  {journal} {\bibinfo
  {journal} {{S}itzungsberichte der {P}reuss {A}kad. {W}issen. {P}hys. {M}ath.
  {K}lasse {S}onderausgabe}\ }\textbf {\bibinfo {volume} {9}},\ \bibinfo
  {pages} {144} (\bibinfo {year} {1931})}\BibitemShut {NoStop}%
\bibitem [{\citenamefont {Liu}\ \emph {et~al.}(2023)\citenamefont {Liu},
  \citenamefont {Vahdat}, \citenamefont {Huang}, \citenamefont {Theodorou},
  \citenamefont {Nie},\ and\ \citenamefont {Anandkumar}}]{SB_nvidia}%
  \BibitemOpen
  \bibfield  {author} {\bibinfo {author} {\bibfnamefont {G.-H.}\ \bibnamefont
  {Liu}}, \bibinfo {author} {\bibfnamefont {A.}~\bibnamefont {Vahdat}},
  \bibinfo {author} {\bibfnamefont {D.-A.}\ \bibnamefont {Huang}}, \bibinfo
  {author} {\bibfnamefont {E.~A.}\ \bibnamefont {Theodorou}}, \bibinfo {author}
  {\bibfnamefont {W.}~\bibnamefont {Nie}}, \ and\ \bibinfo {author}
  {\bibfnamefont {A.}~\bibnamefont {Anandkumar}},\ }\href@noop {} {\bibfield
  {journal} {\bibinfo  {journal} {arXiv preprint arXiv:2302.05872}\ } (\bibinfo
  {year} {2023})}\BibitemShut {NoStop}%
\bibitem [{\citenamefont {Ho}\ \emph {et~al.}(2020)\citenamefont {Ho},
  \citenamefont {Jain},\ and\ \citenamefont {Abbeel}}]{ddpm}%
  \BibitemOpen
  \bibfield  {author} {\bibinfo {author} {\bibfnamefont {J.}~\bibnamefont
  {Ho}}, \bibinfo {author} {\bibfnamefont {A.}~\bibnamefont {Jain}}, \ and\
  \bibinfo {author} {\bibfnamefont {P.}~\bibnamefont {Abbeel}},\ }in\ \href
  {https://proceedings.neurips.cc/paper_files/paper/2020/file/4c5bcfec8584af0d967f1ab10179ca4b-Paper.pdf}
  {\emph {\bibinfo {booktitle} {Advances in Neural Information Processing
  Systems}}},\ Vol.~\bibinfo {volume} {33},\ \bibinfo {editor} {edited by\
  \bibinfo {editor} {\bibfnamefont {H.}~\bibnamefont {Larochelle}}, \bibinfo
  {editor} {\bibfnamefont {M.}~\bibnamefont {Ranzato}}, \bibinfo {editor}
  {\bibfnamefont {R.}~\bibnamefont {Hadsell}}, \bibinfo {editor} {\bibfnamefont
  {M.}~\bibnamefont {Balcan}}, \ and\ \bibinfo {editor} {\bibfnamefont
  {H.}~\bibnamefont {Lin}}}\ (\bibinfo  {publisher} {Curran Associates, Inc.},\
  \bibinfo {year} {2020})\ pp.\ \bibinfo {pages} {6840--6851}\BibitemShut
  {NoStop}%
\bibitem [{\citenamefont {Peyr{\'e}}\ \emph {et~al.}(2019)\citenamefont
  {Peyr{\'e}}, \citenamefont {Cuturi} \emph {et~al.}}]{peyre2019computational}%
  \BibitemOpen
  \bibfield  {author} {\bibinfo {author} {\bibfnamefont {G.}~\bibnamefont
  {Peyr{\'e}}}, \bibinfo {author} {\bibfnamefont {M.}~\bibnamefont {Cuturi}},
  \emph {et~al.},\ }\href@noop {} {\bibfield  {journal} {\bibinfo  {journal}
  {Foundations and Trends{\textregistered} in Machine Learning}\ }\textbf
  {\bibinfo {volume} {11}},\ \bibinfo {pages} {355} (\bibinfo {year}
  {2019})}\BibitemShut {NoStop}%
\bibitem [{\citenamefont {Andreassen}\ \emph {et~al.}(2019)\citenamefont
  {Andreassen}, \citenamefont {Komiske}, \citenamefont {Metodiev},
  \citenamefont {Nachman},\ and\ \citenamefont
  {Thaler}}]{andreassen_anders_2019_3548091}%
  \BibitemOpen
  \bibfield  {author} {\bibinfo {author} {\bibfnamefont {A.}~\bibnamefont
  {Andreassen}}, \bibinfo {author} {\bibfnamefont {P.}~\bibnamefont {Komiske}},
  \bibinfo {author} {\bibfnamefont {E.}~\bibnamefont {Metodiev}}, \bibinfo
  {author} {\bibfnamefont {B.}~\bibnamefont {Nachman}}, \ and\ \bibinfo
  {author} {\bibfnamefont {J.}~\bibnamefont {Thaler}},\ }\href {\doibase
  10.5281/zenodo.3548091} {\enquote {\bibinfo {title} {{Pythia/Herwig + Delphes
  Jet Datasets for OmniFold Unfolding}},}\ } (\bibinfo {year}
  {2019})\BibitemShut {NoStop}%
\bibitem [{\citenamefont {Bahr}\ \emph {et~al.}(2008)\citenamefont {Bahr} \emph
  {et~al.}}]{Bahr:2008pv}%
  \BibitemOpen
  \bibfield  {author} {\bibinfo {author} {\bibfnamefont {M.}~\bibnamefont
  {Bahr}} \emph {et~al.},\ }\href {\doibase 10.1140/epjc/s10052-008-0798-9}
  {\bibfield  {journal} {\bibinfo  {journal} {Eur. Phys. J.}\ }\textbf
  {\bibinfo {volume} {C58}},\ \bibinfo {pages} {639} (\bibinfo {year}
  {2008})},\ \Eprint {http://arxiv.org/abs/0803.0883} {arXiv:0803.0883
  [hep-ph]} \BibitemShut {NoStop}%
\bibitem [{\citenamefont {Bellm}\ \emph {et~al.}(2016)\citenamefont {Bellm}
  \emph {et~al.}}]{Bellm:2015jjp}%
  \BibitemOpen
  \bibfield  {author} {\bibinfo {author} {\bibfnamefont {J.}~\bibnamefont
  {Bellm}} \emph {et~al.},\ }\href {\doibase 10.1140/epjc/s10052-016-4018-8}
  {\bibfield  {journal} {\bibinfo  {journal} {Eur. Phys. J.}\ }\textbf
  {\bibinfo {volume} {C76}},\ \bibinfo {pages} {196} (\bibinfo {year}
  {2016})},\ \Eprint {http://arxiv.org/abs/1512.01178} {arXiv:1512.01178
  [hep-ph]} \BibitemShut {NoStop}%
\bibitem [{\citenamefont {Bellm}\ \emph {et~al.}(2017)\citenamefont {Bellm}
  \emph {et~al.}}]{Bellm:2017bvx}%
  \BibitemOpen
  \bibfield  {author} {\bibinfo {author} {\bibfnamefont {J.}~\bibnamefont
  {Bellm}} \emph {et~al.},\ }\href@noop {} {\  (\bibinfo {year} {2017})},\
  \Eprint {http://arxiv.org/abs/1705.06919} {arXiv:1705.06919 [hep-ph]}
  \BibitemShut {NoStop}%
\bibitem [{ATL(2014)}]{ATL-PHYS-PUB-2014-021}%
  \BibitemOpen
  \href {https://cds.cern.ch/record/1966419} {\emph {\bibinfo {title} {{ATLAS
  Run 1 Pythia8 tunes}}}},\ \bibinfo {type} {Tech. Rep.}\ \bibinfo {number}
  {ATL-PHYS-PUB-2014-021}\ (\bibinfo  {institution} {CERN},\ \bibinfo {address}
  {Geneva},\ \bibinfo {year} {2014})\BibitemShut {NoStop}%
\bibitem [{\citenamefont {Sj{\"o}strand}\ \emph {et~al.}(2008)\citenamefont
  {Sj{\"o}strand}, \citenamefont {Mrenna},\ and\ \citenamefont
  {Skands}}]{Sjostrand:2007gs}%
  \BibitemOpen
  \bibfield  {author} {\bibinfo {author} {\bibfnamefont {T.}~\bibnamefont
  {Sj{\"o}strand}}, \bibinfo {author} {\bibfnamefont {S.}~\bibnamefont
  {Mrenna}}, \ and\ \bibinfo {author} {\bibfnamefont {P.~Z.}\ \bibnamefont
  {Skands}},\ }\href {\doibase 10.1016/j.cpc.2008.01.036} {\bibfield  {journal}
  {\bibinfo  {journal} {Comput. Phys. Commun.}\ }\textbf {\bibinfo {volume}
  {178}},\ \bibinfo {pages} {852} (\bibinfo {year} {2008})},\ \Eprint
  {http://arxiv.org/abs/0710.3820} {arXiv:0710.3820 [hep-ph]} \BibitemShut
  {NoStop}%
\bibitem [{\citenamefont {Sj{\"o}strand}\ \emph {et~al.}(2006)\citenamefont
  {Sj{\"o}strand}, \citenamefont {Mrenna},\ and\ \citenamefont
  {Skands}}]{Sjostrand:2006za}%
  \BibitemOpen
  \bibfield  {author} {\bibinfo {author} {\bibfnamefont {T.}~\bibnamefont
  {Sj{\"o}strand}}, \bibinfo {author} {\bibfnamefont {S.}~\bibnamefont
  {Mrenna}}, \ and\ \bibinfo {author} {\bibfnamefont {P.~Z.}\ \bibnamefont
  {Skands}},\ }\href {\doibase 10.1088/1126-6708/2006/05/026} {\bibfield
  {journal} {\bibinfo  {journal} {JHEP}\ }\textbf {\bibinfo {volume} {05}},\
  \bibinfo {pages} {026} (\bibinfo {year} {2006})},\ \Eprint
  {http://arxiv.org/abs/hep-ph/0603175} {arXiv:hep-ph/0603175 [hep-ph]}
  \BibitemShut {NoStop}%
\bibitem [{\citenamefont {Sj{\"o}strand}\ \emph {et~al.}(2015)\citenamefont
  {Sj{\"o}strand}, \citenamefont {Ask}, \citenamefont {Christiansen},
  \citenamefont {Corke}, \citenamefont {Desai}, \citenamefont {Ilten},
  \citenamefont {Mrenna}, \citenamefont {Prestel}, \citenamefont {Rasmussen},\
  and\ \citenamefont {Skands}}]{Sjostrand:2014zea}%
  \BibitemOpen
  \bibfield  {author} {\bibinfo {author} {\bibfnamefont {T.}~\bibnamefont
  {Sj{\"o}strand}}, \bibinfo {author} {\bibfnamefont {S.}~\bibnamefont {Ask}},
  \bibinfo {author} {\bibfnamefont {J.~R.}\ \bibnamefont {Christiansen}},
  \bibinfo {author} {\bibfnamefont {R.}~\bibnamefont {Corke}}, \bibinfo
  {author} {\bibfnamefont {N.}~\bibnamefont {Desai}}, \bibinfo {author}
  {\bibfnamefont {P.}~\bibnamefont {Ilten}}, \bibinfo {author} {\bibfnamefont
  {S.}~\bibnamefont {Mrenna}}, \bibinfo {author} {\bibfnamefont
  {S.}~\bibnamefont {Prestel}}, \bibinfo {author} {\bibfnamefont {C.~O.}\
  \bibnamefont {Rasmussen}}, \ and\ \bibinfo {author} {\bibfnamefont {P.~Z.}\
  \bibnamefont {Skands}},\ }\href {\doibase 10.1016/j.cpc.2015.01.024}
  {\bibfield  {journal} {\bibinfo  {journal} {Comput. Phys. Commun.}\ }\textbf
  {\bibinfo {volume} {191}},\ \bibinfo {pages} {159} (\bibinfo {year}
  {2015})},\ \Eprint {http://arxiv.org/abs/1410.3012} {arXiv:1410.3012
  [hep-ph]} \BibitemShut {NoStop}%
\bibitem [{\citenamefont {de~Favereau}\ \emph {et~al.}(2014)\citenamefont
  {de~Favereau}, \citenamefont {Delaere}, \citenamefont {Demin}, \citenamefont
  {Giammanco}, \citenamefont {Lemaître}, \citenamefont {Mertens},\ and\
  \citenamefont {Selvaggi}}]{deFavereau:2013fsa}%
  \BibitemOpen
  \bibfield  {author} {\bibinfo {author} {\bibfnamefont {J.}~\bibnamefont
  {de~Favereau}}, \bibinfo {author} {\bibfnamefont {C.}~\bibnamefont
  {Delaere}}, \bibinfo {author} {\bibfnamefont {P.}~\bibnamefont {Demin}},
  \bibinfo {author} {\bibfnamefont {A.}~\bibnamefont {Giammanco}}, \bibinfo
  {author} {\bibfnamefont {V.}~\bibnamefont {Lemaître}}, \bibinfo {author}
  {\bibfnamefont {A.}~\bibnamefont {Mertens}}, \ and\ \bibinfo {author}
  {\bibfnamefont {M.}~\bibnamefont {Selvaggi}} (\bibinfo {collaboration}
  {DELPHES 3}),\ }\href {\doibase 10.1007/JHEP02(2014)057} {\bibfield
  {journal} {\bibinfo  {journal} {JHEP}\ }\textbf {\bibinfo {volume} {02}},\
  \bibinfo {pages} {057} (\bibinfo {year} {2014})},\ \Eprint
  {http://arxiv.org/abs/1307.6346} {arXiv:1307.6346 [hep-ex]} \BibitemShut
  {NoStop}%
\bibitem [{\citenamefont {Cacciari}\ \emph {et~al.}(2008)\citenamefont
  {Cacciari}, \citenamefont {Salam},\ and\ \citenamefont
  {Soyez}}]{Cacciari:2008gp}%
  \BibitemOpen
  \bibfield  {author} {\bibinfo {author} {\bibfnamefont {M.}~\bibnamefont
  {Cacciari}}, \bibinfo {author} {\bibfnamefont {G.~P.}\ \bibnamefont {Salam}},
  \ and\ \bibinfo {author} {\bibfnamefont {G.}~\bibnamefont {Soyez}},\ }\href
  {\doibase 10.1088/1126-6708/2008/04/063} {\bibfield  {journal} {\bibinfo
  {journal} {JHEP}\ }\textbf {\bibinfo {volume} {04}},\ \bibinfo {pages} {063}
  (\bibinfo {year} {2008})},\ \Eprint {http://arxiv.org/abs/0802.1189}
  {arXiv:0802.1189 [hep-ph]} \BibitemShut {NoStop}%
\bibitem [{\citenamefont {Cacciari}\ \emph {et~al.}(2012)\citenamefont
  {Cacciari}, \citenamefont {Salam},\ and\ \citenamefont
  {Soyez}}]{Cacciari:2011ma}%
  \BibitemOpen
  \bibfield  {author} {\bibinfo {author} {\bibfnamefont {M.}~\bibnamefont
  {Cacciari}}, \bibinfo {author} {\bibfnamefont {G.~P.}\ \bibnamefont {Salam}},
  \ and\ \bibinfo {author} {\bibfnamefont {G.}~\bibnamefont {Soyez}},\ }\href
  {\doibase 10.1140/epjc/s10052-012-1896-2} {\bibfield  {journal} {\bibinfo
  {journal} {Eur. Phys. J.}\ }\textbf {\bibinfo {volume} {C72}},\ \bibinfo
  {pages} {1896} (\bibinfo {year} {2012})},\ \Eprint
  {http://arxiv.org/abs/1111.6097} {arXiv:1111.6097 [hep-ph]} \BibitemShut
  {NoStop}%
\bibitem [{\citenamefont {Cacciari}\ and\ \citenamefont
  {Salam}(2006)}]{Cacciari:2005hq}%
  \BibitemOpen
  \bibfield  {author} {\bibinfo {author} {\bibfnamefont {M.}~\bibnamefont
  {Cacciari}}\ and\ \bibinfo {author} {\bibfnamefont {G.~P.}\ \bibnamefont
  {Salam}},\ }\href {\doibase 10.1016/j.physletb.2006.08.037} {\bibfield
  {journal} {\bibinfo  {journal} {Phys. Lett.}\ }\textbf {\bibinfo {volume}
  {B641}},\ \bibinfo {pages} {57} (\bibinfo {year} {2006})},\ \Eprint
  {http://arxiv.org/abs/hep-ph/0512210} {arXiv:hep-ph/0512210 [hep-ph]}
  \BibitemShut {NoStop}%
\bibitem [{\citenamefont {Thaler}\ and\ \citenamefont
  {Van~Tilburg}(2011)}]{Thaler:2010tr}%
  \BibitemOpen
  \bibfield  {author} {\bibinfo {author} {\bibfnamefont {J.}~\bibnamefont
  {Thaler}}\ and\ \bibinfo {author} {\bibfnamefont {K.}~\bibnamefont
  {Van~Tilburg}},\ }\href {\doibase 10.1007/JHEP03(2011)015} {\bibfield
  {journal} {\bibinfo  {journal} {JHEP}\ }\textbf {\bibinfo {volume} {03}},\
  \bibinfo {pages} {015} (\bibinfo {year} {2011})},\ \Eprint
  {http://arxiv.org/abs/1011.2268} {arXiv:1011.2268 [hep-ph]} \BibitemShut
  {NoStop}%
\bibitem [{\citenamefont {Thaler}\ and\ \citenamefont
  {Van~Tilburg}(2012)}]{Thaler:2011gf}%
  \BibitemOpen
  \bibfield  {author} {\bibinfo {author} {\bibfnamefont {J.}~\bibnamefont
  {Thaler}}\ and\ \bibinfo {author} {\bibfnamefont {K.}~\bibnamefont
  {Van~Tilburg}},\ }\href {\doibase 10.1007/JHEP02(2012)093} {\bibfield
  {journal} {\bibinfo  {journal} {JHEP}\ }\textbf {\bibinfo {volume} {02}},\
  \bibinfo {pages} {093} (\bibinfo {year} {2012})},\ \Eprint
  {http://arxiv.org/abs/1108.2701} {arXiv:1108.2701 [hep-ph]} \BibitemShut
  {NoStop}%
\bibitem [{\citenamefont {Larkoski}\ \emph {et~al.}(2014)\citenamefont
  {Larkoski}, \citenamefont {Marzani}, \citenamefont {Soyez},\ and\
  \citenamefont {Thaler}}]{Larkoski:2014wba}%
  \BibitemOpen
  \bibfield  {author} {\bibinfo {author} {\bibfnamefont {A.~J.}\ \bibnamefont
  {Larkoski}}, \bibinfo {author} {\bibfnamefont {S.}~\bibnamefont {Marzani}},
  \bibinfo {author} {\bibfnamefont {G.}~\bibnamefont {Soyez}}, \ and\ \bibinfo
  {author} {\bibfnamefont {J.}~\bibnamefont {Thaler}},\ }\href {\doibase
  10.1007/JHEP05(2014)146} {\bibfield  {journal} {\bibinfo  {journal} {JHEP}\
  }\textbf {\bibinfo {volume} {05}},\ \bibinfo {pages} {146} (\bibinfo {year}
  {2014})},\ \Eprint {http://arxiv.org/abs/1402.2657} {arXiv:1402.2657
  [hep-ph]} \BibitemShut {NoStop}%
\bibitem [{\citenamefont {Dasgupta}\ \emph {et~al.}(2013)\citenamefont
  {Dasgupta}, \citenamefont {Fregoso}, \citenamefont {Marzani},\ and\
  \citenamefont {Salam}}]{Dasgupta:2013ihk}%
  \BibitemOpen
  \bibfield  {author} {\bibinfo {author} {\bibfnamefont {M.}~\bibnamefont
  {Dasgupta}}, \bibinfo {author} {\bibfnamefont {A.}~\bibnamefont {Fregoso}},
  \bibinfo {author} {\bibfnamefont {S.}~\bibnamefont {Marzani}}, \ and\
  \bibinfo {author} {\bibfnamefont {G.~P.}\ \bibnamefont {Salam}},\ }\href
  {\doibase 10.1007/JHEP09(2013)029} {\bibfield  {journal} {\bibinfo  {journal}
  {JHEP}\ }\textbf {\bibinfo {volume} {09}},\ \bibinfo {pages} {029} (\bibinfo
  {year} {2013})},\ \Eprint {http://arxiv.org/abs/1307.0007} {arXiv:1307.0007
  [hep-ph]} \BibitemShut {NoStop}%
\bibitem [{fjc()}]{fjcontrib}%
  \BibitemOpen
  \href@noop {} {\enquote {\bibinfo {title} {Fastjet contrib},}\ }\bibinfo
  {howpublished} {\url{https://fastjet.hepforge.org/contrib/}}\BibitemShut
  {NoStop}%
\bibitem [{\citenamefont {He}\ \emph {et~al.}(2016)\citenamefont {He},
  \citenamefont {Zhang}, \citenamefont {Ren},\ and\ \citenamefont
  {Sun}}]{he2016deep}%
  \BibitemOpen
  \bibfield  {author} {\bibinfo {author} {\bibfnamefont {K.}~\bibnamefont
  {He}}, \bibinfo {author} {\bibfnamefont {X.}~\bibnamefont {Zhang}}, \bibinfo
  {author} {\bibfnamefont {S.}~\bibnamefont {Ren}}, \ and\ \bibinfo {author}
  {\bibfnamefont {J.}~\bibnamefont {Sun}},\ }in\ \href@noop {} {\emph {\bibinfo
  {booktitle} {Proceedings of the IEEE conference on computer vision and
  pattern recognition}}}\ (\bibinfo {year} {2016})\ pp.\ \bibinfo {pages}
  {770--778}\BibitemShut {NoStop}%
\bibitem [{\citenamefont {Xu}\ \emph {et~al.}(2015)\citenamefont {Xu},
  \citenamefont {Wang}, \citenamefont {Chen},\ and\ \citenamefont
  {Li}}]{DBLP:journals/corr/XuWCL15}%
  \BibitemOpen
  \bibfield  {author} {\bibinfo {author} {\bibfnamefont {B.}~\bibnamefont
  {Xu}}, \bibinfo {author} {\bibfnamefont {N.}~\bibnamefont {Wang}}, \bibinfo
  {author} {\bibfnamefont {T.}~\bibnamefont {Chen}}, \ and\ \bibinfo {author}
  {\bibfnamefont {M.}~\bibnamefont {Li}},\ }\href
  {http://arxiv.org/abs/1505.00853} {\bibfield  {journal} {\bibinfo  {journal}
  {CoRR}\ }\textbf {\bibinfo {volume} {abs/1505.00853}} (\bibinfo {year}
  {2015})},\ \Eprint {http://arxiv.org/abs/1505.00853} {1505.00853}
  \BibitemShut {NoStop}%
\bibitem [{\citenamefont {Paszke}\ \emph {et~al.}(2019)\citenamefont {Paszke},
  \citenamefont {Gross}, \citenamefont {Massa}, \citenamefont {Lerer},
  \citenamefont {Bradbury}, \citenamefont {Chanan}, \citenamefont {Killeen},
  \citenamefont {Lin}, \citenamefont {Gimelshein}, \citenamefont {Antiga} \emph
  {et~al.}}]{pytorch}%
  \BibitemOpen
  \bibfield  {author} {\bibinfo {author} {\bibfnamefont {A.}~\bibnamefont
  {Paszke}}, \bibinfo {author} {\bibfnamefont {S.}~\bibnamefont {Gross}},
  \bibinfo {author} {\bibfnamefont {F.}~\bibnamefont {Massa}}, \bibinfo
  {author} {\bibfnamefont {A.}~\bibnamefont {Lerer}}, \bibinfo {author}
  {\bibfnamefont {J.}~\bibnamefont {Bradbury}}, \bibinfo {author}
  {\bibfnamefont {G.}~\bibnamefont {Chanan}}, \bibinfo {author} {\bibfnamefont
  {T.}~\bibnamefont {Killeen}}, \bibinfo {author} {\bibfnamefont
  {Z.}~\bibnamefont {Lin}}, \bibinfo {author} {\bibfnamefont {N.}~\bibnamefont
  {Gimelshein}}, \bibinfo {author} {\bibfnamefont {L.}~\bibnamefont {Antiga}},
  \emph {et~al.},\ }\href@noop {} {\bibfield  {journal} {\bibinfo  {journal}
  {Advances in neural information processing systems}\ }\textbf {\bibinfo
  {volume} {32}} (\bibinfo {year} {2019})}\BibitemShut {NoStop}%
\bibitem [{\citenamefont {Abadi}\ \emph {et~al.}(2016)\citenamefont {Abadi},
  \citenamefont {Barham}, \citenamefont {Chen}, \citenamefont {Chen},
  \citenamefont {Davis}, \citenamefont {Dean}, \citenamefont {Devin},
  \citenamefont {Ghemawat}, \citenamefont {Irving}, \citenamefont {Isard} \emph
  {et~al.}}]{tensorflow}%
  \BibitemOpen
  \bibfield  {author} {\bibinfo {author} {\bibfnamefont {M.}~\bibnamefont
  {Abadi}}, \bibinfo {author} {\bibfnamefont {P.}~\bibnamefont {Barham}},
  \bibinfo {author} {\bibfnamefont {J.}~\bibnamefont {Chen}}, \bibinfo {author}
  {\bibfnamefont {Z.}~\bibnamefont {Chen}}, \bibinfo {author} {\bibfnamefont
  {A.}~\bibnamefont {Davis}}, \bibinfo {author} {\bibfnamefont
  {J.}~\bibnamefont {Dean}}, \bibinfo {author} {\bibfnamefont {M.}~\bibnamefont
  {Devin}}, \bibinfo {author} {\bibfnamefont {S.}~\bibnamefont {Ghemawat}},
  \bibinfo {author} {\bibfnamefont {G.}~\bibnamefont {Irving}}, \bibinfo
  {author} {\bibfnamefont {M.}~\bibnamefont {Isard}},  \emph {et~al.},\ }in\
  \href@noop {} {\emph {\bibinfo {booktitle} {OSDI}}},\ Vol.~\bibinfo {volume}
  {16}\ (\bibinfo {year} {2016})\ pp.\ \bibinfo {pages} {265--283}\BibitemShut
  {NoStop}%
\bibitem [{\citenamefont {Topsoe}(2000)}]{850703}%
  \BibitemOpen
  \bibfield  {author} {\bibinfo {author} {\bibfnamefont {F.}~\bibnamefont
  {Topsoe}},\ }\href {\doibase 10.1109/18.850703} {\bibfield  {journal}
  {\bibinfo  {journal} {IEEE Transactions on Information Theory}\ }\textbf
  {\bibinfo {volume} {46}},\ \bibinfo {pages} {1602} (\bibinfo {year}
  {2000})}\BibitemShut {NoStop}%
\bibitem [{\citenamefont {Gras}\ \emph {et~al.}(2017)\citenamefont {Gras},
  \citenamefont {H\"oche}, \citenamefont {Kar}, \citenamefont {Larkoski},
  \citenamefont {L\"onnblad}, \citenamefont {Pl\"atzer}, \citenamefont
  {Si\'odmok}, \citenamefont {Skands}, \citenamefont {Soyez},\ and\
  \citenamefont {Thaler}}]{Gras:2017jty}%
  \BibitemOpen
  \bibfield  {author} {\bibinfo {author} {\bibfnamefont {P.}~\bibnamefont
  {Gras}}, \bibinfo {author} {\bibfnamefont {S.}~\bibnamefont {H\"oche}},
  \bibinfo {author} {\bibfnamefont {D.}~\bibnamefont {Kar}}, \bibinfo {author}
  {\bibfnamefont {A.}~\bibnamefont {Larkoski}}, \bibinfo {author}
  {\bibfnamefont {L.}~\bibnamefont {L\"onnblad}}, \bibinfo {author}
  {\bibfnamefont {S.}~\bibnamefont {Pl\"atzer}}, \bibinfo {author}
  {\bibfnamefont {A.}~\bibnamefont {Si\'odmok}}, \bibinfo {author}
  {\bibfnamefont {P.}~\bibnamefont {Skands}}, \bibinfo {author} {\bibfnamefont
  {G.}~\bibnamefont {Soyez}}, \ and\ \bibinfo {author} {\bibfnamefont
  {J.}~\bibnamefont {Thaler}},\ }\href {\doibase 10.1007/JHEP07(2017)091}
  {\bibfield  {journal} {\bibinfo  {journal} {JHEP}\ }\textbf {\bibinfo
  {volume} {07}},\ \bibinfo {pages} {091} (\bibinfo {year} {2017})},\ \Eprint
  {http://arxiv.org/abs/1704.03878} {arXiv:1704.03878 [hep-ph]} \BibitemShut
  {NoStop}%
\bibitem [{\citenamefont {Bright-Thonney}\ and\ \citenamefont
  {Nachman}(2019)}]{Bright-Thonney:2018mxq}%
  \BibitemOpen
  \bibfield  {author} {\bibinfo {author} {\bibfnamefont {S.}~\bibnamefont
  {Bright-Thonney}}\ and\ \bibinfo {author} {\bibfnamefont {B.}~\bibnamefont
  {Nachman}},\ }\href {\doibase 10.1007/JHEP03(2019)098} {\bibfield  {journal}
  {\bibinfo  {journal} {JHEP}\ }\textbf {\bibinfo {volume} {03}},\ \bibinfo
  {pages} {098} (\bibinfo {year} {2019})},\ \Eprint
  {http://arxiv.org/abs/1810.05653} {arXiv:1810.05653 [hep-ph]} \BibitemShut
  {NoStop}%
\bibitem [{\citenamefont {Mikuni}\ \emph {et~al.}(2023)\citenamefont {Mikuni},
  \citenamefont {Nachman},\ and\ \citenamefont {Pettee}}]{PhysRevD.108.036025}%
  \BibitemOpen
  \bibfield  {author} {\bibinfo {author} {\bibfnamefont {V.}~\bibnamefont
  {Mikuni}}, \bibinfo {author} {\bibfnamefont {B.}~\bibnamefont {Nachman}}, \
  and\ \bibinfo {author} {\bibfnamefont {M.}~\bibnamefont {Pettee}},\ }\href
  {\doibase 10.1103/PhysRevD.108.036025} {\bibfield  {journal} {\bibinfo
  {journal} {Phys. Rev. D}\ }\textbf {\bibinfo {volume} {108}},\ \bibinfo
  {pages} {036025} (\bibinfo {year} {2023})}\BibitemShut {NoStop}%
\end{thebibliography}%
\bibliographystyle{apsrev4-1}

\clearpage

\end{document}